\documentclass[11pt,preprint]{aastex}
\usepackage{graphicx}
\usepackage{soul,color}
\usepackage{txfonts}
\usepackage[normalem]{ulem} 

\newcommand{\rph}{\ensuremath{r_{\mathrm{ph}}}}
\newcommand{\Feddi}{\ensuremath{F_{\mathrm{Edd,\infty}}}}
\newcommand{\Fedd}{\ensuremath{F_{\mathrm{Edd}}}}
\newcommand{\Ftdi}{\ensuremath{F_{\mathrm{TD,\infty}}}}
\newcommand{\Ftd}{\ensuremath{F_{\mathrm{TD}}}}
\newcommand{\Tbb}{\ensuremath{T_{\mathrm{bb}}}}
\newcommand{\Tbbi}{\ensuremath{T_{\mathrm{bb,\infty}}}}
\newcommand{\Rinf}{\ensuremath{R_{\infty}}}
\newcommand{\Finf}{\ensuremath{F_{\infty}}}

\newcommand{\Aunit}{\ensuremath{\mathrm{km^{2}\,kpc^{-2}}}}
\newcommand{\fluxunit}{\ensuremath{\mathrm{ergs\,cm^{-2}\,s^{-2}}}}
\newcommand{\Msun}{\ensuremath{M_{\odot}}}
\begin{document}

\title{The Equation of State from Observed Masses and Radii of Neutron Stars}

\author{Andrew W. Steiner}
\affil{Department of Physics \& Astronomy, National
Superconducting Cyclotron Laboratory, and the Joint Institute for
Nuclear Astrophysics, Michigan State University, East  Lansing, MI}
\author{James M. Lattimer}
\affil{Department of Physics \& Astronomy, Stony Brook University, 
Stony Brook, NY}
\and
\author{Edward F. Brown}
\affil{Department of Physics \& Astronomy, National
Superconducting Cyclotron Laboratory, and the Joint Institute for
Nuclear Astrophysics, Michigan State University, East  Lansing, MI}
\email{steinera@pa.msu.edu; lattimer@mail.astro.sunysb.edu; ebrown@pa.msu.edu}

\begin{abstract}
We determine an empirical dense matter equation of state from a
heterogeneous dataset of six neutron stars: three type I X-ray
bursters with photospheric radius expansion, studied by \"Ozel et
al., and three transient low-mass X-ray binaries. We critically assess
the mass and radius determinations from the X-ray burst sources and
show explicitly how systematic uncertainties, such as the photospheric
radius at touchdown, affect the most probable masses and radii.
We introduce a parameterized equation of state and use a Markov Chain
Monte Carlo algorithm within a Bayesian framework to determine nuclear
parameters such as the incompressibility and the density dependence of
the bulk symmetry energy. Using this framework we show, for the first
time, that these parameters, predicted solely on the basis of
astrophysical observations, all lie in ranges expected from nuclear
systematics and laboratory experiments. We find significant
constraints on the mass-radius relation for neutron stars, and hence
on the pressure-density relation of dense matter. The predicted
symmetry energy and the equation of state near the saturation density
are soft, resulting in relatively small neutron star radii around
11--12~km for $M=1.4\,\Msun$. The predicted equation of state stiffens
at higher densities, however, and our preferred model for X-ray bursts
suggests that the neutron star maximum mass is relatively large,
1.9--2.2$\,\Msun$. Our results imply that several commonly used
equations of state are inconsistent with observations.
\end{abstract}

\keywords{dense matter --- stars: neutron --- X-rays: 
binaries --- X-rays: bursts }


\section{Introduction}

The masses and radii of neutron stars are determined by the
pressure-energy density relation (equation of state; EOS hereafter) of
cold dense matter using the familiar Tolman-Oppenheimer-Volkov
\citep[TOV;][]{tolman:1939,oppenheimer39} relativistic stellar
structure equations. Within tens of seconds after birth, the neutron
star is cold (meaning temperature much less than Fermi energy) and
deleptonized (meaning in beta equilibrium with no trapped neutrinos);
as a result, for a given EOS the mass and radius of the star depend
only on the central density. Inversion of the structure equations,
given simultaneous mass and radius measurements, can therefore
constrain the pressure-density relation, although the quality of the
constraints are very sensitive to the mass and radius measurement
uncertainties.

A host of observable phenomena and experimental information is
becoming available~\citep[for a recent review,
  see][]{Lattimer2006Neutron-Star-Ob}: on the observational side,
pulsar timing, thermal emission from cooling neutron stars, surface
explosions, and gravity wave emissions are some promising areas for
measurements of mass and radius; on the experimental side, heavy-ion
collisions, giant dipole resonances, and parity-violating electron
scattering are some promising techniques for measuring the density
dependence of the pressure of nuclear matter. These efforts are
complementary, and determining an EOS from the many disparate
measurements is a challenging task.

This paper consists of two parts. In the first, we consider
simultaneous mass and radius information from astrophysical
observations of X-ray bursts and thermal emissions from quiescent low-mass X-ray binaries. We critically
assess the mass and radius constraints determined from X-ray bursts
that may reach the Eddington limit (\S~\ref{s.pre-bursts})~
\citep{vanParadijs79,vanParadijs82,Paczynski86,vanparadijs.ea:very_energetic}.
Heretofore, these bursts have been interpreted assuming that the
photospheric radius is equal to the stellar radius at ``touchdown'',
when the effective temperature reaches a
maximum~\citep{Ozel06,Ozel09,Guver10,Guver10b,Ozel10}. If true, rather
stringent constraints on mass and radius are predicted. Using the most
probable values for the observed flux and angular emission area,
however, does not result in real-valued masses and radii.
\citet{Ozel10} argues for rejecting much of the observed phase space
on this basis and thus obtains tight constraints on masses and radii.
We show that this model is not internally consistent, but that
consistency can be regained by relaxing the assumption concerning the
effective photospheric radius. With this modification, the inferred
neutron star radii moderately increase and confidence intervals for
predicted masses and radii become substantially larger than those
previously quoted~\citep{Ozel06,Ozel09,Guver10,Guver10b,Ozel10}.

We then discuss (\S~\ref{sec:spectral-fits}) mass and radius estimates
from thermal emission from quiescent
low-mass X-ray binaries (LMXBs hereafter). Neutron stars in quiescent low-mass
X-ray binaries may be used to obtain angular emission areas, and
thereby mass and radius information, from their quiescent thermal
emission. In total, we have mass and radius constraints for six
neutron stars: three bursting sources with photospheric radius
expansion bursts and three quiescent neutron star transients.

While not all of the uncertainties involved in constraining the masses
and radii of neutron stars are under control, it is important to {\em
  quantify} the constraints on the EOS which are implied by the
observations. In the second part of this paper (\S~\ref{sect:bayes}),
we use a Bayesian analysis to combine these mass and radius
constraints to determine empirical pressure-density and neutron star
mass-radius relations. We include constraints to the EOS from
causality, the observed minimum value of the neutron star maximum
mass, and the observed maximum pulsar spin frequency. We employ a
parameterized EOS (\S~\ref{sec:eos-parameterization}) that is
compatible with laboratory measurements, and use it with a Markov
Chain Monte Carlo (MC) algorithm to jointly fit these mass-radius
constraints (\S~\ref{sec:bayes-results}). We show, for the first time,
using astrophysical observations alone, that values of the nuclear
parameters such as the incompressibility $K$, bulk symmetry energy
$S_v$, and the density dependence of the symmetry energy $\gamma$, all
lie in ranges expected from nuclear systematics and laboratory
experiments. Furthermore, our results imply that the most likely
neutron star radius is relatively small, of order 11--12~km for
neutron stars with masses near $1.4\,\Msun$, so that the predicted EOS
is relatively soft in the density range 1--3 times the nuclear
saturation density. Implications for the maximum mass sensitively
depend on assumptions concerning X-ray burst models and are discussed
in \S~\ref{sec:bayes-results2}. In \S 5 we discuss astrophysical and
nuclear physical implications of our results, and compare our methods
and results with other studies.

\section{Mass and radius constraints from photospheric radius
  expansion bursts}\label{s.pre-bursts}

Although a few dozen neutron star masses have been determined very
accurately (to within a few percent) in binaries containing pulsars
\citep[for a recent compilation, see][]{Lattimer05b},
no radius information is available for these systems. For systems in
which the neutron star accretes matter from a nearby companion,
nuclear processes in the crust and envelope of the neutron star
provide additional observables that can be used to constrain its mass
and radius. Our discussion here initially follows that of
\citet{Ozel06}, and is provided to establish a
framework for the subsequent assessment of this method in
\S\S~\ref{sec:mc-pre} and \ref{s.rph-touchdown}.

Type I X-ray bursts are the result of thermally unstable helium (or in
some cases, hydrogen) ignition in the accreted envelope of a neutron
star~\citep[for a review, see][]{strohmayer03:review}. The ignition
generates a thermonuclear explosion that is observed as an X-ray burst
with a rapid rise ($\sim 1\,\mathrm{s}$) followed by a slower
decay ($\sim 10\textrm{--}100\,\mathrm{s}$). With the discovery of
significant spectral softening during some bursts it quickly became
apparent that significant radial expansion of the photosphere can
occur during powerful X-ray bursts: if the burst is sufficiently luminous,
radiation pressure drives the photosphere outwards to larger radii, in
some cases substantially so. About 20\% of X-ray bursts show
evidence for photospheric radius expansion \citep[hereafter
  PRE;][]{Galloway08}. The inference that radius expansion
occurred spurred many
\citep[e.g.,][]{paczynski83,Ebisuzaki1983Mass-loss-from-} to construct
models of extended, radiation-dominated neutron star envelopes. There
is widespread agreement from these calculations that during a radius
expansion burst, the flux at the photosphere approaches (to within a
few percent) the Eddington value. The convective zone is expected to
reach the photosphere during a powerful burst
\citep{woosley.heger.ea:models,Weinberg2005Exposing-the-Nu} thereby
polluting the accreted material with heavier nuclei synthesized during
the burst.

During a typical PRE burst, the flux rapidly increases, peaks, and
then decreases. While the flux $F_\infty$ is near maximum, the
blackbody temperature $\Tbbi$ at first decreases, then increases to a
maximum before decreasing again. (The $\infty$ subscript indicates
that the quantities are observed at the Earth and therefore differ
from their value in a reference frame local to the emission.) During
this time, the normalized area $F_{\infty}/\Tbbi^{4}$ increases to a
maximum value and then decreases as $\Tbbi$ increases. The point at
which $\Tbbi$ reaches a maximum (and the normalized area typically
stops decreasing and becomes constant) is thought to be when the
photosphere ``touches down'' at the stellar radius. We shall denote
the observed flux, measured at this time, as $\Ftdi$. Following this
point, the angular emission area remains roughly constant as both
$F_{\infty}$ and $\Tbbi$ slowly decrease. At the peak of the
expansion, the low blackbody temperature puts much of the emitted flux
below the bandpass for many X-ray instruments; in extreme cases this
can create the appearance of a ``precursor'' burst
\citep{hoffman.ea:sas_observation,Zand10}.
\citet{Damen1990X-ray-bursts-wi} argued that to minimize systematic
errors in determining the Eddington flux, the measurement should be
made at touchdown, when the the temperature is at a maximum and the
photosphere has presumably just retreated to near the quiescent
stellar radius.

At touchdown, the observed flux is expected to equal the \emph{local},
i.e., as measured in a frame co-moving with the photosphere, Eddington
value
\begin{equation}\label{eq.Fedd}
\Ftdi \simeq \frac{G M c}{\kappa D^2} \sqrt{1 - 2
\beta(\rph)} \equiv \Fedd \sqrt{1 - 2\beta(\rph)}.
\end{equation}
Here $\beta(r)=GM/(rc^2)$, $\kappa$ is the opacity, and $\rph$ is the
photospheric radius at the time this flux is evaluated. \citet{Ozel06}
argued that at touchdown, i.e., when $\Tbb$ reaches a maximum, $\rph =
R$.  For clarity, when we refer to the Eddington flux in the remainder
of the paper, we shall mean $ \Fedd \equiv GMc/(\kappa D^{2})$. This
definition is independent of the stellar radius, and is a true
limiting flux: $\Ftdi \le \Fedd$, with equality holding if $\rph \gg
R$. Finally, for Thomson scattering in a hydrogen-helium plasma,
$\kappa \approx 0.2(1+X)\,\mathrm{cm^{2}\,g^{-1}}$, where $X$ is the
mass fraction of hydrogen.

\citet{Galloway2008Biases-for-neut} examined a large sample of PRE
bursts. They found that in all cases for which the inclination was not
extremely high, the touchdown flux was within a factor of 1.6 of the
peak flux, consistent with it occurring when the photosphere had
retreated. The source EXO~0748--676, which was used in previous
analysis~\citep {Ozel06} because of its claimed redshift
measurement~\citep{Cottam02}, has a high inclination (it is a
``dipper''); as a result, the observed touchdown flux may be obscured
by the disk, and indeed $\Ftdi$ is much less than the observed maximum
flux for this source. The distance to EXO~0748--676 is also not well
known, and for these reasons we omit this source in our analysis
below.

In the latter part of the burst the ratio $F_{\infty}/\Tbbi^{4}$ is
observed to be roughly constant. This allows one to define a
normalized angular surface area,
\begin{equation}\label{eq.a}
A \equiv \frac{\Finf}{\sigma \Tbbi^4} =  f_c^{-4} 
\left( \frac{R}{D}\right)^2 \left(1-2\beta\right)^{-1}.
\end{equation}
Here $\Finf$ and $\Tbbi$ are the flux and blackbody temperature as
measured by a distant observer in the late phase of the burst, when
$A$ is roughly constant, $\beta = GM/(Rc^{2})$ is the compactness, $D$
is the distance to the source, and $f_c\equiv
T_{\mathrm{bb}}/T_{\mathrm{eff}}$ is a color correction factor that
accounts for the departure of the spectrum from a blackbody.

If a distance to the star can be estimated with sufficient precision,
the observed touchdown flux (when the blackbody temperature reaches a
maximum following the PRE) and the inferred apparent angular area
measured late in the burst can be converted into an estimate of the
stellar mass and radius. It has been claimed that 1-$\sigma$
uncertainties of less than 10\% in both mass and radius are possible
\citep{Ozel06,Ozel09,Ozel10}. Systematic uncertainties not included in
the analyses of~\citet{Ozel06} and~\citet{Ozel09} include the
composition of the accreted material and the effects of the neutron
star atmosphere on the spectral shape, although these are included in
later analyses \citep{Guver10,Ozel10}. A key question is whether the value of $
\beta(\rph)$ in equation~(\ref{eq.Fedd}), which is determined by
$\rph$, is the same as the value of $\beta$ in equation~(\ref{eq.a}).
In the following discussion, we shall first assume that these $\beta$
values are the same, i.e., $\rph=R$, as would be the case if the
photosphere has indeed just retreated to the stellar radius. We will
demonstrate that for EXO~1745--248, 4U~1608--522, and 4U~1820--30 the
most probable observed values of $A$ and $\Feddi$ do not lead to
real-valued solutions for $M$ and $R$. Indeed,~\citet{Ozel10} notes
that forcing a real-valued solution for mass and radius by Monte Carlo
sampling within the probability distributions of the observables
$\Ftdi$, $A$, and $D$ reduces the uncertainties in the mass and radius to
values smaller than those of the measurements. We shall explore in
\S~\ref{s.rph-touchdown} a different interpretation, which is that
$\rph > R$ when $\Ftdi$ is evaluated, so that $\Ftdi \lesssim \Fedd$.
We show that relaxing the assumption $\rph = R$ generally yields
real-valued solutions for $M$ and $R$ for the most probable values of
the observables.

We combine the observed quantities
$\Ftdi$ and $A$, along with a measurement of the distance $D$ and
theoretical estimates of $f_{c}$ and $\kappa$, into two parameters,
\begin{eqnarray}
\alpha &\equiv& \frac{\Ftdi}{\sqrt{A}} 
\frac{\kappa D}{c^{3}f_{c}^{2}}, \label{eq:agr1}\\
\gamma &\equiv& \frac{Ac^{3}f_{c}^{4}}{\Ftdi\kappa}. \label{eq:agr}
\end{eqnarray}
If we then make the assumption, following \citet{Ozel06}, that $\Ftdi
= \Fedd\sqrt{1-2\beta}$ in equation~(\ref{eq.Fedd}), we find
\begin{eqnarray}
\alpha    &=& \beta(1-2\beta), \label{eq:agr1-a} \\
\gamma &=& \frac{R}{\beta(1-2\beta)^{3/2}}.    \label{eq:agr-a}
\end{eqnarray}
Solving eq.~(\ref{eq:agr1-a}) for $\beta$, we then solve
eq.~(\ref{eq:agr-a}) for the radius and mass:
\begin{eqnarray}
\beta &=& \frac{1}{4}\pm\frac{1}{4}\sqrt{1-8\alpha},\label{eq:beta}\\
R &=& \alpha\gamma\sqrt{1-2\beta},\label{eq:rb}\\
M &=& \frac{\beta R c^2}{G}.\label{eq:brm}
\end{eqnarray}
Note that $\gamma$ is independent of $D$. An important consequence of
eq.~(\ref{eq:beta}) is that for both $M$ and $R$ to be real, we must
have $\alpha \le 1/8$. Since $\alpha$ is determined, however, from the
observables $\Ftdi$, $A$, and $D$, as well as the estimated parameters
$\kappa$ and $f_c$, the inferred value of $\alpha$ does not
necessarily satisfy this mathematical limit. \emph{This condition
serves as a check on the validity of the assumptions made in the
modeling of the radius expansion bursts.}

\subsection{Monte Carlo analysis of photospheric radius expansion burst 
data}\label{sec:mc-pre}

Observational information for three Type-I X-ray bursters with
PRE bursts~\citep{Ozel09,Guver10,Guver10b} is presented in Table
\ref{tab:ozel}, along with associated uncertainties. We have
reexamined the uncertainties for each object as described in
Appendix~\ref{s.observations}; in some cases our values for the
uncertainties differ from those used previously.

\begin{deluxetable}{lccc}
\tablecolumns{4}
\tablewidth{0pc}
\tablecaption{Observational values for Type I X-ray burst sources 
used in this paper; see Appendix~\ref{s.observations} 
for details about how the uncertainties were determined. 
\label{tab:ozel}}
\tablehead{
\colhead{Quantity} & \colhead{EXO 1745--248} & \colhead{4U 1608--522} &
\colhead{4U 1820--30} }
\startdata
$D~({\rm kpc})$ & $6.3 \pm 0.6$ & 
$5.8 \pm2.0$\tablenotemark{a} & $8.2 \pm 0.7$\tablenotemark{b} \\
$A~({\rm km}^{2}~{\rm kpc}^{-2})$ & $1.17 \pm 0.13$
& $3.246 \pm 0.024$ & $0.9198 \pm 0.0186$ \\
$\Ftdi~(10^{-8}~{\rm ergs~cm}^{-2}~{\rm s}^{-1})$ &
$6.25 \pm 0.2$ & $15.41\pm 0.65$ & $5.39 \pm 0.12$ \\
$\alpha$\tablenotemark{c} & $0.131\pm0.017$ & $0.179\pm0.062 $ 
& $0.166\pm0.015$ \\
$\gamma~({\rm km})$\tablenotemark{c} & $101.7\pm11.8$ & $114.5\pm4.9$ 
& $92.7\pm2.8$ \\
${R_\infty}=
\alpha\gamma~({\rm km})$\tablenotemark{c} & $13.4\pm2.0$ 
& $20.5\pm7.1$ & $15.4\pm1.4$ \\
\enddata
\tablerefs{\citealt{Ozel06,Ozel09,Guver10,Guver10b}}
\tablenotetext{a}{The distance distribution for 4U 1608--522
was cut off below 3.9~kpc. }
\tablenotetext{b}{The distance uncertainty for 4U 1820-30 was approximated 
by~\citet{Guver10b} as a boxcar with halfwidth 1.4 kpc.}
\tablenotetext{c}{
Assuming $f_c=1.4$ and $X=0$ with $\Delta f_c=\Delta X=0$. Errors are
computed assuming uncorrelated Gaussian errors for $D$, $\Ftdi$ and
$A$.}
\end{deluxetable}

We first fix $f_c=1.4$ and $X=0$ as was done in previous work for
EXO~1745--248~\citep{Ozel09}. We observe in this case from Table
\ref{tab:ozel} that no real-valued solution of
equations~(\ref{eq:beta})--(\ref{eq:brm}) is possible for any of the
sources because $\alpha > 1/8$ for the central values of the
observables. Moreover, it is not consistent with X-ray burst models to
fix the color-correction factor to $f_c=1.4$ and the hydrogen
abundance to $X=0$, and indeed these restrictions were relaxed in
subsequent analyses of 4U~1608--522 and 4U~1820--30~\citep{Guver10,
  Guver10b}. Atmosphere models \citep{Madej04} suggest a range $1.33 <
f_c < 1.81$; these are consistent with earlier calculations
\citep{London1986Model-atmospher,Ebisuzaki1988The-Difference-}. The
largest values of $f_{c}$ are approached as the flux approaches the
Eddington limit. In the tail of the burst, the flux is lower and
$f_{c}$ does not vary strongly. We choose to select the value of $f_c$
in our Monte Carlo analysis from a boxcar distribution centered at
$f_{c}=1.4$ with an uncertainty of $0.07$. This selection is
comparable to the boxcar distribution with $f_c=1.35\pm0.05$ used
previously \citep{Guver10, Guver10b}. (In the following, we denote a
boxcar uncertainty in $X$ as $\Delta X$ and a Gaussian uncertainty
with $\sigma_X$.) In addition, as described in
Appendix~\ref{s.exo1745} and \ref{s.4u1608}, we find insufficient
information from the observations to exclude any possible
values of $X$ for EXO~1745--248 and 4U~1608--522, so we choose a
uniform distribution with $0<X<0.7$, which is also assumed by
\citet{Guver10, Guver10b}. The source 4U~1820--30 is an ultracompact
and accretes He-rich fuel; see Appendix~\ref{s.4u1820}.

If values of $f_c$ and $X$ or the observables $\Ftdi$, $A$, or $D$ are
selected at random from their respective probability distributions,
some combinations will satisfy the condition $\alpha<1/8$. In this
way, using only these acceptable combinations, most probable values
$\hat\alpha$ and $\hat\gamma$ can be established using Monte Carlo
sampling. (Here and below, we define $\hat{X}$ to be the most probable
value of $X$ obtained from a MC simulation, and the uncertainties on
this quantity are determined by selecting the region surrounding the
most probable value that includes 68\% of the total MC weight.) Values
for $M$ and $R$, and their uncertainties, can then be determined from
$\hat\alpha$ and $\hat\gamma$ using
equations~(\ref{eq:beta})--(\ref{eq:brm}). But the fraction of
accepted realizations is then very small as shown in
Table~\ref{t.pre-compare}. The entries in the first group of this
table give the most probable values $\hat\alpha$, $\hat\gamma$, and
$\hat R_{\infty}$ using $\rph=R$ and values for $\Ftdi, A,$ and $D$
from their probability distributions summarized in
Table~\ref{tab:ozel} and, for $f_c$ and $X$, from previous discussion.
For each quantity, the uncertainty is determined by creating a
histogram for the indicated MC realizations, sorting the bins by
decreasing weight, and selecting the range which encloses 68\% of the
sum of all bins. We identify $R_{\infty} = [\Finf/(4\pi\sigma
  T_{\mathrm{eff,}\infty}^{4})]^{1/2} = R/\sqrt{1-2\beta} =
\alpha\gamma$ and impose no \emph{a priori} restriction for the value
of $\alpha$. The most probable value of $\alpha$, $\hat\alpha$, is
observed to be approximately a factor $1+\bar X$ larger than the value
of $\alpha$ in Table \ref{tab:ozel}, where $\bar X$ is the average
value of $X$ in its assumed range, {\it i.e.} 1.35 for EXO~1745--248
and 4U~1608--522, and 1.0 for 4U 1820-30. For all three sources,
$\hat\alpha > 1/8$ by several standard deviations. 

\begin{deluxetable}{llll}
\tiny
\tablecolumns{4} \tablewidth{0pc} \tablecaption{\label{t.pre-compare}
Comparison of Monte Carlo realizations for PRE bursts employing data
from Table \ref{tab:ozel}. In the third group, we use the values and 
distributions of $F_{\mathrm{TD}}$, $A$, $D$, $f_c$, and $X$ 
from~\citet{Ozel09,Guver10,Guver10b}. For all other entries, 
we use $0<X<0.7$ for the
hydrogen mass fraction, except for 4U~1820--30, for which $X = 0$
and we use $1.33 < f_{c} < 1.47$ for the color correction factor.}
\tablehead{ \colhead{}
& \colhead{EXO~1745--248} & \colhead{4U~1608--522} &
\colhead{4U~1820--30} } \startdata
\cutinhead{$\Ftdi = \Fedd[1-2\beta(R)]^{1/2}$; $\alpha$ unrestricted}
$\hat\alpha$ & 
\phn$0.165^{+0.045}_{-0.028}$ &
\phn$0.229^{+0.054}_{-0.038}$ & 
\phn$0.167^{+0.018}_{-0.017}$ \\
$\hat\gamma~(\rm{km})$ & 
$70.4^{+19.2}_{-14.7}$ & 
$80.2^{+18.8}_{-15.5}$ & 
$80.9^{+21.2}_{-3.56}$ \\
$\hat\Rinf~(\rm{km})$ & 
$13.2^{+1.7}_{-1.6}$ & 
$20.5^{+5.7}_{-5.5}$ & 
$15.3^{+1.7}_{-1.5}$ \\
\cutinhead{$\Ftdi = \Fedd[1-2\beta(R)]^{1/2}$; 
$\alpha < 1/8$ restriction}
$\hat\alpha$ & 
\phn\phn$0.122^{+0.003}_{-0.007}$ &
\phn$0.123^{+0.003}_{-0.005}$ & 
\phn\phn$0.123^{+0.003}_{-0.004}$ \\
$\hat\gamma~(\rm{km})$ & 
$107.9^{+14.2}_{-14.2}$ &
$127.1^{+8.0}_{-7.9}$ & 
$109.3^{+5.3}_{-8.1}$ \\
$\hat\Rinf~(\rm{km})$ & 
\phn$12.7^{+1.6}_{-1.6}$ & 
$15.2^{+0.85}_{-0.81}$ & 
\phn$13.2^{+0.78}_{-1.1}$ \\
Points accepted &
4.4 \% & 
0.24 \% &
0.44 \% \\
\cutinhead{$\Ftdi = \Fedd[1-2\beta(R)]^{1/2}; 
\alpha<1/8$; Original inputs from Ozel et al.}
$\hat\alpha$ & 
\phn$0.122^{+0.003}_{-0.003}$ &
\phn$0.122^{+0.003}_{-0.003}$ & 
\phn$0.124^{+0.001}_{-0.001}$ \\
$\hat\gamma~(\rm{km})$ & 
$109.0^{+4.2}_{-4.2}$ &
$113.1^{+5.0}_{-5.0}$ & 
$104.0^{+1.8}_{-1.8}$ \\
$\hat\Rinf~(\rm{km})$ & 
$13.3^{+0.4}_{-0.4}$ & 
$13.8^{+0.5}_{-0.5}$ & 
$12.9^{+0.2}_{-0.2}$ \\
Points accepted &
13 \% & 
0.18 \% &
$1.5{\times}10^{-6}$ \% \\
\cutinhead{$\Ftdi = \Fedd$; $\alpha < 3^{-3/2} = 0.192$ restriction}
$\hat\alpha$ & 
\phn$0.165^{+0.026}_{-0.018}$ &
\phn$0.180^{+0.012}_{-0.023}$ & 
\phn$0.167^{+0.016}_{-0.016}$ \\
$\hat\gamma~(\rm{km})$ & 
$79.0^{+17.2}_{-12.4}$ &
$97.0^{+15.4}_{-14.4}$ & 
$91.0^{+13.8}_{-10.6}$ \\
$\hat\Rinf~(\rm{km})$ & 
$13.1^{+1.8}_{-1.6}$ & 
$15.3^{+2.8}_{-1.3}$ & 
$15.0^{+1.8}_{-1.4}$ \\
Points accepted &
65 \% & 
15 \% &
91 \% \\
\cutinhead{$\Fedd[1-2\beta(R)]^{1/2}<\Ftdi<\Fedd; q^3+p^2<0$ restriction}
$\hat\alpha$ & 
\phn$0.143^{+0.022}_{-0.018}$ &
\phn$0.155^{+0.018}_{-0.019}$ & 
\phn$0.158^{+0.014}_{-0.015}$ \\
$\hat\gamma~(\rm{km})$ & 
$87.1^{+17.5}_{-13.8}$ &
$103.0^{+16.6}_{-13.6}$ & 
$104.8^{+5.6}_{-17.0}$ \\
$\hat\Rinf~(\rm{km})$ & 
$13.0^{+1.8}_{-1.6}$ & 
$15.3^{+2.3}_{-1.4}$ & 
$15.0^{+1.6}_{-1.6}$ \\
Points accepted &
32 \% & 
5.8 \% &
37 \% \\
\enddata
\end{deluxetable}

\normalsize

In the second group of Table~\ref{t.pre-compare}, we repeat the
analysis, but this time only accept realizations for which $\alpha \le
1/8$. Note that at most 4\% of the realizations are accepted. This
implies, at a minimum, that the assumptions in this model are
incomplete. The most probable value for $\alpha$ decreases and now
lies approximately 1-$\sigma $ below the value 1/8, and the size of
confidence interval for $\hat\alpha$ is now significantly reduced
compared to the unrestricted case, by factors of 4--18. {\it The net
  effect is to pre-select the value 1/8 for $\hat\alpha$ irrespective
  of the observed values of $\Ftdi$, $A$, and $D$.} Another
consequence of the selective rejection of most of the Monte Carlo
realizations is to increase the most probable value of $\gamma$ and
decreases the most probable value of $\Rinf$, and to greatly decrease
their uncertainties; the latter result was already noted
by~\citet{Ozel10}. In addition, the resulting values of $\bar\gamma$
and $\bar R_\infty$ for the different sources are also ``herded'' into
similar values \citep{Ozel09,Guver10,Guver10b}. The third group
summarizes the MC results obtained using the exact same observational
inputs found in those references. Note that for 4U 1820--30, only
about 1 realization out of 100 million is now accepted! The larger
acceptance rate for EXO 1745--248 is due to the assumption that $X=0$
which gives the smallest possible $\alpha$. Analyses of X-ray bursts
with this model result in certain values for $M$ and $R$ that are
nearly independent of $\alpha$ and $\gamma$, and are therefore almost
completely independent of the observables $\Ftdi, A$, and $D$, which
seems unrealistic. The fact that only a small fraction of realizations
are accepted may indicate that the underlying model contains
systematic errors or is unphysical, a point we shall explore in the
next section.

\begin{figure}
\includegraphics[width=3.3in]{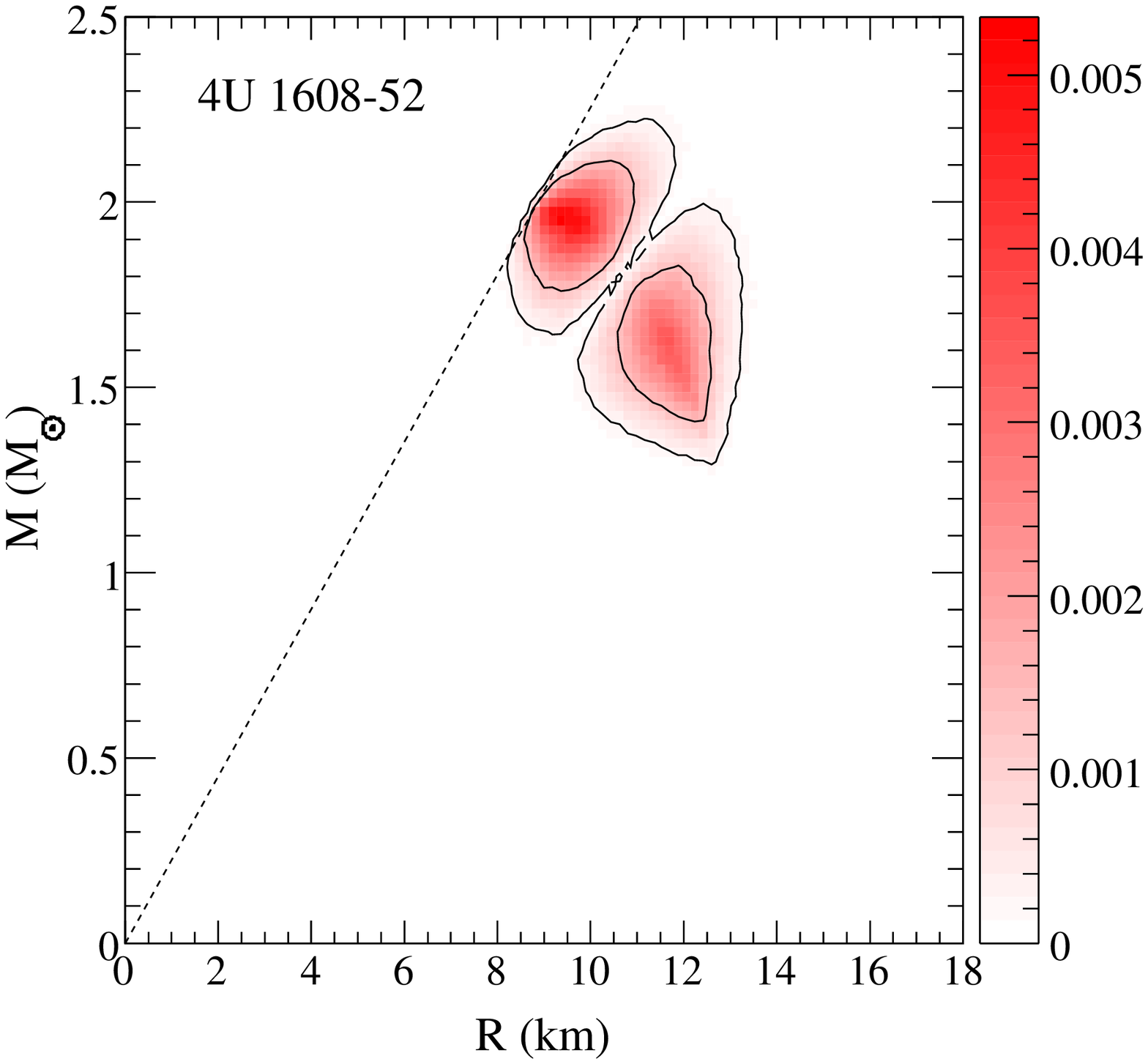}
\includegraphics[width=3.3in]{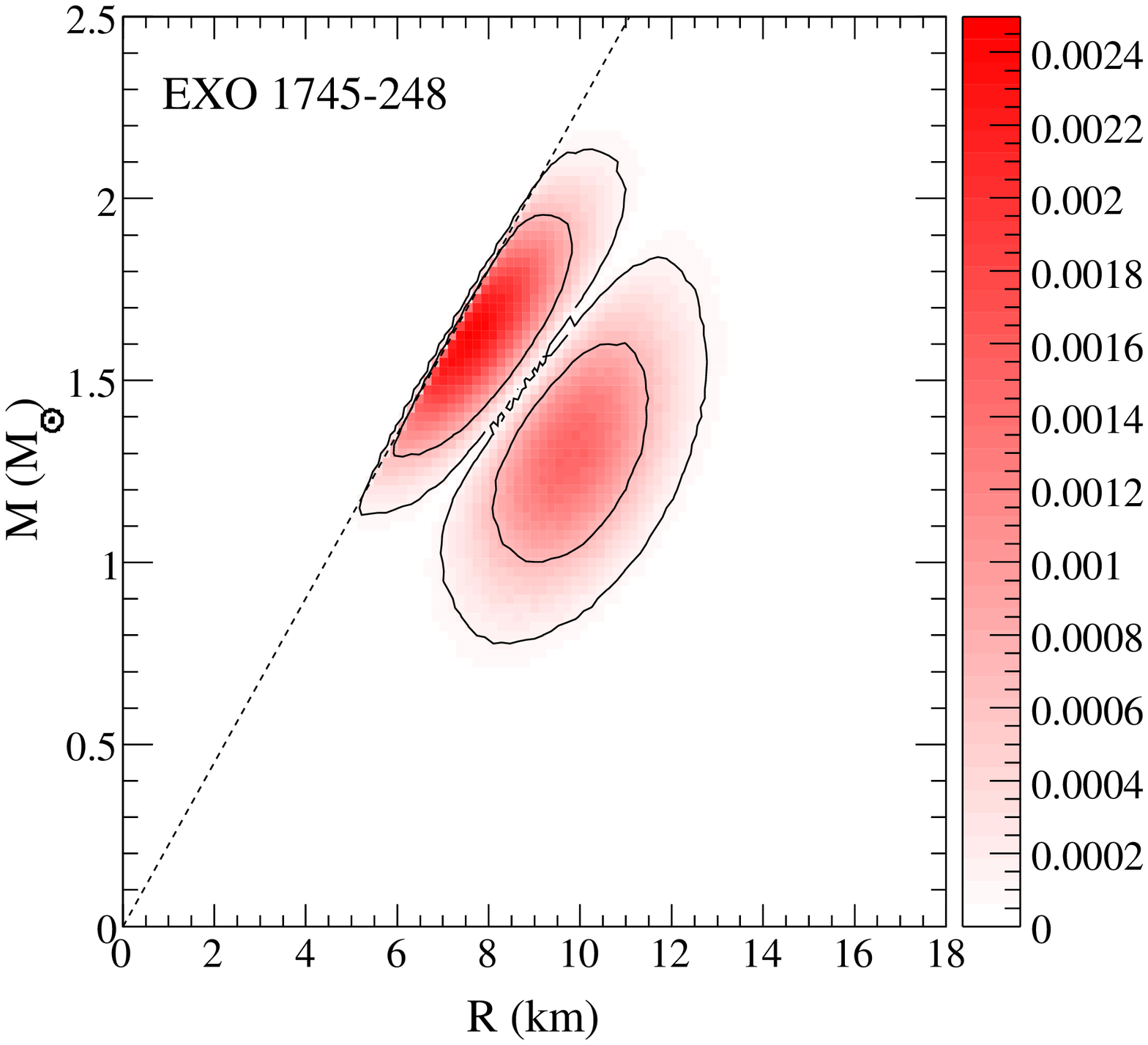}
\includegraphics[width=3.3in]{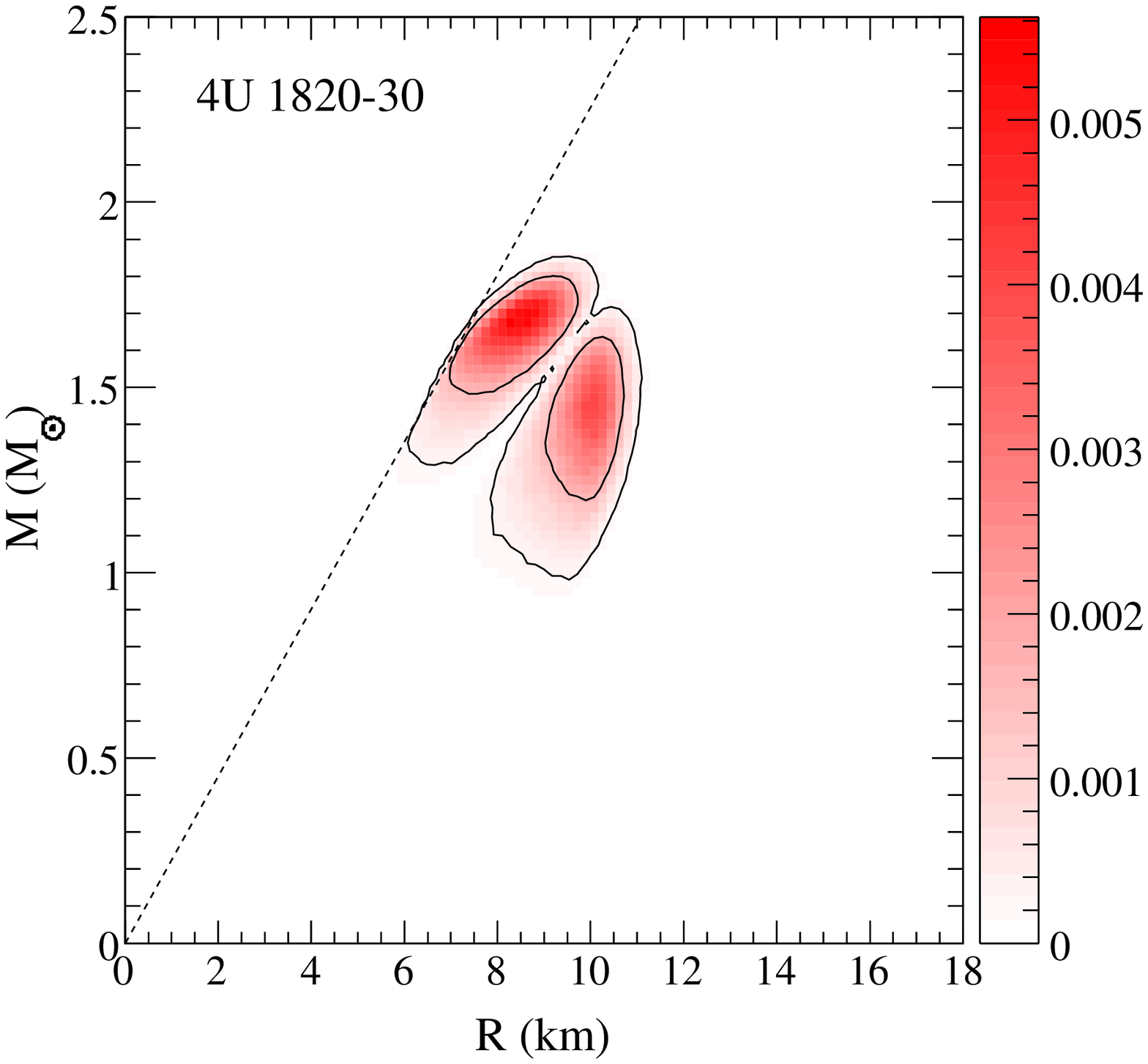}
\caption{Mass-radius probability distributions for Type I X-ray bursts
assuming that the photospheric radius and the stellar radius are
identical. The causal limit $\beta=1/2.94$ is indicated with a dashed
line. These plots correspond to the results shown in the second group
in Table~\ref{t.pre-compare}.  The solid curves indicate the 68\% and
95\% confidence boundaries while the shading level reflects the
relative probabilities. All distributions, $P_i$, are normalized so 
that $\int P_i~dM~dR = 1$.}
\label{fig:pre}
\end{figure}

Figure~\ref{fig:pre} displays the Monte Carlo probability
distributions for the masses and radii of EXO~1745--248, 4U~1608--522,
and 4U~1820--30 as determined from $\alpha$ and $\gamma$ following
equations~(\ref{eq:beta})--(\ref{eq:brm}), and with the restriction
$\alpha \le 1/8$. There are two peaks in the distributions because of
the quadratic in equation~(\ref{eq:beta}). We also impose that $\beta
< 1/2.94 \approx 0.34 $~\citep{Glendenning92}, which is based on
requiring a subluminal sound speed throughout the star~\citep[see
  also][]{Lattimer2006Neutron-Star-Ob}. This causes the rejection of a
small fraction of realizations from the higher-redshift region. For
all three sources, our confidence intervals are larger than computed
in previous works~\citep{Ozel09,Guver10,Guver10b}. The origin of this
distinction is because we use larger variations $f_c$ and, for EXO~1745--248, in $X$, and we employ a different
uncertainty in $D$ for 4U 1820--30.  A detailed discussion of the
treatment of the uncertainties in the observables for each object is
given in the Appendix.

When using a Monte Carlo scheme to determine uncertainties, we find that the two error ranges in Fig.\ \ref{fig:pre} are equally populated, while \citet{Ozel09}, \citet{Guver10}, \citet{Guver10b}, and \citet{Ozel10}, transforming probabilities using a Jacobian, indicate a much higher probability for the higher redshift solution.  This difference is caused by a numerical error in their Jacobian (\"Ozel, priv.\ comm.); when the correct Jacobian is used, both integration methods agree.

\subsection{Alternate interpretations of the location of the 
photospheric radius at touchdown}\label{s.rph-touchdown}

The lack of real-valued solutions for $M$ and $R$ for the
most probable values of the observables,
and indeed, the rejection of the vast majority of Monte Carlo realizations,
motivates us to consider another possibility; namely, that at
``touchdown'' the photosphere is still extended.
To explore this situation, we first consider the extreme possibility
that $\rph\gg R$ at the point identified as touchdown from the maximum
in $\Tbbi$. In this case $\Ftdi = \Fedd$ and is independent of the
stellar radius $R$. With this assumption, $\alpha$
(eq.~[\ref{eq:agr1}]) and $\gamma$ (eq.~[\ref{eq:agr}]) are now
related to $\beta$ and $R$ via
\begin{eqnarray}
\label{eq:agr1_2}
\alpha    &=&  \beta\sqrt{1-2\beta}, \\
\gamma &=& \frac{R}{\beta(1-2\beta)} .
\label{eq:agr_2}
\end{eqnarray}
Defining a quantity $\theta = \cos^{-1}(1-54\alpha^2)$, the
expressions for the compactness, radius, and mass are then
\begin{eqnarray}
\beta_1 &=& {1\over6}\left[1+\sqrt{3}\sin(\theta/3)-
\cos(\theta/3)\right],\label{eq:beta_21}\\
\beta_2 &=& {1\over6}\left[1+2\cos(\theta/3)\right],\label{eq:beta_22}\\
R &=& \alpha\gamma\sqrt{1-2\beta_{1,2}}.\label{eq:rb_2}\\
M &=& \frac{c^2}{G}\alpha^2\gamma,\label{eq:brm_2}
\end{eqnarray}
Obviously, $M$ is real for all $\alpha,\gamma > 0$. For
$\alpha<3^{-3/2} \simeq 0.192$, $\theta$ is real and there are 3 real
roots for $\beta$ and $R$. Only two of these, $0\le\beta_1\le1/3$ 
and $1/3\le\beta_2\le1/2$,
are physically meaningful: the other root is negative.  When $\alpha>3^{-3/2}$
there is one real root for $\beta$ and two imaginary ones, but the
real root is negative.  

\begin{figure}
\includegraphics[width=3.3in]{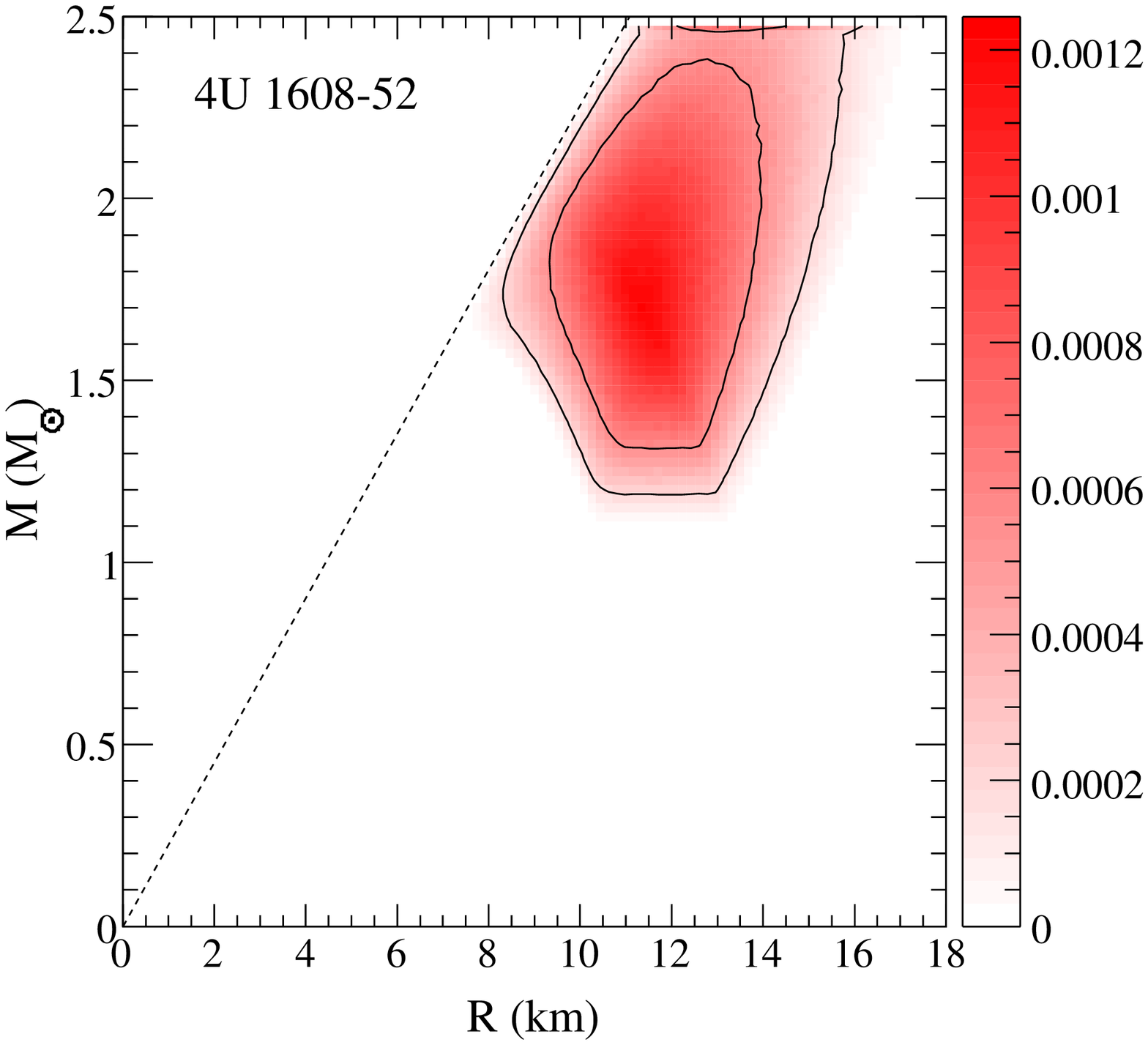}
\includegraphics[width=3.3in]{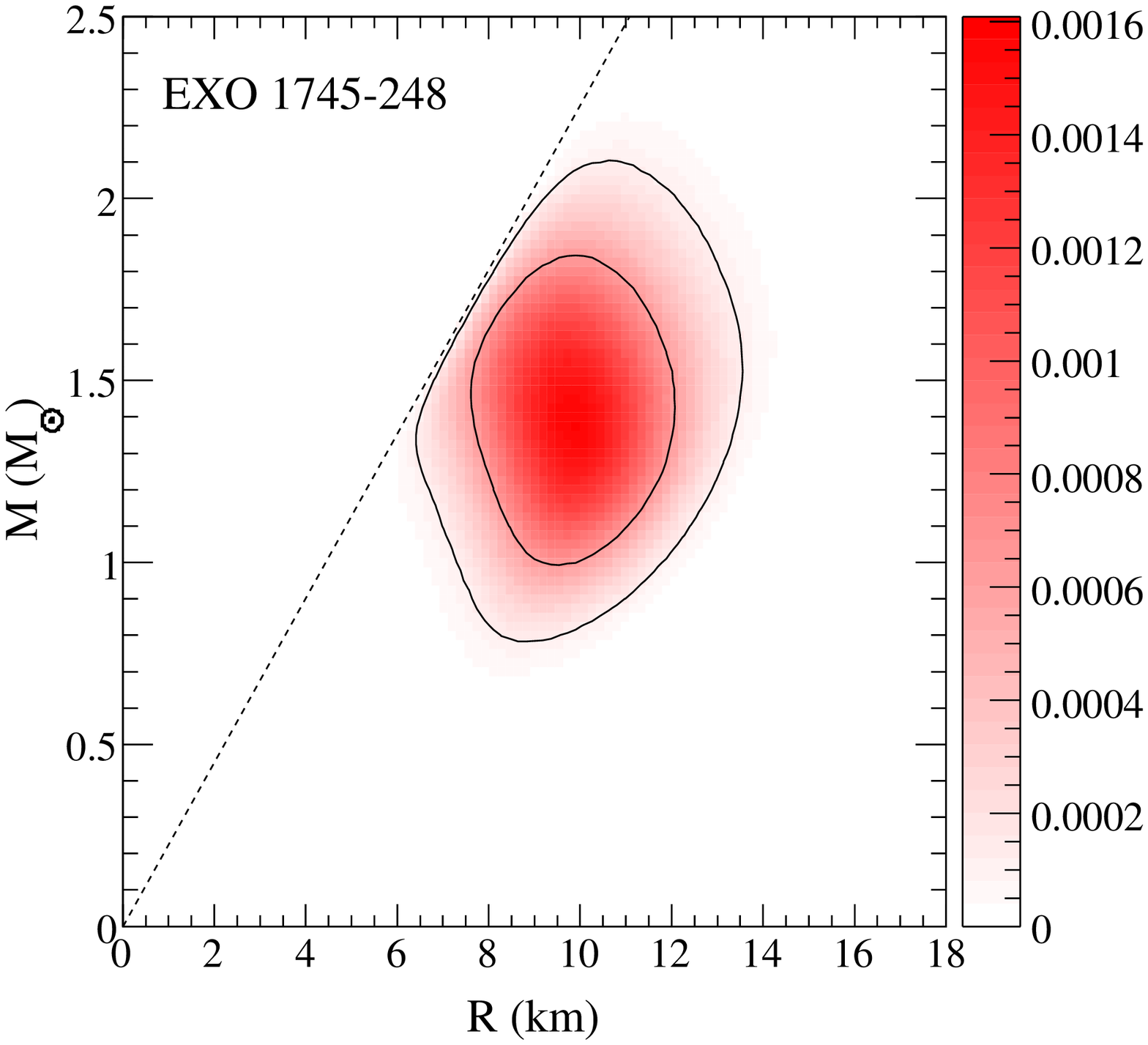}
\includegraphics[width=3.3in]{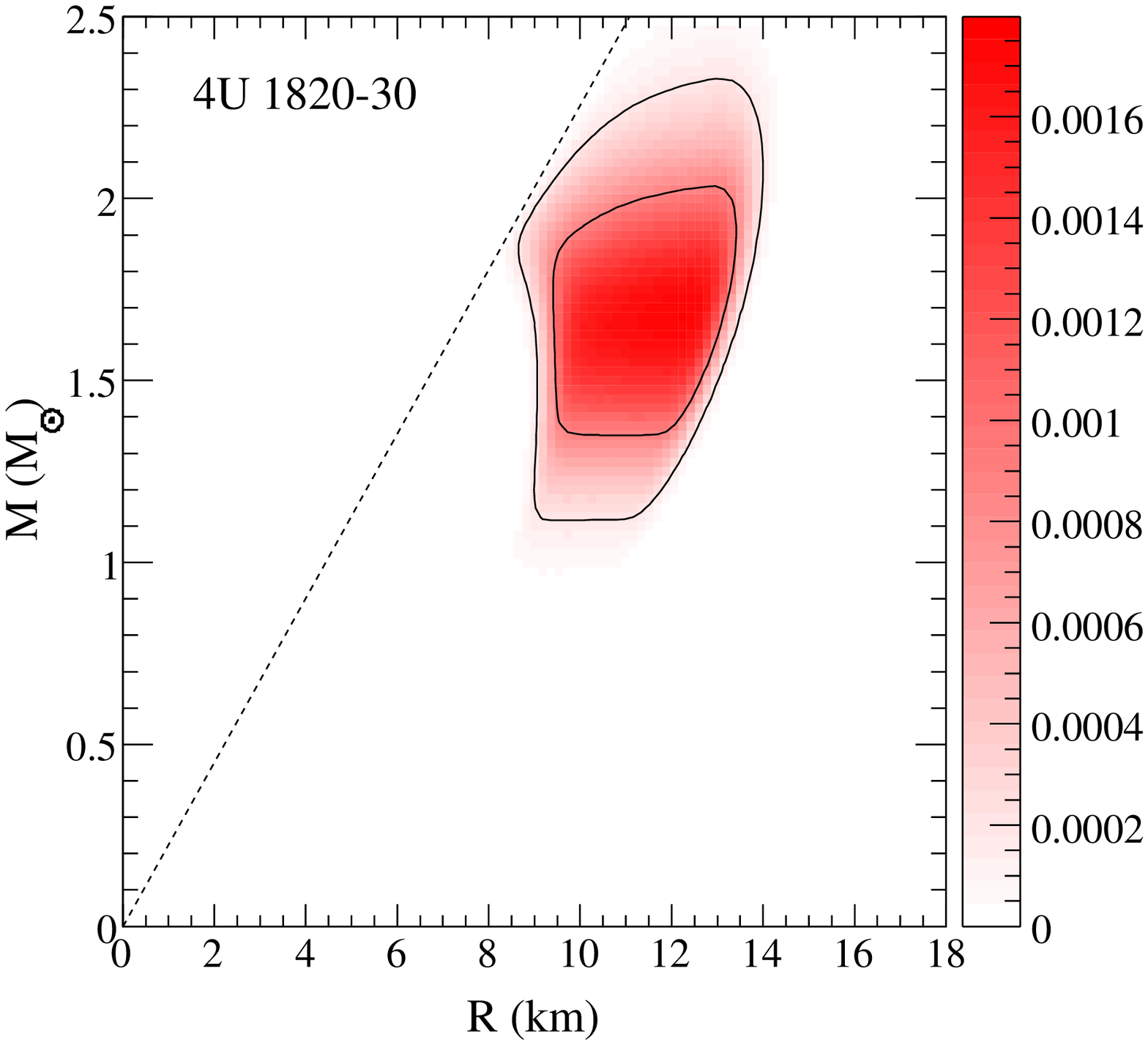}
\caption{Mass-radius probability distributions for Type I X-ray bursts
  assuming that $\rph\gg R$. The causal limit $\beta=1/2.94$ is
  indicated with a dashed line. These plots correspond to the results
  shown in the fourth group in Table~\ref{t.pre-compare}. The solid
  curves indicate the 68\% and 95\% confidence boundaries while the
  shading level reflects the relative probabilities. All
  distributions, $P_i$, are normalized so that $\int P_i~dM~dR = 1$.}
\label{fig:pre2}
\end{figure}

From the top group in Table~\ref{t.pre-compare}, we can see that both
4U~1820--30 and EXO~1745--248 have most-probable values of $\alpha$
that satisfy the constraint for positive real values of $\beta$ and
$R$, and 4U~1608--522 lies less than 1-$\sigma$ above this limit.
Therefore, we now find that a much larger fraction of the Monte Carlo
realizations are accepted when selecting values for $\Ftdi$, $A$, $D$,
$f_c$, and $X$ from their probability distributions. This is shown in
the fourth group of Table~\ref{t.pre-compare}, for which we use the
model given in equations~(\ref{eq:agr1_2})--(\ref{eq:agr_2}) and
impose the restriction $\alpha < 0.192$. In contrast to the case in
which $\rph=R$, the uncertainties in $\hat\alpha$ and $\hat\gamma$ are
not as strongly diminished. Moreover, the values for $\hat\alpha$ are
no longer nearly the same for the three sources. While in principle
each accepted realization results in two $M,R$ values, one
corresponding to $\beta_1$ and the other to $\beta_2$, nearly all the
$\beta_2$ realizations are rejected on the basis of causality when
$\beta_2>1/2.94$. Figure~\ref{fig:pre2} displays the probability
distributions for the fourth group of runs listed in Table~\ref
{t.pre-compare}, as well as their 68\% and 95\% contours. The error
contours are larger than those shown in Fig. (\ref{fig:pre}) and
considerably larger than determined
previously~\citep{Ozel06,Ozel09,Guver10,Guver10b}.

It is perhaps unphysical to make the extreme assumption $\rph\gg R$
when $T_{\mathrm{eff}}$ reaches a maximum. But because of the relative
consistency of the solutions and the much larger fraction of MC points
accepted in this case, it seems reasonable to consider the possibility
that at touchdown, that is, when $\Tbbi$ reaches a maximum and the
normalization reaches a minimum, the photospheric radius may in fact
be larger than the quiescent stellar radius. We can write $\alpha =
\beta\sqrt{1-2\beta}\sqrt{1-\beta_{\mathrm{ph}}}$, where $
\beta_{\mathrm{ph}} = GM/(\rph c^{2}) < \beta$. Defining the quantity
$h=2R/\rph$, the quantities $\alpha$ and $\gamma$ now become
\begin{eqnarray}
\label{eq:agr1_3}
\alpha &=& \frac{\Fedd}{\sqrt{A}} \frac{\kappa D}{c^{3}f_{c}^{2}} = 
\beta\sqrt{1-2\beta}\sqrt{1-h\beta}, \\
\gamma &=& \frac{Ac^{3}f_{c}^{4}}{\Fedd\,\kappa} = 
\frac{R}{\beta(1-2\beta)\sqrt{1-h\beta}} .
\label{eq:agr_3}
\end{eqnarray}
Note that $R_\infty=\alpha\gamma$ remains unchanged.
Equation~(\ref{eq:agr1_3}) is a quartic equation for $\beta$.  Two positive
real solutions for $\beta$ exist when $q^2+p^3<0$, where
\begin{eqnarray}
p &=&{1\over6h}\left(\alpha^2-{1\over24h}\right)\,,\\
q &=&-{1\over12h}\left({1\over144h^2}+3a\alpha^2\right)\,,\\
a &=& {1\over6h}\left(1-{3(2+h)^2\over16h}\right)\,;
\label{eq:pq}
\end{eqnarray}
otherwise the only real solution is negative.
Defining the quantities
\begin{eqnarray}
\theta &=& \cos^{-1}\left[\left({q\over-p}\right)^{3/2}\right]\,,\\
v &=& 2\sqrt{-p}\cos({\theta/3})\,,\\
w &=& \sqrt{2v-2a}\,,\\
z_\pm &=& -\left(2v+4a\pm{2b\over w}\right)\,,\\
b &=& {(2+h)^2\over8h^2}\left[1-{2+h\over8h}\right]\,,
\label{eq:theta}
\end{eqnarray}
the two solutions for $\beta$ with values in the range $0\le\beta\le0.5$ are
\begin{equation}
\beta_{\pm} ={2+h\over8h}\pm{w-\sqrt{z_\pm}\over2}\,,
\label{eq:bet}
\end{equation}
where the sign has the same sense in the two occurrences.

\begin{figure}
\includegraphics[width=3.3in]{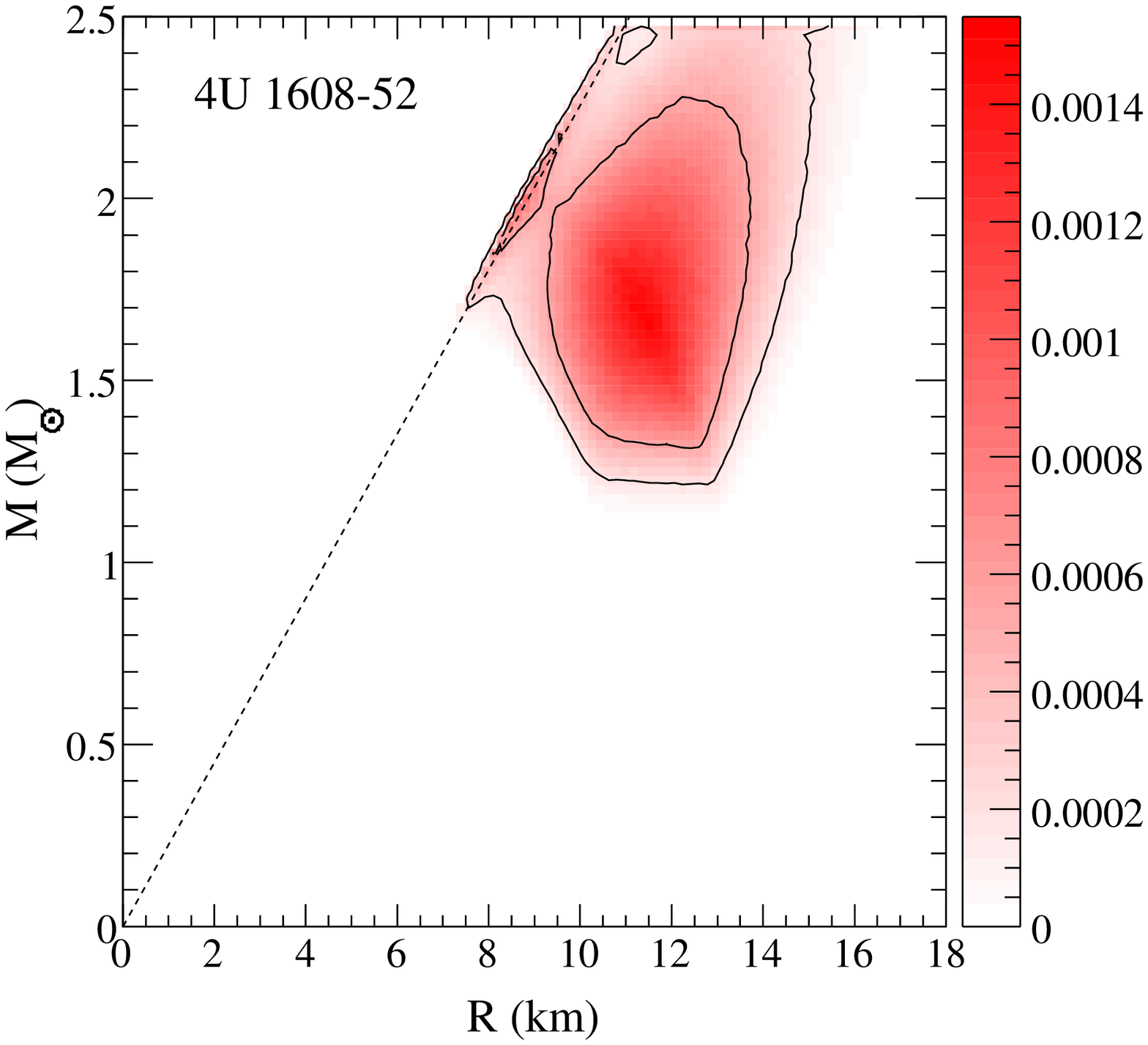}
\includegraphics[width=3.3in]{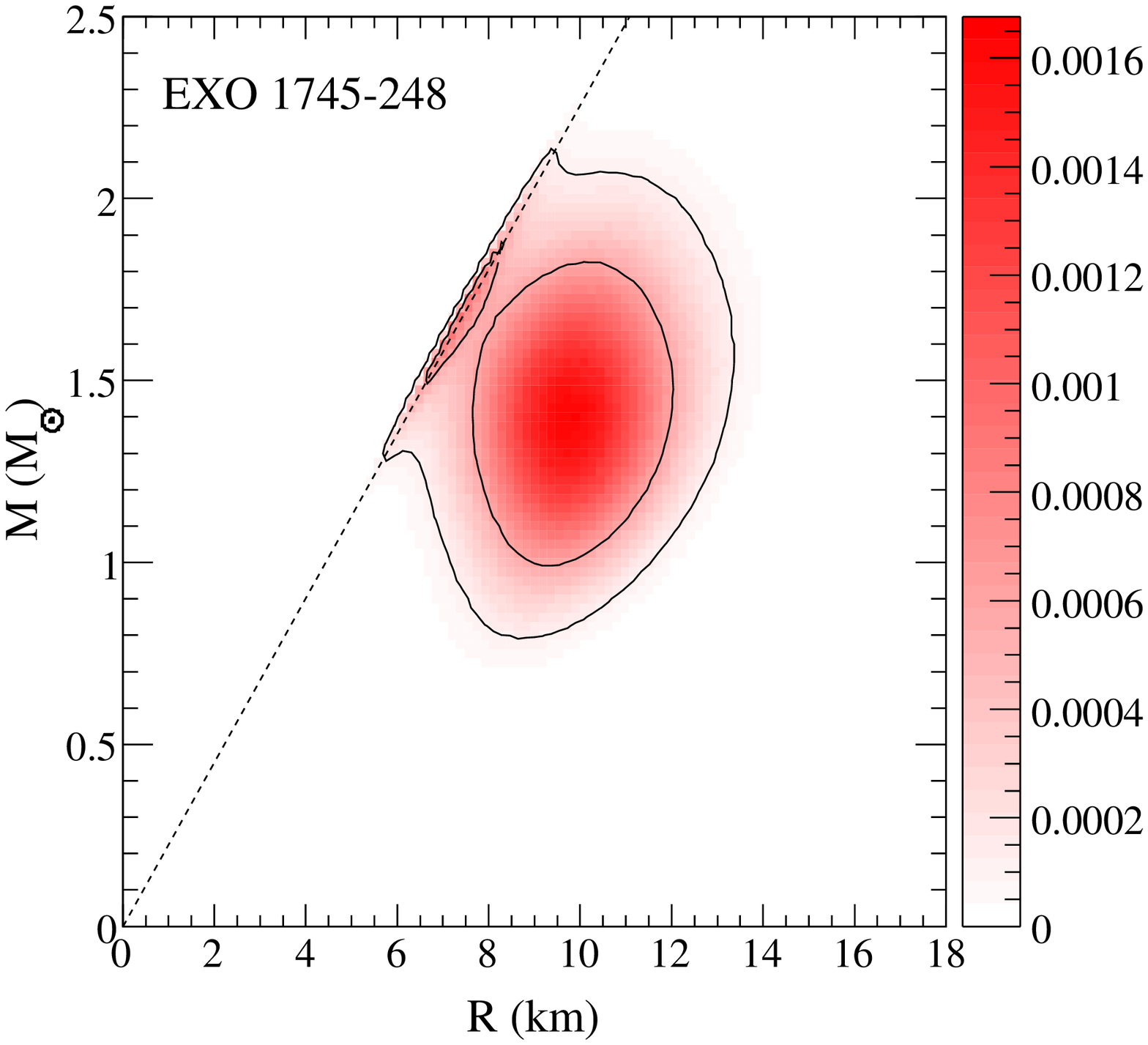}
\includegraphics[width=3.3in]{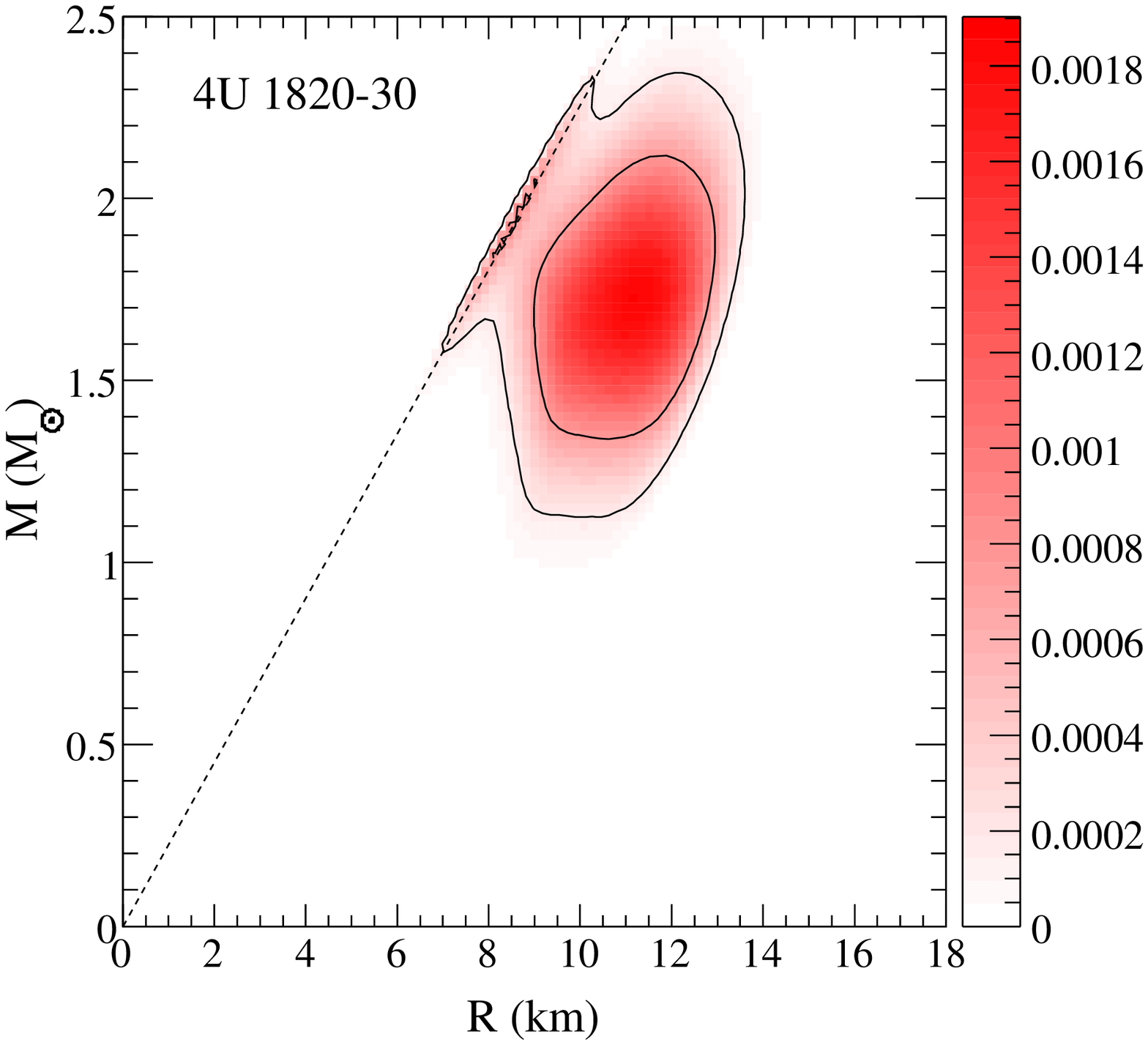}
\caption{ Mass-radius probability distributions for Type I X-ray
bursts assuming a uniform distribution in $h = 2R/\rph$. The
shadings and lines have the same meaning as in Fig.~\ref{fig:pre2}.}
\label{fig:pre3}
\end{figure}

Figure~\ref{fig:pre3} displays the resulting probability distributions
assuming that $h$ is uniformly distributed in the range $0<h<2$, i.e.,
$R<\rph<\infty$. The error contours are much larger than those
determined in the $h=2$ ($\rph=R$) case and slightly larger than those
in the $h=0$ ($\rph\gg R$) cases. Most of the solutions for $\beta_+$
are rejected because they lead to acausal combinations of $M$ and $R$.
The fifth group in Table~\ref{t.pre-compare} presents the associated
values of $\alpha$, $\gamma$, and $R_{\infty}$. These results are
nearly identical to those of Figure~\ref{fig:pre2} because small
values of the photospheric radius are strongly disfavored by the requirement
that the masses and radii are real-valued. We will therefore use the
$\rph \gg R$ probability distributions in $M$ and $R$ for the Bayesian
estimation of the EOS in Section \ref{sect:bayes}, and note that these
results will be essentially identical to what one would obtain without
assuming a particular value for the radius of the photosphere, as long
as it is not {\em a priori} assumed to be equal to $R$.

Using the number of accepted MC realizations as a guide, we can also
obtain an estimate for the lower limit of the location of the
photosphere at touchdown. As is common in Bayesian analysis, we use
the ratio 0.1 as a guide; models for which the fraction of accepted
realizations is less than 0.1 are rejected. This implies that $\rph >
5.0 R$ for 4U~1608--522, $\rph > 1.1 R$ for EXO~1745--248, and $\rph >
1.4 R$ for 4U~1820--30. This analysis appears to strongly disfavor an
interpretation of the touchdown radius $\rph=R$ and we find it likely
that the photospheric radius is extended at touchdown. At the same
time, there is little difference between the results assuming $h=0$ or
a uniform range $0<h<2$. It would then seem justified that we, in our
remaining analysis, reject the interpretation $\rph=R$, and make use
of the $\rph \gg R, h=0$ interpretation. Nonetheless, we will also
include results for the $\rph=R$ interpretation in the work below for
comparison to previous studies. We will show that some aspects of the
EOS constraints are dependent on whether one assumes that $\rph=R$ or
$\rph \gg R$.

\section{Mass and radius constraints from thermal spectra}
\label{sec:spectral-fits}

The masses and radii of isolated neutron stars or ones in transient
LMXBs can be inferred from spectral modeling if their distances are
accurately determined. Relatively accurate distances are known for
several accreting neutron star transients located in globular
clusters. Although the uncertainties for any individual source are
large, it is productive to use the ensemble of observations to improve
constraints on the dense matter equation of state. In addition, future
mass and radius measurements from thermal sources are expected to
tighten such constraints further.

Many neutron stars are in transients, for which the accretion of matter
proceeds intermittently, with episodes of accretion separated by long
periods of quiescence. While the neutron star accretes, compression of
matter in the crust induces nuclear reactions~\citep{haensel90a} that
release heat. When the accretion ceases, the heated crust cools,
resulting in an observable thermal
luminosity~\citep{brown98:transients}. Because the timescale for
heavier nuclei to sink below the photosphere is short ($\sim
10\,\mathrm{s}$;~\citealt{bildsten92}), and these systems show no
evidence, such as pulsations or cyclotron spectral features, for a
significant magnetic field, the spectra can be fitted with
well-understood unmagnetized hydrogen atmosphere spectra
\citep{zavlin96,rutledge.ea.99:refit,Heinke06}. As a result, the
observed X-ray spectra can be used to reliably infer an apparent
angular emitting area, and, possibly, the surface gravity
\citep{Heinke06}. Such objects with accurately determined distances
(such as those in globular clusters) can be used to estimate masses
and radii. 

The spectra of isolated cooling neutron stars with well-determined
distances, such as RX~J1856--3754 discovered by \citet{Walter96}, have
a much larger signal-to-noise than those of neutron star transients.
The interpretation of their spectra is complicated, however, by the
potentially strong magnetic field. The distance to RX~J1856--3754 has
been controversial. \citet{Walter02} gave a parallax distance
$D=117\pm12$ pc based on HST observations from 1996-98, but
\citet{vanKerkwijk07} later found $D=161\pm16$ pc using new HST data
from 2002-4. Recently, \citet{Walter10} reanalyzed the new data and
found $D=122\pm13$ pc, in good agreement with their older estimate. A
multi-wavelength spectral fit by~\citet{Pons02} for non-magnetic heavy
element atmospheres (which they argued gave the best fits to the X-ray
and optical spectra of RX~J1845--3754) yielded
$R_\infty/D=0.13\pm0.01\,\mathrm{km\,pc^{-1}}$ and $0.3<z<0.4$.
\citet{burwitz01} argued, however, that these models predict spectral
features such as absorption edges that are not apparent in the data.
In addition, the existence of a bow shock \citep{vankerkwijk01} and
the detection of pulsations in the X-ray flux~\citep{tiengo07} and
their period derivative \citep{vankerkwijk08} imply this star has a
magnetic field of order $10^{13}\,\mathrm{G}$. Nevertheless, there
could be many reasons why spectral features are washed-out from a
rotating, highly magnetized, object. A two-temperature blackbody
model, in which the X-ray flux is primarily emitted by a
high-temperature region and the optical flux by a low-temperature
region, gives $R_\infty/D\simeq0.14\,\mathrm{ km\,pc^{-1}}$,
approximately the same value determined by \citet{Pons02} and
\citet{burwitz03}. Motivated by the large distance determined
by~\citet{vanKerkwijk07}, which implied radii incompatible with
neutron star equations of state,~\citet{ho07} developed a magnetized
neutron star atmosphere model with a condensed surface and a trace
amount of H remaining in the atmosphere. This model resulted in
$R_\infty/D\simeq0.12\,\mathrm{ km\,pc^{-1}}$. The origin of the trace
H in the atmosphere is unknown, and its mass must be finely-tuned to
explain the magnitude of the optical flux. The analysis of
\citet{Pons02} coupled with the confirmed distance determination of
\citet{Walter10} yields $M=1.7\pm0.3\, \Msun$ and
$R=11.5\pm1.2\,\mathrm{km}$, but this model neither accounts for a
large magnetic field or the observed lack of spectral lines. Because
of this, we do not include this source in our baseline results
although we have performed a separate analysis to determine if our
results would be appreciably affected.

\begin{figure}
\includegraphics[width=3.5in]{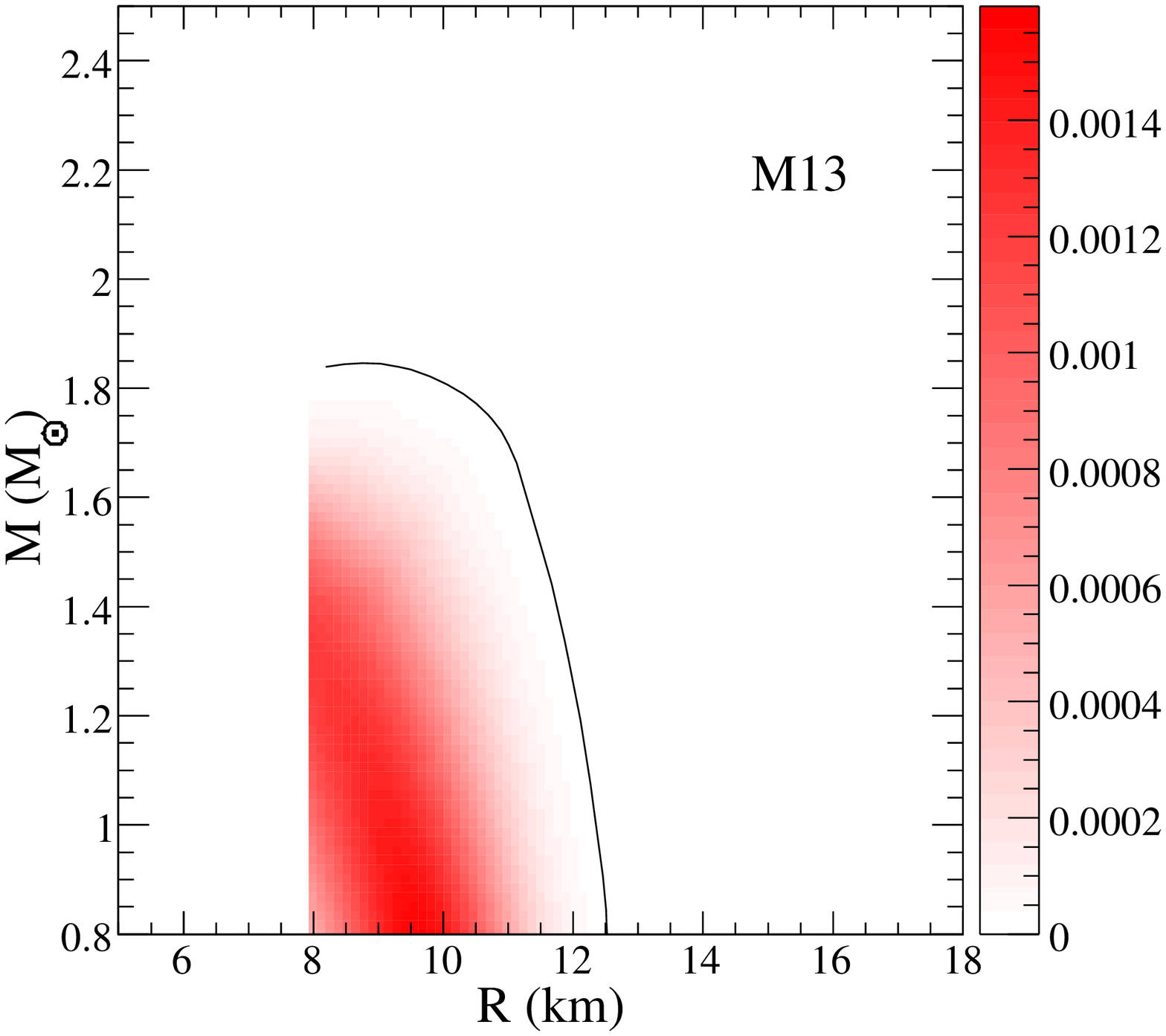}
\includegraphics[width=3.5in]{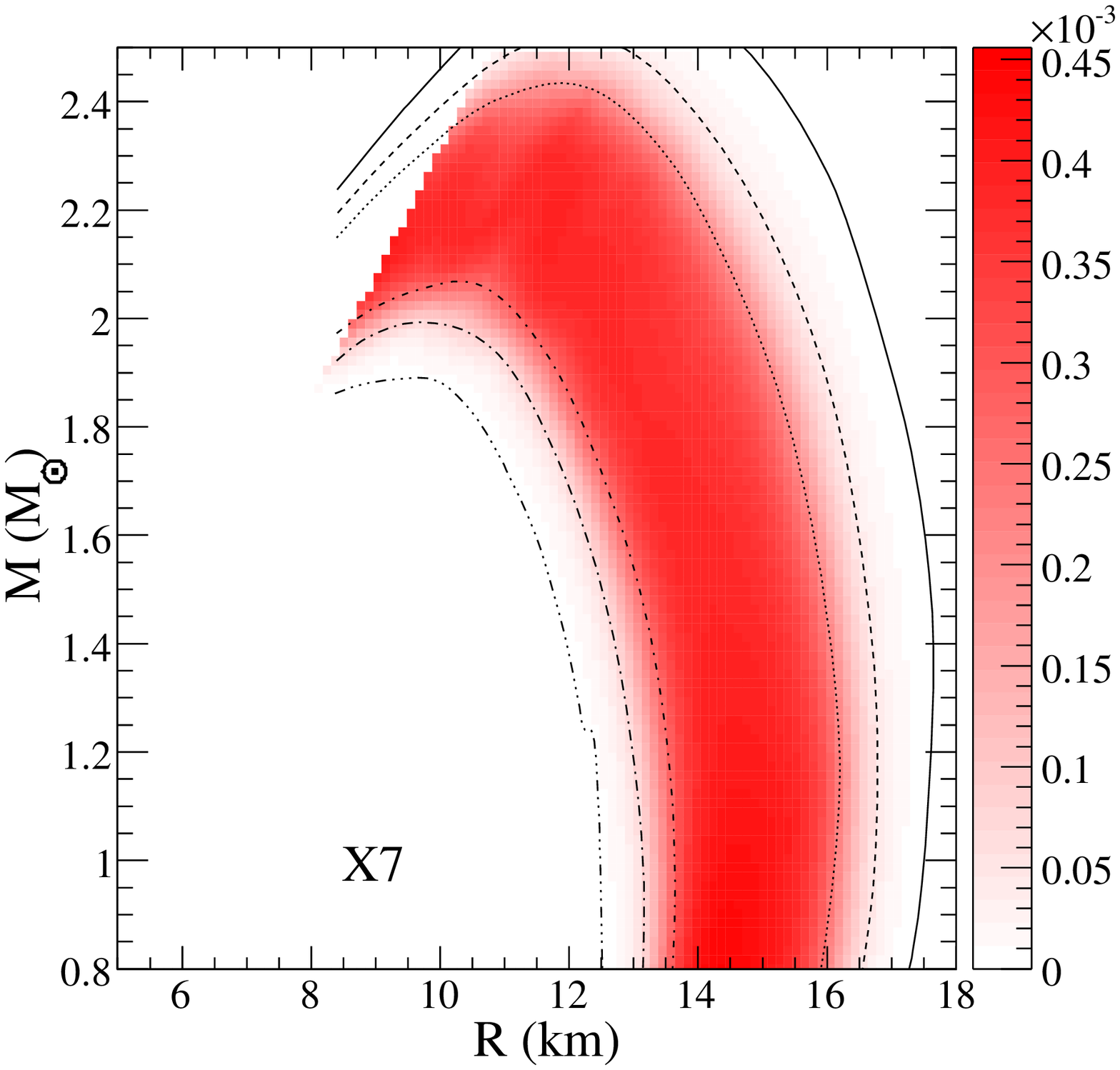}
\includegraphics[width=3.5in]{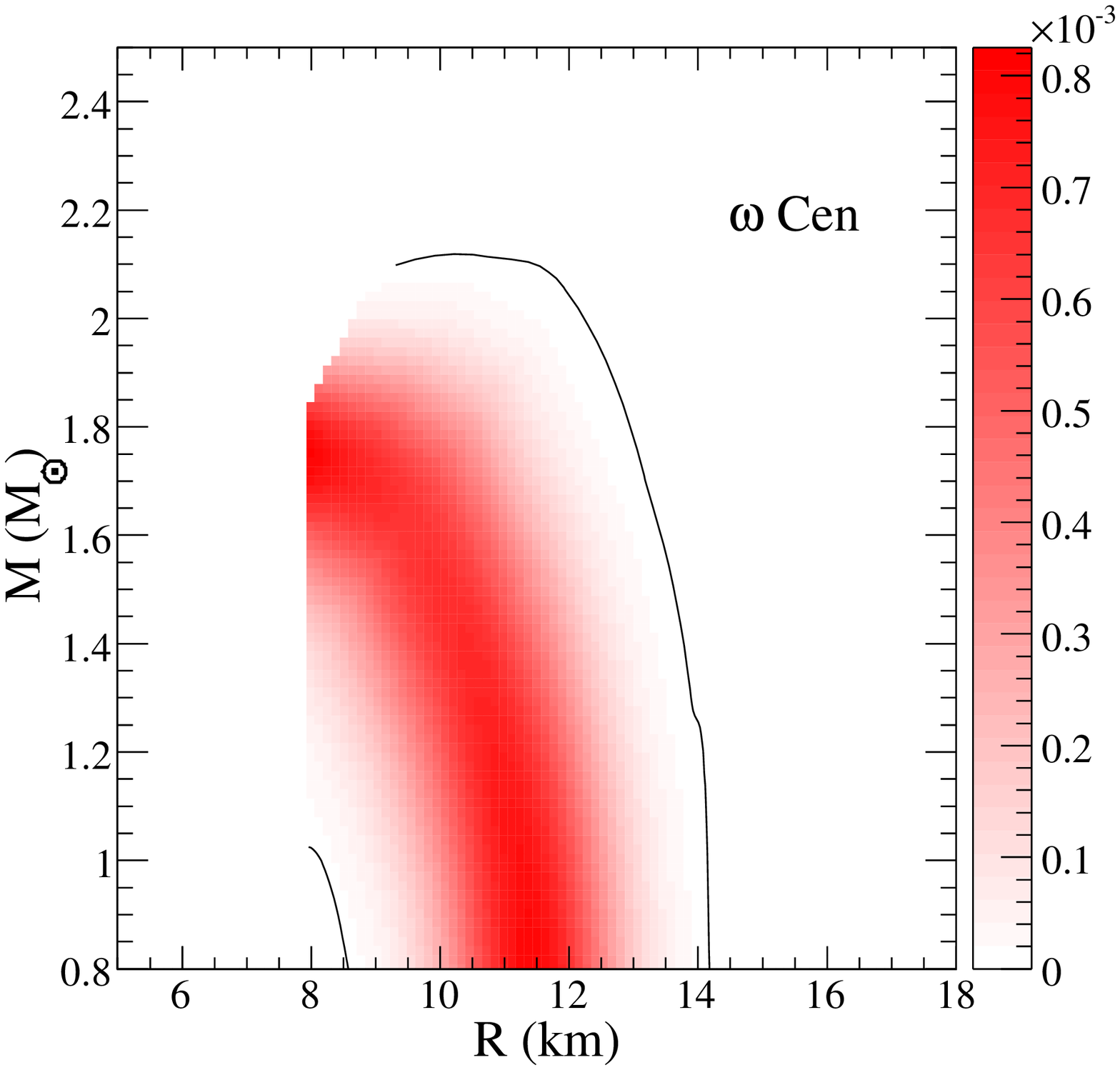}
\caption{%
  (Upper left panel) The solid curve is the 99\% probability
  contour in the mass-radius plane for the LMXRB in M13
  from~\citet{Webb07}. The red shading shows the probability
  distribution adopted in this paper.
  (Upper right panel) The various
  lines are the 68\%, 90\%, and 99\% probability contours for X7 in 47
  Tuc from~\citet{Heinke06}. The red shading shows the probability
  distribution adopted in this paper. 
  (Lower left panel) The same as
  the upper left panel except for $\omega$ Cen. 
  All distributions, $P_i$, are normalized so that
  $\int P_i~dM~dR = 1$.}
\label{fig:other}
\end{figure}

Some of the strongest constraints on  properties of transient neutron stars are for
X7 in the globular cluster 47 Tuc~\citep{Heinke06}, and for the neutron
stars in the globular clusters $\omega$~Cen and M13~\citep{Webb07}. The
99\% contours which define the $M$ and $R$ values inferred by the
atmosphere models in Fig.~8 of~\citet{Webb07} and the 68\%, 90\%, and
99\% contours inferred by the atmosphere models in Fig.~2 of
\citet{Heinke06} are shown in Fig.~\ref{fig:other}. For use in
constraining the EOS, we need to create probability distributions for
these observations in the $M,R$ plane. While one can roughly represent
the published constraints on $M$ and $R$ in terms of a single
constraint on $R_{\infty}$, this is accurate only for smaller masses.
To construct a better representation of the data, we first note that
the likelihood contours are nearly perpendicular to lines emanating
from the origin of the $M,R$ plane. To construct a probability
distribution, we assume that the probability is distributed as a
Gaussian with a width determined by the spacing of the contours across
a line going through the origin. Figure~\ref{fig:other} shows a
comparison between the resulting probability distributions used in
this work and contours obtained in the original references.

\section{Application of statistical methods to constraints on the EOS}
\label{sect:bayes}

As is clear from the preceding discussion, not all of the
uncertainties involved in constraining the masses and radii of neutron
stars are under control. Nevertheless, it is interesting to understand
what these observations may imply for the EOS. Furthermore, it is
important to {\em quantify} their implications for the EOS in order to
motivate future observational work that will reduce these
uncertainties. 

In this section, we apply a Bayesian analysis to the data described
above. We will first briefly review the formalism
(\S~\ref{sec:bayesian-analysis}), and then develop
(\S~\ref{sec:eos-parameterization}) a general parameterization of the
EOS. The output probability distribution for the EOS parameters and
the EOS itself are presented in \S~\ref{sec:bayes-results}. The most
probable masses and radii are presented in
\S~\ref{sec:bayes-results2}.  Even though the results of
\S~\ref{s.pre-bursts} strongly suggest that $\rph = R$ is disfavored,
we include results for both $\rph = R$ and $\rph \gg R$ 
to simplify comparison to previous studies.

\subsection{Bayesian analysis}\label{sec:bayesian-analysis}

Bayes theorem can be formulated as~\citep[see, e.g.,][]{Grinstead97}
\begin{equation}
P({\cal M}|D) = \frac{P(D|{\cal M}) P({\cal M})}{P(D)} \, ,
\end{equation}
where $P({\cal M})$ is the prior probability of the model ${\cal M}$
without any information from the data $D$, $P(D)$ is the prior
probability of the data $D$, $P(D|{\cal M})$ is the conditional
probability of the data $D$ given the model ${\cal M}$, and $P({\cal
  M}|D)$ is the conditional probability of the model ${\cal M}$ given
the data $D$. This latter quantity, $P({\cal M}|D)$ is what we want to
obtain, namely, the probability that a given model is correct given
the data.

For many non-overlapping models ${\cal M}_i$ which
exhaust the total model space ${\cal M}$, this relation can be
rewritten
\begin{equation}
P({\cal M}_i|D) = \frac{P(D|{\cal M}_i) P({\cal M}_i)}{\sum_j
  P(D|{\cal M}_j) P({\cal M}_j)} \, . \label{eq:bayes2}
\end{equation}
For our problem, the model space consists of all of the parameters for
the equation of state (EOS), $p_{i=1,\ldots,N_p}$, plus values for all
of the masses of the neutron stars for which we have data,
$M_{i=1,\ldots,N_M}$. From the parameters, $p_i$, we can construct the
EOS and solve the TOV equations to get a radius $R_i$ for each of the
neutron star masses $M_i$. 
To be more concise in the following, we refer to our model 
${\cal M}(p_1,p_2,...,p_{N_p},M_1,M_2,...,M_{N_M})$ as
${\cal M}(p_{1{\ldots}N_p},M_{1{\ldots}N_M})$.
Applied to this specific problem,
eq.~(\ref{eq:bayes2}) is
\begin{eqnarray}
\lefteqn{P\left[{\cal M}(p_{1{\ldots}N_p},M_{1{\ldots}N_M})|D\right] = 
 P\left[D|{\cal M}(p_{1{\ldots}N_p},M_{1{\ldots}N_M})\right] }\nonumber\\
 &\times&
 P\left[{\cal M}(p_{1{\ldots}N_p},M_{1{\ldots}N_M})\right] 
\left\{\int P[D|{\cal M}] P[{\cal M}]~d^{N} {\cal M}
\right\}^{-1}  \label{eq:bfull}
\end{eqnarray}
where $N=N_p+N_M$ is the dimensionality of our model space. The total
number of EOS parameters is $N_p = 8$ and the total number of neutron
stars in our data set is $N_M=6$. 

We construct our data $D$ as a set of $N_M$ probability distributions,
${\cal D}_i(M,R)$ in the $(M,R)$ plane, which are all normalized to
unity, i.e.
\begin{equation}
 \int_{M_{\mathrm{low}}}^{M_{\mathrm{high}}} dM~
\int_{R_{\mathrm{low}}}^{R_{\mathrm{high}}} dR~{\cal D}_i(M,R) = 
1~\forall~i \, .
\end{equation}
This normalization ensures that the data set for each neutron star is
treated on an equal footing. We choose $M_{\mathrm{low}} = 0.8\,\Msun$
because current core-collapse supernova simulations fail to generate
lower neutron star masses. The remaining limits, $M_{\mathrm{high}} =
2.5\,\Msun$, $R_{\mathrm{low}} = 5\,\mathrm{km}$, and
$R_{\mathrm{high}} = 18\,\mathrm{km}$ are extreme enough to ensure
that they have no impact on our final results. None of the probability
distributions ${\cal D}$ inferred from the data described above has a
significant probability for neutron stars outside of these ranges.
Note also that the results for the neutron stars in M13 and
$\omega$~Cen are cutoff in~\citet{Webb07} for radii below 8~km so our
distributions also exhibit the same cutoff.

In order to apply equation~(\ref{eq:bfull}) to our problem, we assume that
$P(D|{\cal M})$, the conditional probability of the data given the
model, is proportional to the product over the probability
distributions ${\cal D}_i$ evaluated at the masses which are chosen in
the model and evaluated at the radii which are determined from the
model, i.e.
\begin{equation}
P\left[D|{\cal M}(p_{1{\ldots}N_p},M_{1{\ldots}N_M})\right] 
\propto
\prod_{i=1,\ldots,N_M} \left. {\cal D}_i(M,R) \right|_{M=M_i,R=R(M_i)}
\label{eq:maxlike}
\end{equation}
This implicitly assumes that all of the data distributions ${\cal
  D}_i$ are independent of each other and also independent of the
model assumptions and prior distributions. Another required input for
equation~(\ref{eq:bfull}) is the prior distribution. We assume that
the prior distribution is uniform in all of the $n_p+n_M$ model
parameters, except for a few physical constraints on the parameter
space described below. Taking a uniform distribution just means that
the $P({\cal M})$ terms cancel from equation~(\ref{eq:bfull}) and the
integration limits become the corresponding prior parameter limits.

In the maximum likelihood formalism, the function $P\left[D|{\cal
    M}(p_{1{\ldots}N_p},M_{1{\ldots}N_M})\right] $ is equivalent to
the likelihood function, and maximizing the likelihood function gives
the best fit to the data. In the case that the probability
distributions represent Gaussian uncertainties, 
then the best fit is a equivalent to a
least-squares fit~\citep{Bevington02}. In Bayesian inference, model
parameters are determined using marginal estimation, where the
posterior probability for a model parameter $p_j$ is given by
\begin{equation}
P[p_j|D](p_j) = \frac{1}{V} 
\int P[D|{\cal M}]~d p_1~d p_2~\ldots~d p_{j-1}~d p_{j+1}~\ldots
~d p_{N_p}~d M_1~d M_2~\ldots~d M_{N_M}
\label{eq:expect}
\end{equation}
where $V$ is the denominator in equation~(\ref{eq:bfull}), without the
model priors that determine the integration limits. The
one-dimensional function $P[p_j|D](p_j)$ represents the probability
that the $j$-th parameter will take a particular value given the
observational data. Our problem thus boils down to computing integrals
of the form in equation~(\ref{eq:expect}). We use the
Metropolis-Hastings algorithm to construct a Markov chain to simulate
the distribution $P\left[D|{\cal
    M}(p_{1{\ldots}N_p},M_{1{\ldots}N_M})\right]$. For each point, we
generate the EOS parameters $p_i$ and the neutron star masses $M_i$
from a uniform distribution within limits that are described below.
The TOV equations are solved and this generates a $R(M)$ curve and the
radii for each neutron star, $R_i$. From these 6 masses and radii, the
weight function $P\left[D|{\cal M}\right]$ is computed from
eq.~(\ref{eq:maxlike}) and the point is rejected or accepted using the
Metropolis algorithm. In order to compute posterior probabilities
$P[p_j|D](p_j)$, we construct several histograms of the integrand $
P[D|{\cal M}]$ from all of the points that were accepted. To
construct the 1-$\sigma$ regions, we sort the histogram bins by
decreasing probability, and select the first N bins which exhaust 68\%
of the total weight. A similar procedure is used for the 2-$\sigma$
regions. In order to constrain the full EOS of neutron star matter, we
histogram the value of the pressure predicted by each EOS in the
Markov chain on a fixed grid of energy density. To create the
predicted curve, $R(M)$, we construct a histogram of the predicted
radii for each EOS in the Markov chain on a fixed mass grid.

This analysis is easily extensible to a different number of EOS
parameters or a different number of neutron star data sets. The only
issue is that of computer time: the TOV equations must be solved for
each point in the model space, and enough points must be selected to
cover the model space fully. Another advantage is the explicit
presence of the prior distributions, $P({\cal M})$. Although we have
set these distributions to unity for this work, future work will
utilize these terms to examine the impact of constraints on the
equation of state from terrestrial experiments.

\subsection{Parameterization of the EOS}\label{sec:eos-parameterization}

We divide the EOS into four energy density regimes. The region below
the transition energy density $\varepsilon_{\mathrm{trans}} \approx
\varepsilon_0/2$ is the crust, for which we use the EOS
of~\citet{Baym71} and~\citet{Negele83}. Here $\varepsilon_0$ is the
nuclear saturation energy density; it is convenient to remember that
the nuclear saturation baryon density $0.16\,\mathrm{fm^{-3}}$
corresponds to an energy density of $\approx
160\,\mathrm{MeV\,fm^{-3}}$ and a mass density of $\approx 2.7\times
10^{14}\mathrm{\,g\,cm^{-3}}$. For $\varepsilon_{\mathrm{trans}} <
\varepsilon < \varepsilon_1$, we use a schematic expression
representing charge-neutral uniform baryonic matter in beta
equilibrium that is compatible with laboratory data. Finally, two
polytropic pressure-density relations are used in the regions
$\varepsilon_1<\varepsilon<\varepsilon_2$ and
$\varepsilon>\varepsilon_2$. The densities $\varepsilon_{1}$ and
$\varepsilon_{2}$ are themselves parameters of the model. The
schematic EOS for
$\varepsilon_{\mathrm{trans}}<\varepsilon<\varepsilon_1$ is taken to
be
\begin{equation}
\varepsilon = n_B\left\{m_B+B+\frac{K}{18}(u-1)^2+
\frac{K^{\prime}}{162}(u-1)^3+ (1- 2 x)^2 \left[S_k u^{2/3}+
S_p u^{\gamma} \right]+{3\over4}\hbar c x(3\pi^2n_bx)^{1/3}\right\}
\label{eq:ldeos}
\end{equation}
where $n_B$ is the baryon number density, $m_B$ is the baryon mass,
$u=n_B/n_0$, and $x$ is the proton (electron) fraction. The saturation
number density, $n_0$, is fixed at 0.16~fm$^{-3}$, the binding energy
of saturated nuclear matter, $B$, is fixed at $-$16~MeV, and the
kinetic part of the symmetry energy, $S_k$, is fixed at 17~MeV. The
compressibility $K$, the skewness $K^{\prime}$, the bulk symmetry
energy parameter, $S_v\equiv S_p+S_k$ (where $S_p$ is the potential
part of the symmetry energy), and the density dependence of the
symmetry energy, $\gamma$, are parameters, which we constrain to lie within the ranges specified in Table~\ref{tab:pri}.
These limits operate as constraints in our otherwise trivial prior
distributions, $P({\cal M})$. To avoid bias in our results and to
ensure that our model space is not over-constrained, we have
intentionally made these ranges larger than normally expected from
modern models of the EOS of uniform matter which are fit to laboratory
nuclei. While in principle the crust EOS for each set of EOS
parameters could be different, as described in~\citet{Steiner08}, in
practice the masses and radii are not strongly affected by changes in
the crust at this level. The transition between the crust EOS and the
low-density EOS is typically around half of the nuclear saturation
density and is determined for each parameter set by ensuring
that the energy is minimized as a function of the number density. 
We have opted, at this stage, not to include correlations between
parameters that have been shown to exist from nuclear systematics or
neutron matter calculations. For example, the values of $S_v$ and
$\gamma$ (or, equivalently, $S_v$ and $S_s$, the surface symmetry
parameter) are highly correlated~\citep{Steiner05} in liquid drop mass
formula fits to nuclear masses. Such correlations will be considered
in a future publication.

The last term in equation~(\ref{eq:ldeos}) is due
to electrons.  The proton fraction $x$ is determined as a function of
density by the condition of beta equilibrium
\begin{equation}
{\partial\varepsilon\over\partial x}=\hbar c(3\pi^2n_Bx)^{1/3}-4
\left[S_ku^{2/3}+S_pu^\gamma\right](1-2x)=0\,,
\label{eq:betaeq}
\end{equation}
which has the solution
\begin{equation}
x={1\over4}\left[\left(\sqrt{d+1}+1\right)^{1/3}-
\left(\sqrt{d+1}-1\right)^{1/3}\right]^3\,,
\label{eq:beta0}
\end{equation}
where
\begin{equation}
d={\pi^2n_B\over288}\left({\hbar c\over
\left[S_ku^{2/3}+S_pu^\gamma\right]}\right)^3\,.
\label{eq:d}
\end{equation}
We also include muons in our equation of state, but
they are not included in the above expressions for clarity.

\begin{deluxetable}{lcc}
\tablecolumns{3}
\tablecaption{Prior limits for the EOS parameters}
\tablewidth{0pt}
\tablehead{
\colhead{Quantity} & \colhead{Lower limit} & \colhead{Upper limit}
}
\startdata
$K$ (MeV) & 180 & 280 \\
$K^{\prime}$ (MeV) & -1000 & -200 \\
$S_v$ (MeV) & 28 & 38 \\
$\gamma$ & 0.2 & 1.2 \\
$n_1$ (fm$^{-3}$) & 0.2 & 1.5 \\
$n_2$ (fm$^{-3}$) & 0.2 & 2.0 \\
${\varepsilon}_1$ (MeV~fm$^{-3}$) & 150 & 600 \\
${\varepsilon}_2$ (MeV~fm$^{-3}$) & ${\varepsilon}_1$ & 1600 \\
\enddata
\label{tab:pri}
\end{deluxetable}

For lack of a clear theoretical understanding of the nature of matter
above saturation densities, we parameterize the high-density EOS by a
pair of polytropes. \citet{Read09} have shown that such a
parameterization can effectively model a wide variety of theoretical
model predictions in this density range. Specifically, for energy
densities above a parameterized value $\varepsilon_1$, the EOS is
\begin{equation}
P = K_1 \varepsilon^{1+1/n_1}
\end{equation}
where $P$ is the pressure, $n_1$ is the polytropic index,
and $K_1$ is a coefficient. Note that we are parameterizing
the high-density EOS as polytropes in the total energy density
$\varepsilon$ rather than the number density, and that this EOS is
assumed to be in beta equilibrium so it automatically includes
leptonic contributions. Above a parameterized energy density,
$\varepsilon_2$, we use a second polytrope
\begin{equation}
P = K_2 \varepsilon^{1+1/n_2} \, .
\end{equation}
We choose the transition densities $\varepsilon_{1}$ and
$\varepsilon_{2}$ and the polytropic indices $n_1$ and $n_2$ as
parameters. The coefficients $K_1$ and $K_2$ are determined by
pressure and energy density continuity at the transition densities,
with values limited such that $1600{\rm~MeV
  fm}^{-3}>\varepsilon_2>\varepsilon_1>150\,\mathrm{MeV\,fm^{-3}}$.
Note that $1600\,\mathrm{MeV\,fm^{-3}}$ is either larger than or
nearly as large as the central energy density of most configurations.
In addition, we limit $\varepsilon_1 < 600\,\mathrm{MeV\,fm^{-3}}$, so
that the parameters of the schematic EOS maintain a close connection
to their usual definitions in terms of properties near the saturation
density. Finally, we limit the polytropic indices with $ 0.2 < n_1 <
1.5 $ and $0.2 < n_2 < 2.0$; our results are insensitive to this
choice of limits.

While phase transitions are included in our models because they will
appear like successive polytropes with different indices, some
parameter sets do not imply any phase transition, as they simply model
equations of state which have effective polytrope indices which vary
with density. Models with more than two strong phase transitions will
not be well reproduced by our parameterization, but it is not clear
that these models are particularly realistic.

We thus have 8 total EOS parameters, $K$, $K^{\prime}$, $S_v$,
$\gamma$, $n_1$, $n_2$, $\varepsilon_1$, and $\varepsilon_2$, with
limiting values summarized in Table~\ref{tab:pri}. In addition to
these parameter limits, some combination of parameters must be
rejected because they are unphysical. Unphysical combinations include
those in which 
\begin{enumerate}
\item the maximum mass is smaller than $1.66\,\Msun$, which is
  2-$\sigma$ below the mass of PSR~J1903+0327, $1.74 \pm 0.04\,\Msun$
  ~\citep{Champion08};
\item the EOS becomes acausal below the central density of the maximum
mass star;
\item the EOS is anywhere hydrodynamically unstable, i.e., has a
pressure that decreases with increasing density; and 
\item the maximum mass star has a maximum stable rotation rate less
  than 716~Hz, the spin frequency of the fastest known pulsar, Ter
  5AD~\citep{Hessels06}. The spin frequency at which equatorial
  mass-shedding commences is given to within a few percent
  by~\citep{Haensel09}
\begin{equation}
f_K \simeq 1.08 \left(\frac{M}{ \Msun}\right)^{1/2}
\left({10{\rm~km}\over R}\right)^{3/2}{\rm~kHz}\,.
\label{eq:fk}
\end{equation}
There has been claimed possible evidence for a higher spin frequency
in XTE J1239-285~\citep{Kaaret07}, but this observation does not have
strong statistical significance and has not been confirmed by
subsequent observations.  
\end{enumerate}
In addition to these criteria, during the Monte Carlo
generation of neutron star mass sets for the Bayesian analysis, if one of
the seven masses is larger than the maximum mass for the
selected EOS, that realization is discarded and a new one is selected.

\begin{figure}
\begin{center}
\includegraphics[width=4in]{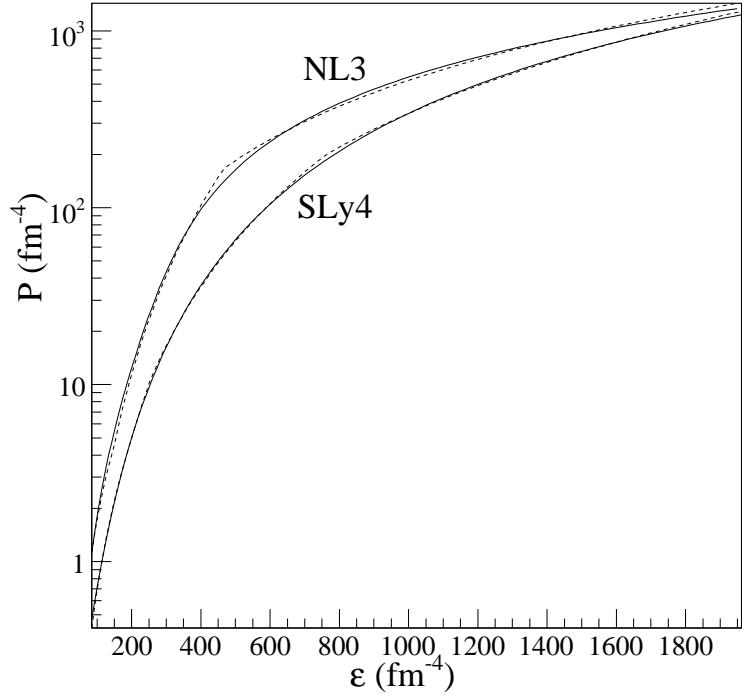}
\end{center}
\caption{Fits of our parametrized EOS (\emph{dotted lines}) to the
Skyrme EOS SLy4 and the field-theoretical EOS NL3(\emph{solid lines}).}
\label{fig:fit}
\end{figure}

To show that this model accurately represents significantly different
EOSs, we fit it to the Skyrme model SLy4 which gives relatively small
neutron star radii (of order 10 km for $M=1.4\,\Msun$), and the
field-theoretical model NL3 which gives rather larger radii, of order
15~km, for the same mass. These fits are shown in Fig.~\ref{fig:fit},
illustrating that the associated EOSs are reproduced to within a few
percent. Our parameterization includes many extreme models, including
those like NL3 with rather large neutron star radii. We will
find below that such models are ruled out, however, by the observational data.

\subsection{EOS Results from the Statistical Analysis}
\label{sec:bayes-results}

\begin{figure}
\includegraphics[width=3.5in]{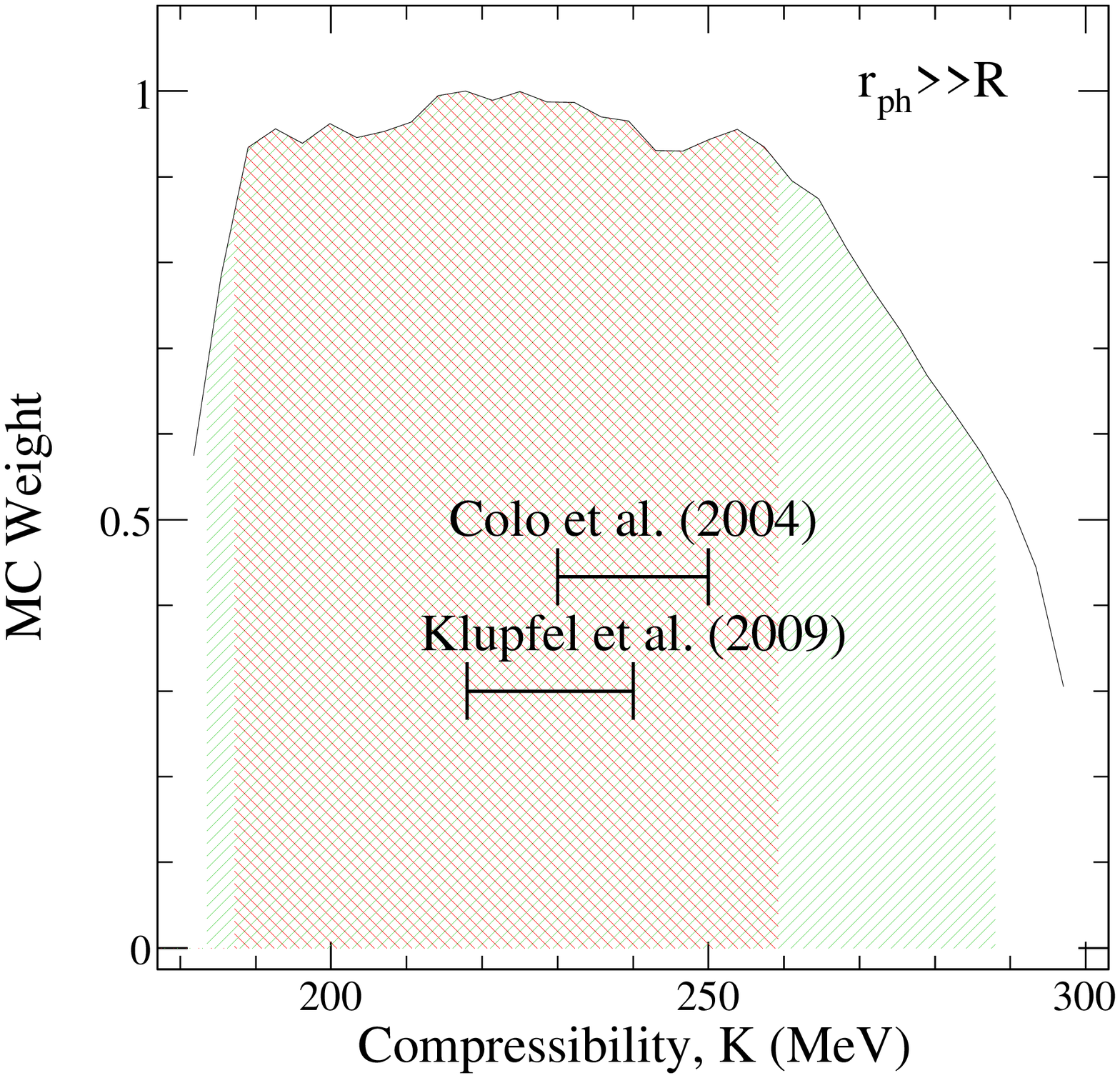}
\includegraphics[width=3.5in]{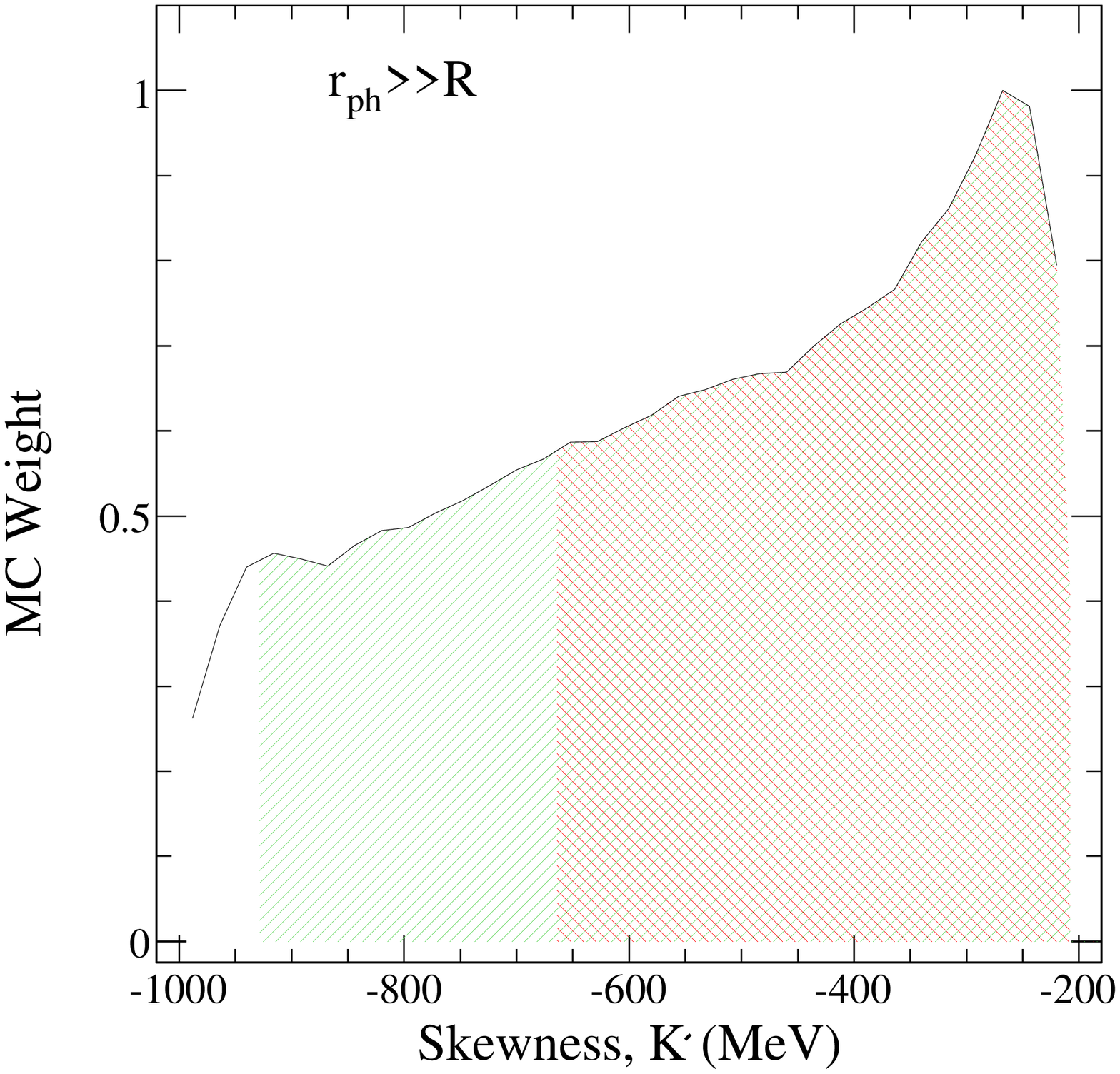}
\includegraphics[width=3.5in]{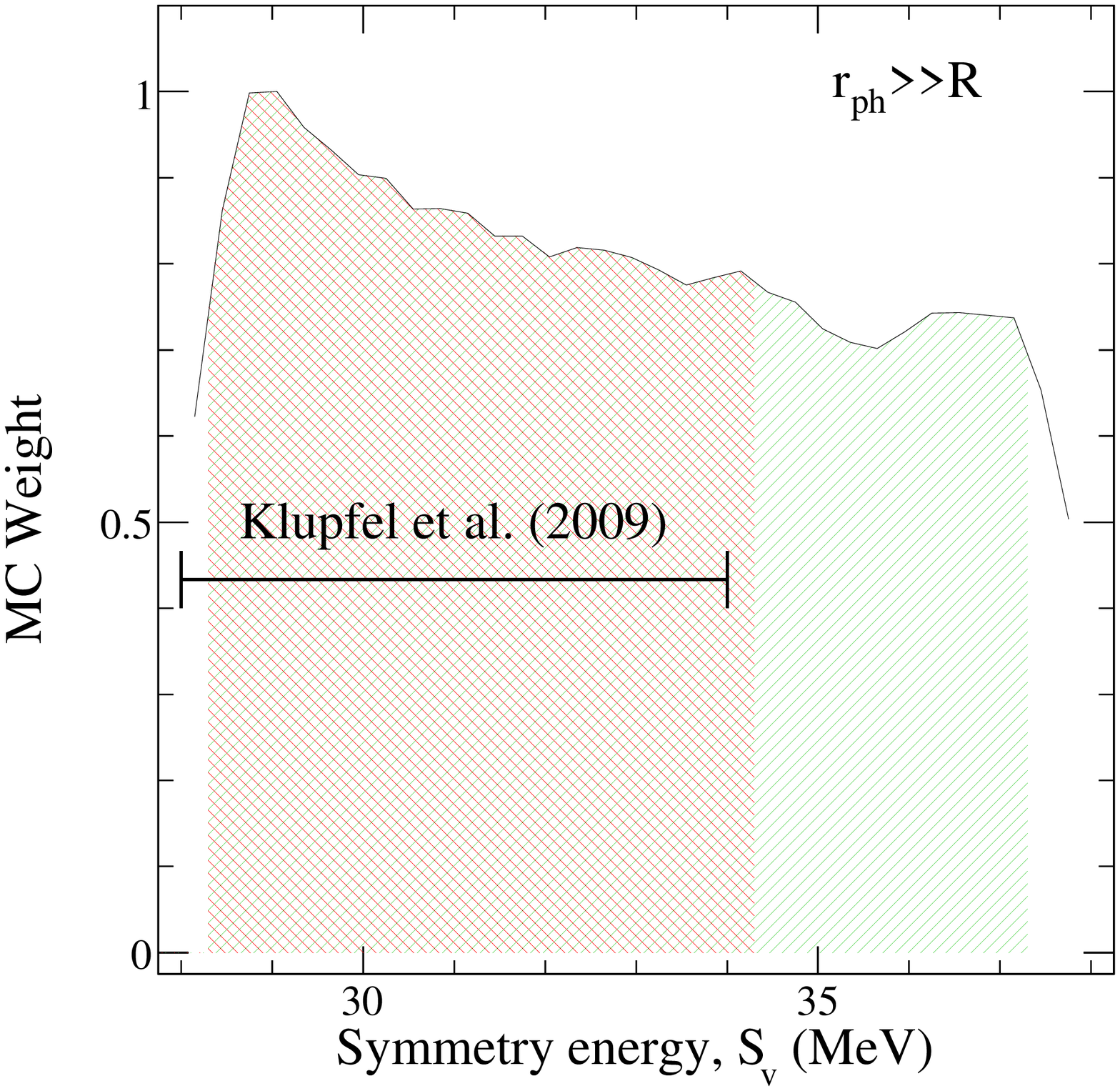}
\includegraphics[width=3.5in]{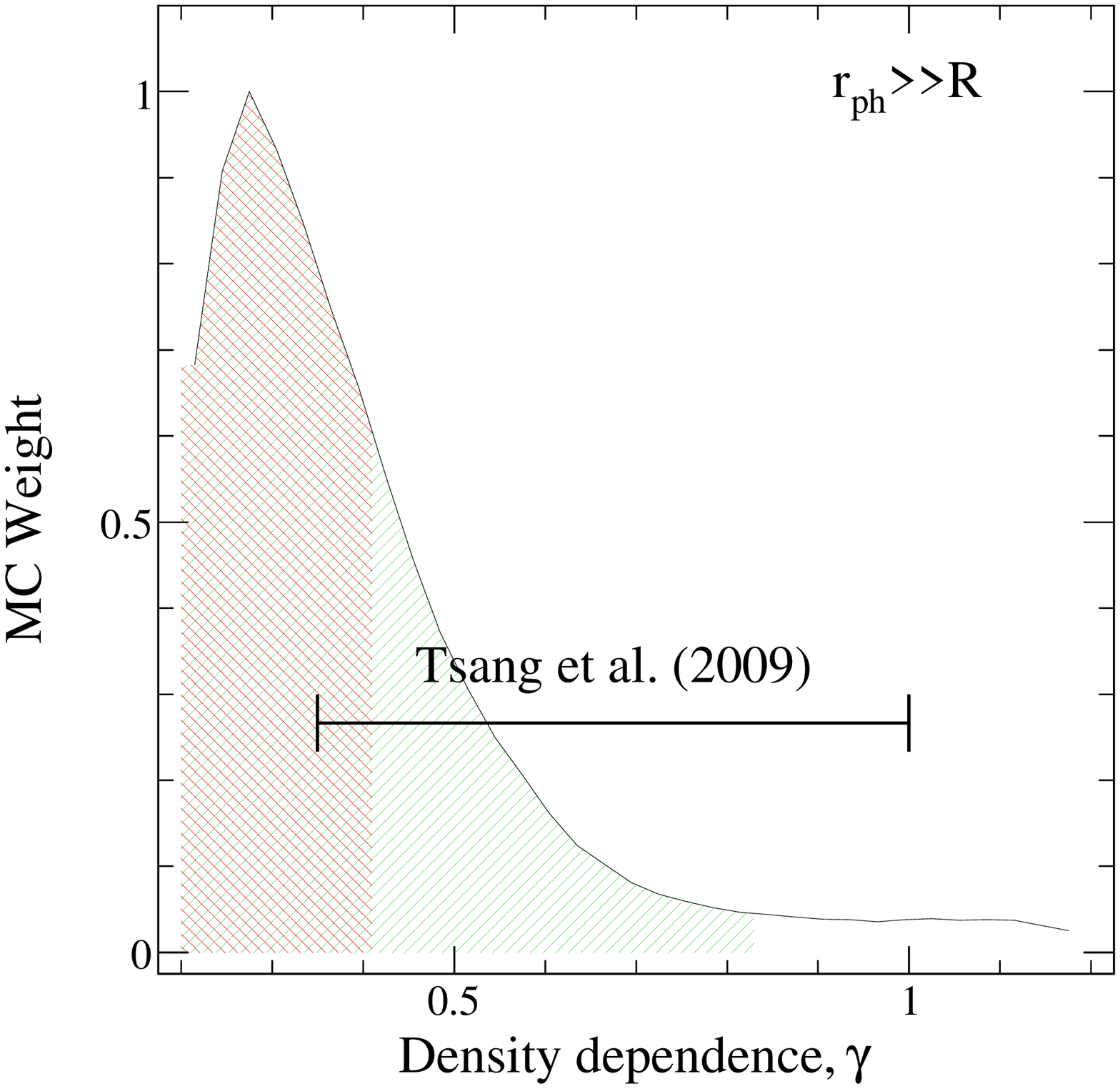}
\caption{Histograms for the compressibility $K$, skewness
  $K^{\prime}$, symmetry energy $S_v$, and density dependence of the
  symmetry energy $\gamma$. The 1-$\sigma$ (\emph{double-hatched}) and
  2-$\sigma$ (\emph{single-hatched}) confidence regions are also
  indicated. These results assume $\rph \gg R$. }
\label{fig:eoslo}
\end{figure}

\begin{figure}
\includegraphics[width=3.5in]{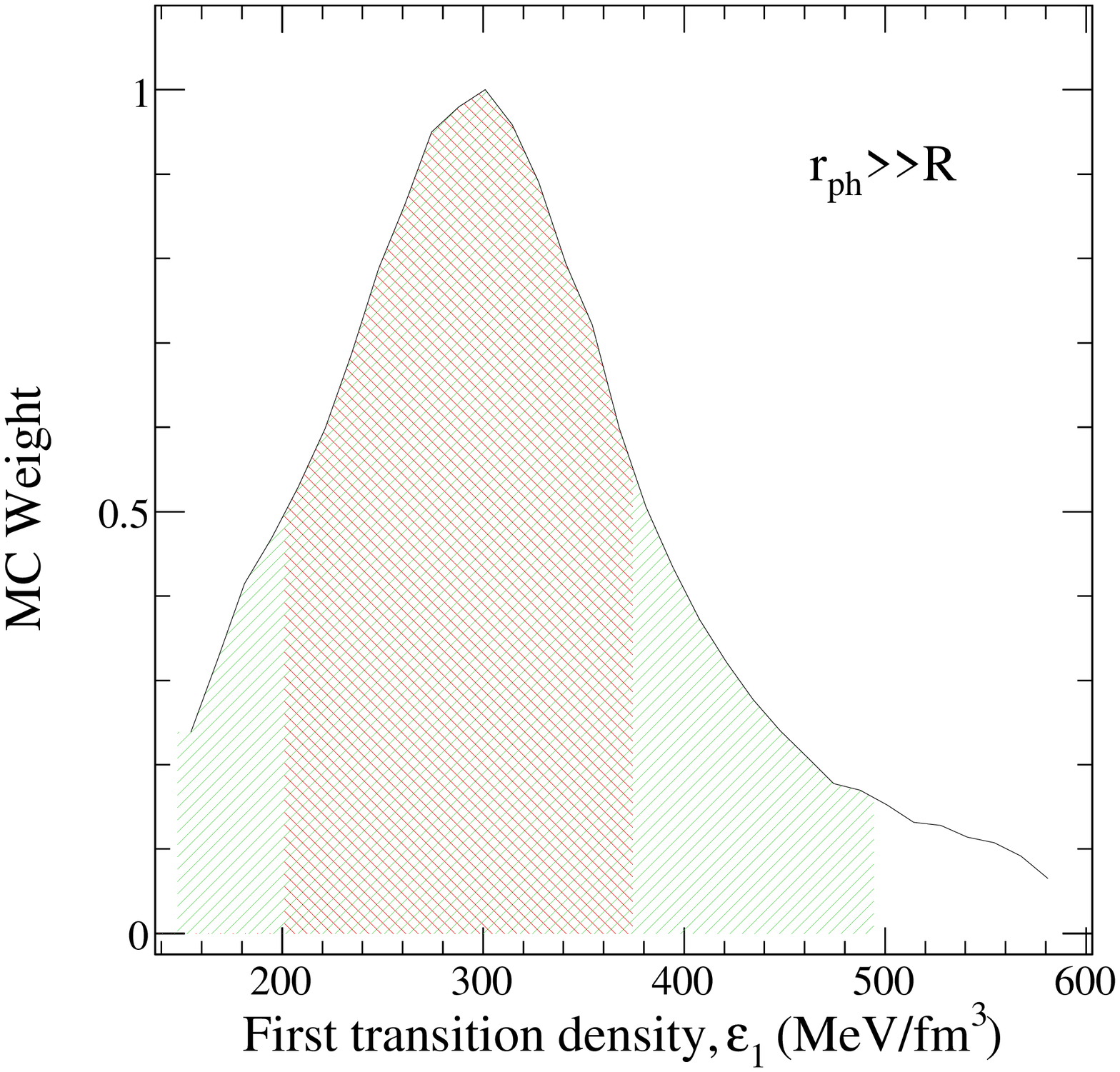}
\includegraphics[width=3.5in]{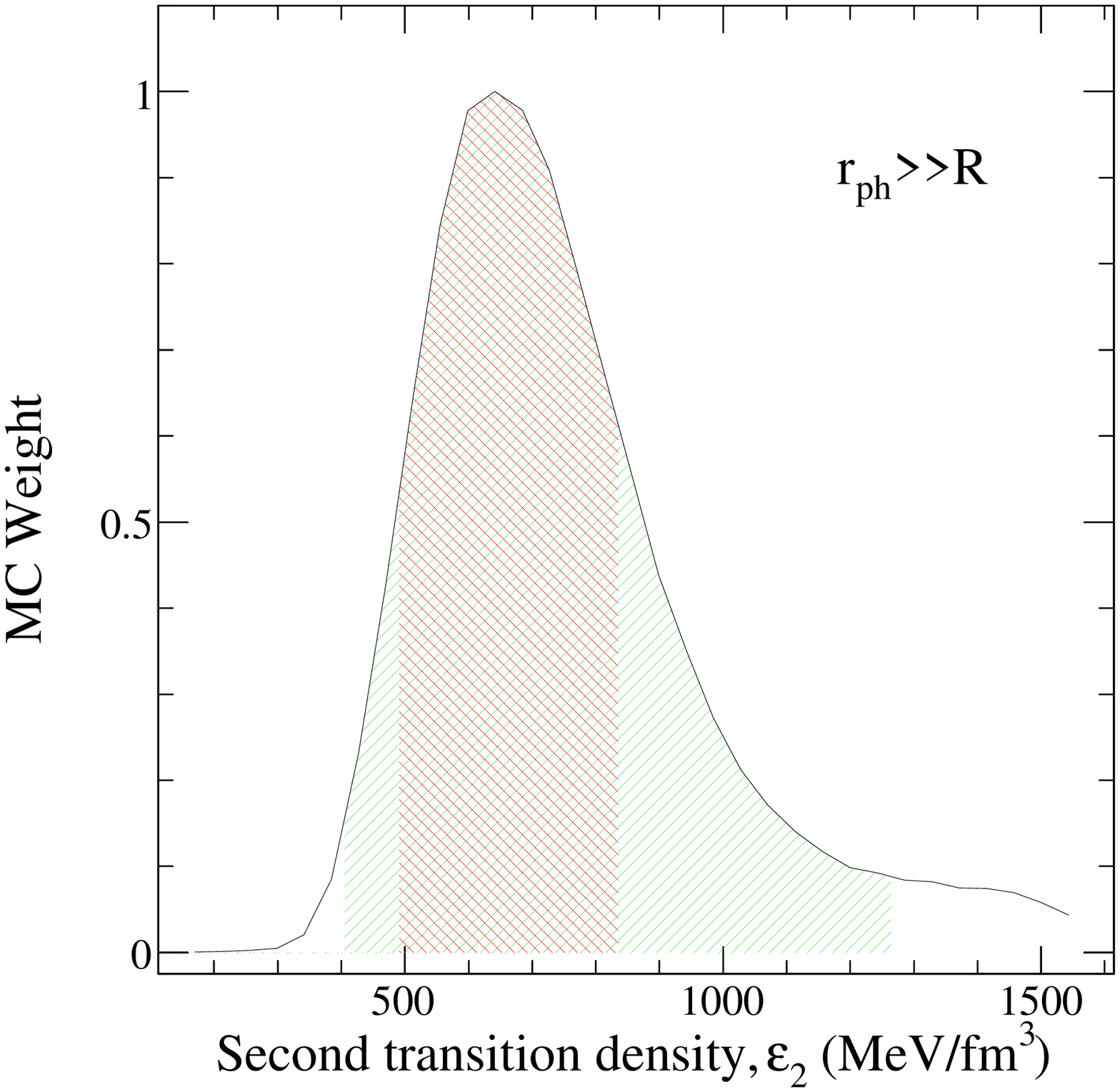}
\includegraphics[width=3.5in]{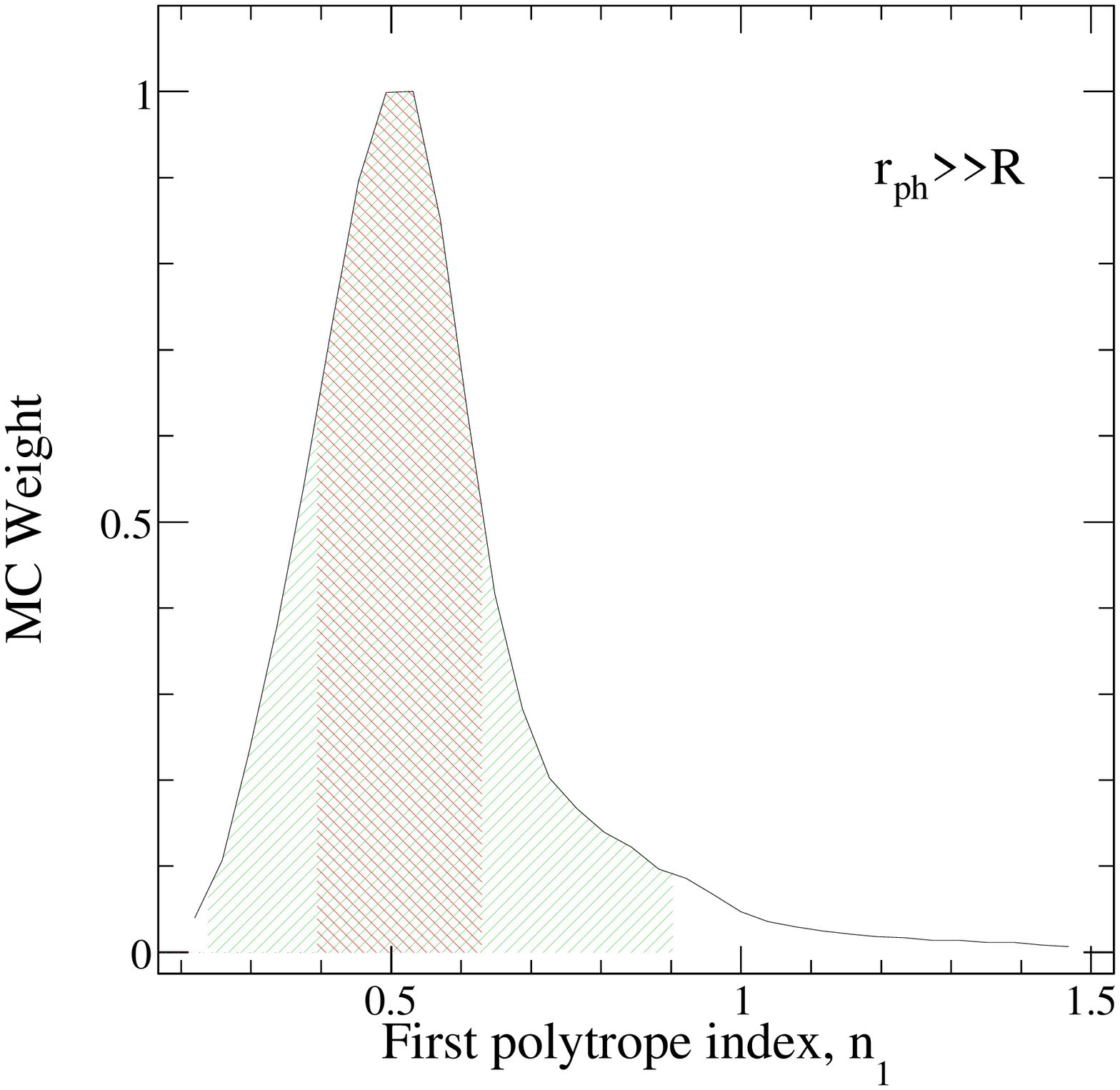}
\includegraphics[width=3.5in]{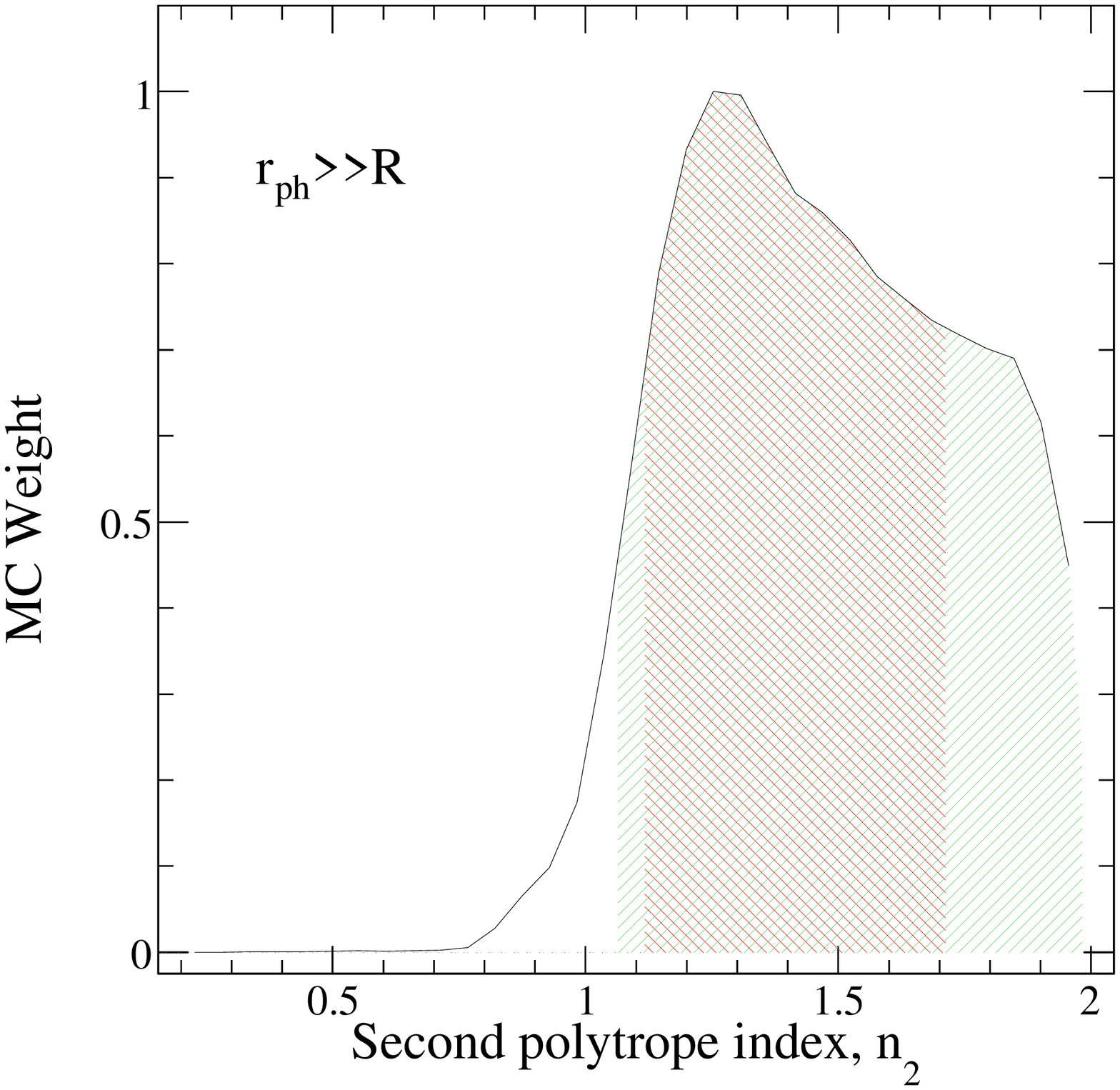}
\caption{Histograms and confidence regions for the transition energy
  densities, $\varepsilon_1$ and $\varepsilon_2$ and the polytropic
  indices $n_1$ and $n_2$, for $\rph \gg R$.}
\label{fig:eoshi}
\end{figure}

We first analyze the EOS implied by observations for the case in which
X-ray bursts are modeled assuming $\rph \gg R$. The histograms for the
EOS parameters are given in Figs.~\ref{fig:eoslo}--\ref{fig:eoshi}.
The double-hatched (red) and single-hatched (green) regions outline
the one- and two-sigma confidence regions which are suggested by the
simulation. The one- and two-sigma parameter limits are summarized in
Table \ref{tab:posteos} along with representative values obtained from
terrestrial experiments. 

It is remarkable that the inferred ranges for the schematic EOS
parameters $K$, $K^{\prime}$, $S_v$, and $\gamma$ are consistent with
the values derived from nuclear experiments. Especially significant is
the inferred low value for $\gamma$, a parameter which controls the
pressure of the EOS in the range of 1--3 $\varepsilon_0$ and therefore
the radii of neutron stars in the mass range
$1<M/\Msun<1.5$~\citep{Lattimer01}. The value of $\gamma$ also
controls the energy and pressure of pure neutron matter.
\citet{Hebeler09} has shown that several recent studies indicate a
convergence in predictions for neutron matter. For the saturation
density used there, $n_0=0.16$ fm$^{-3}$, the mean values of the
neutron matter energy and pressure are 16.3 MeV and 2.5 MeV/fm$^{-3}$,
which imply $S_v=32$ MeV and $\gamma=0.28$ with the parametrization of
Eq. (\ref{eq:ldeos}), well within our predicted ranges.

One must take care in comparing experimental constraints directly to
our results, however. The parameters determined experimentally are
often ``local'' quantities, in the sense that they are properties of
the EOS only at densities close to saturation. Also, the results from
\citet{Tsang09} mostly constrain the EOS in region near half the
saturation density. We have used the schematic EOS parameters over a
somewhat larger range of densities, from 1/2 to, typically, up to 2 or
3 times the saturation density.

The predicted skewness for the schematic equation of state has a
relatively small magnitude, centered at $K^\prime=-280$ MeV. The most
probable value for the transition between the schematic EOS and the
first polytrope is around twice the saturation density,
$\varepsilon_0$. The index of the first polytrope is sharply peaked
around 0.5, which corresponds to a polytropic exponent
$\gamma_1=1+1/n_1\simeq3$ which is rather large. Coupled with the
small magnitude of the skewness, this implies a quite stiff EOS at
supernuclear densities. Finally, note that, typically, $n_2>n_1$,
indicating that the EOS softens at high densities. A summary of the
results of the EOS parameters for both the schematic EOS and the
polytropes are given in Table~\ref{tab:posteos}.

\begin{deluxetable}{llll}
\tablecolumns{4}
\tablewidth{0pc}
\tablecaption{\label{tab:posteos}
Most probable values of the EOS parameters and their associated
1-$\sigma$ uncertainties.
}
\tablehead{
\colhead{Quantity} & \colhead{$\rph \gg R$} & \colhead{$\rph = R$} &
Experiment \\
}
\startdata
\cutinhead{Schematic EOS parameters}
$K$ (MeV) & 
\phn$216^{+43}_{-32}$ & 
\phn$190^{+50}_{-7.2}$ &
230--250~\citep{Colo04} \\
$-K^{\prime}$ (MeV) & 
$280^{+410}_{-72}$ &
$500^{+290}_{-170}$ & \\
$S$ (MeV) &
\phn\phn$29^{+5.4}_{-0.9}$ &
\phn\phn$35^{+2.1}_{-6.3}$ &
28--34~\citep{Klupfel09} \\
$\gamma$ &
\phn$0.26^{+0.15}_{-0.06}$ &
\phn$0.26^{+0.36}_{-0.06}$ &
0.35--1.0~\citep{Tsang09} \\
\cutinhead{High-density EOS parameters}
$n_1$ &
$0.51^{+0.12}_{-0.16}$ &
$0.67^{+0.20}_{-0.20}$ &
\\
$n_2$ &
$1.23^{+0.49}_{-0.16}$ &
$1.23^{+0.59}_{-0.22}$ &
\\
${\varepsilon}_1$ (MeV/fm$^{3}$) &
$290^{+80}_{-110}$ &
$320^{+120}_{-93}$ &
\\
${\varepsilon}_2$ (MeV/fm$^{3}$) &
$620^{+210}_{-170}$ &
$880^{+430}_{-210}$ &
\\
\enddata
\end{deluxetable}

We next consider the results of parameter fitting when the X-ray burst
data is modeled assuming $\rph=R$. Results are summarized and compared
to the previous case in Table \ref{tab:posteos}. Although we don't
show histograms for the parameters in this case, their behavior is
similar to those of Fig. \ref{fig:eoslo} and \ref{fig:eoshi} modulo
the different means and variations of the two cases. It is significant
that the small value for $\gamma$ found for the case $\rph\gg R$ is
duplicated for $\rph=R$, supporting a weak density dependence for the
symmetry energy and a consequent small estimate for neutron star radii
in the mass range $1\textrm{--}1.5\,\Msun$. We note that~\citet{Ozel10} also
concluded, on the basis of observations of X-ray bursts, that the
implied neutron star radii were relatively small. \citet{Ozel10}
obtained radii approximately 1~km smaller than we do, however, for $\rph=R$
(which are already 0.5--1~km smaller than those we obtain for
$\rph\gg R$) because their analysis favors the high-redshift solution,
and because they do not impose the same causality constraint on their Monte
Carlo sampling.

\begin{figure}
\includegraphics[width=3.5in]{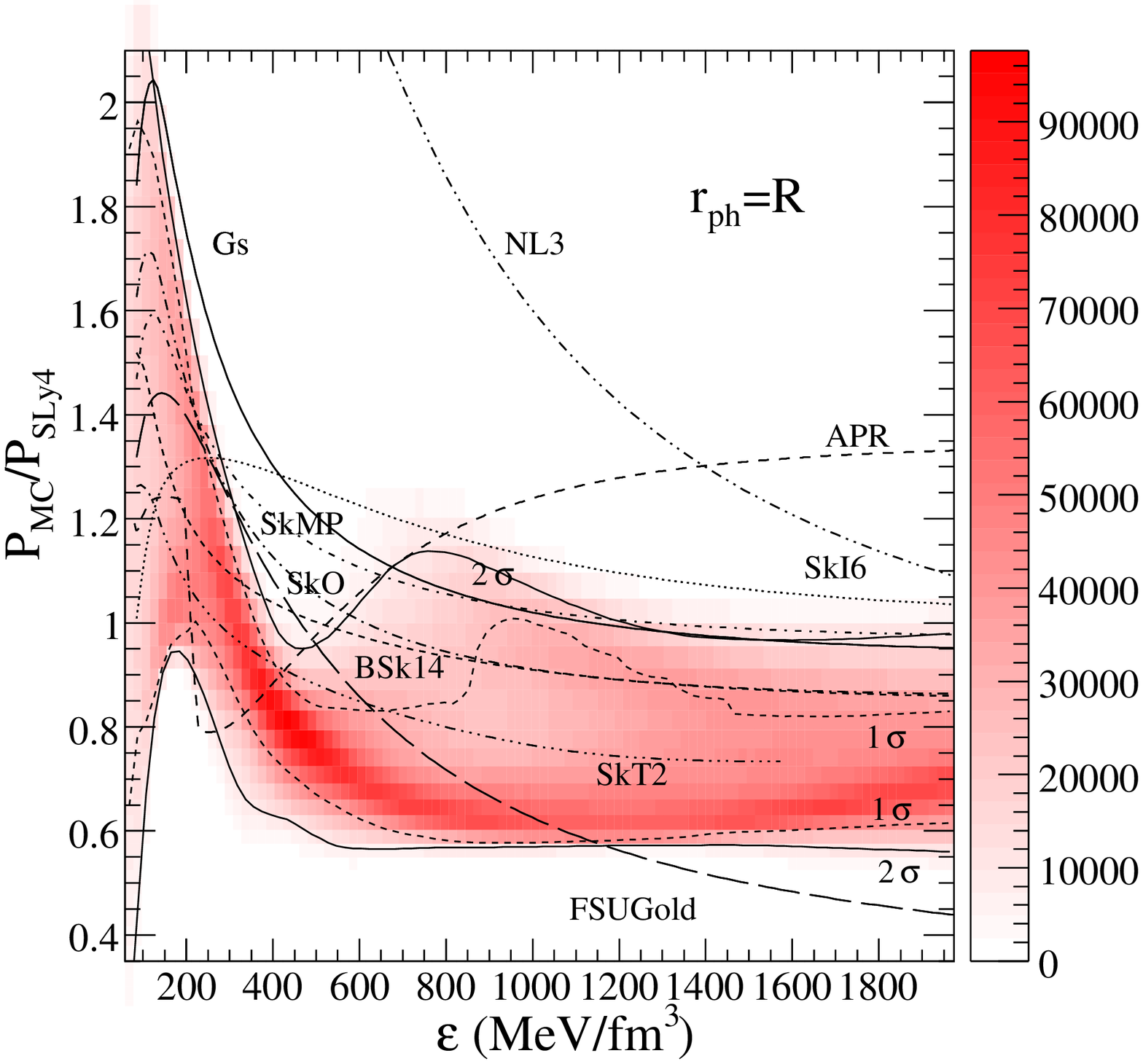}
\includegraphics[width=3.5in]{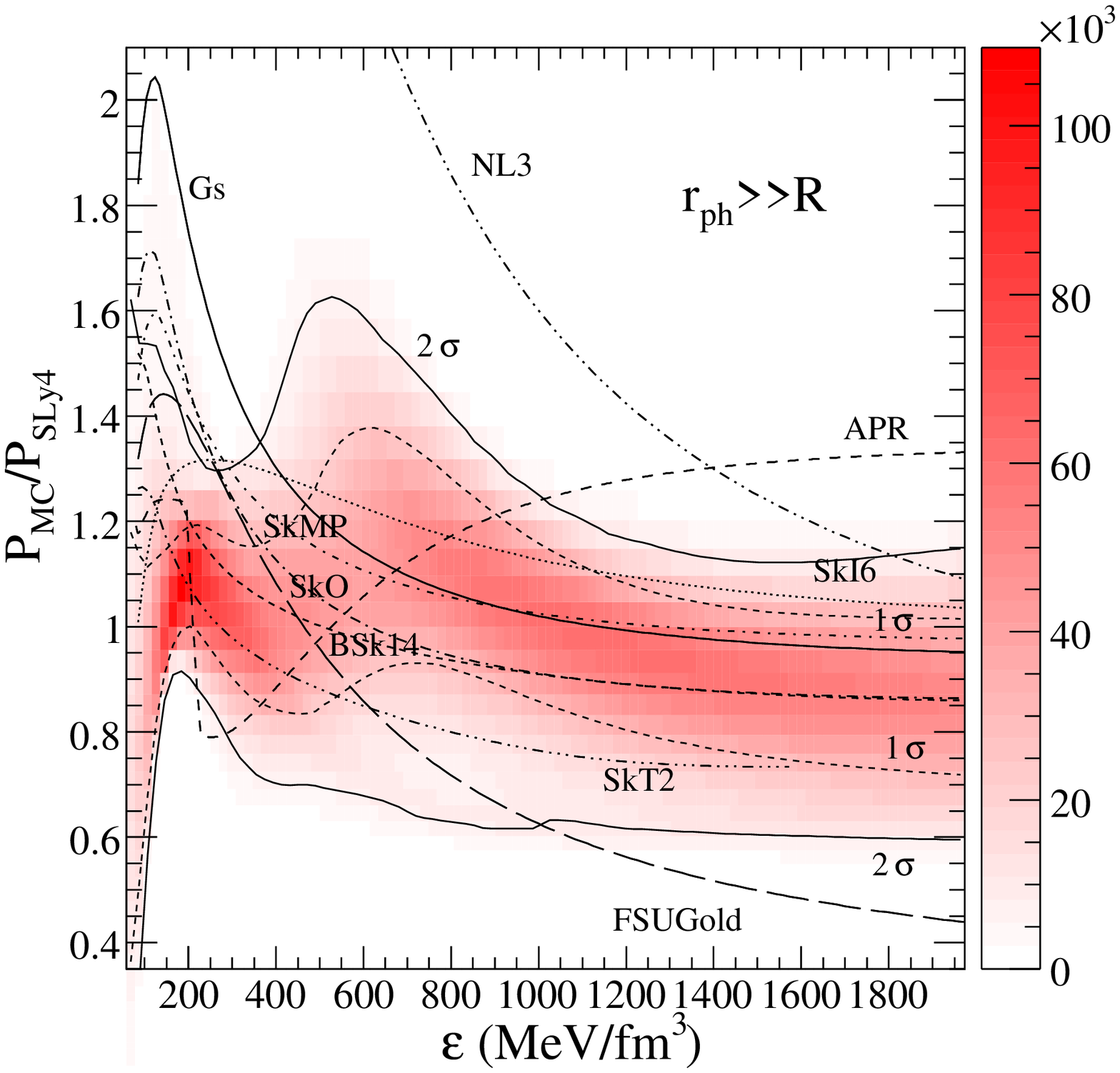}
\includegraphics[width=3.5in]{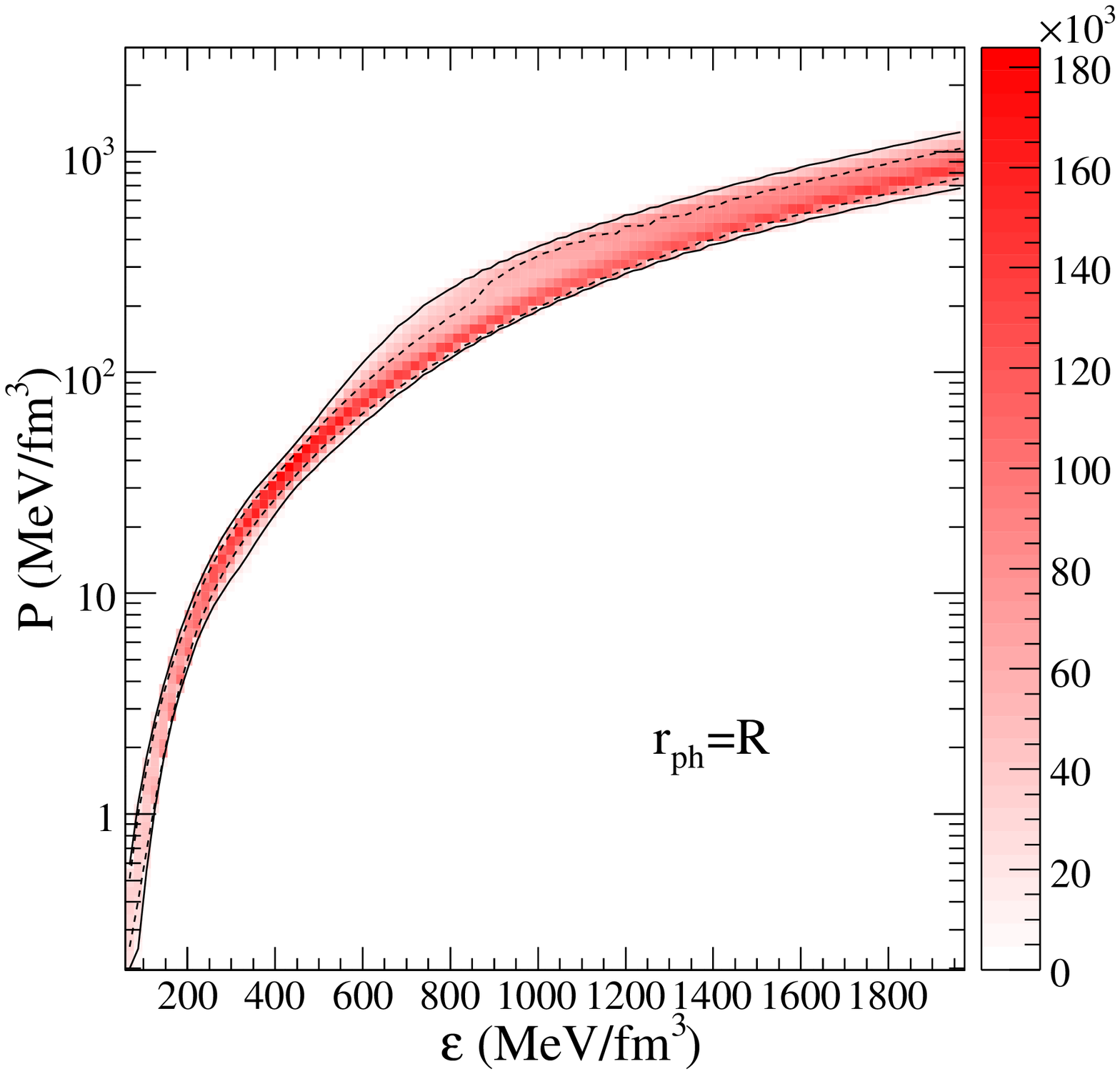}
\includegraphics[width=3.5in]{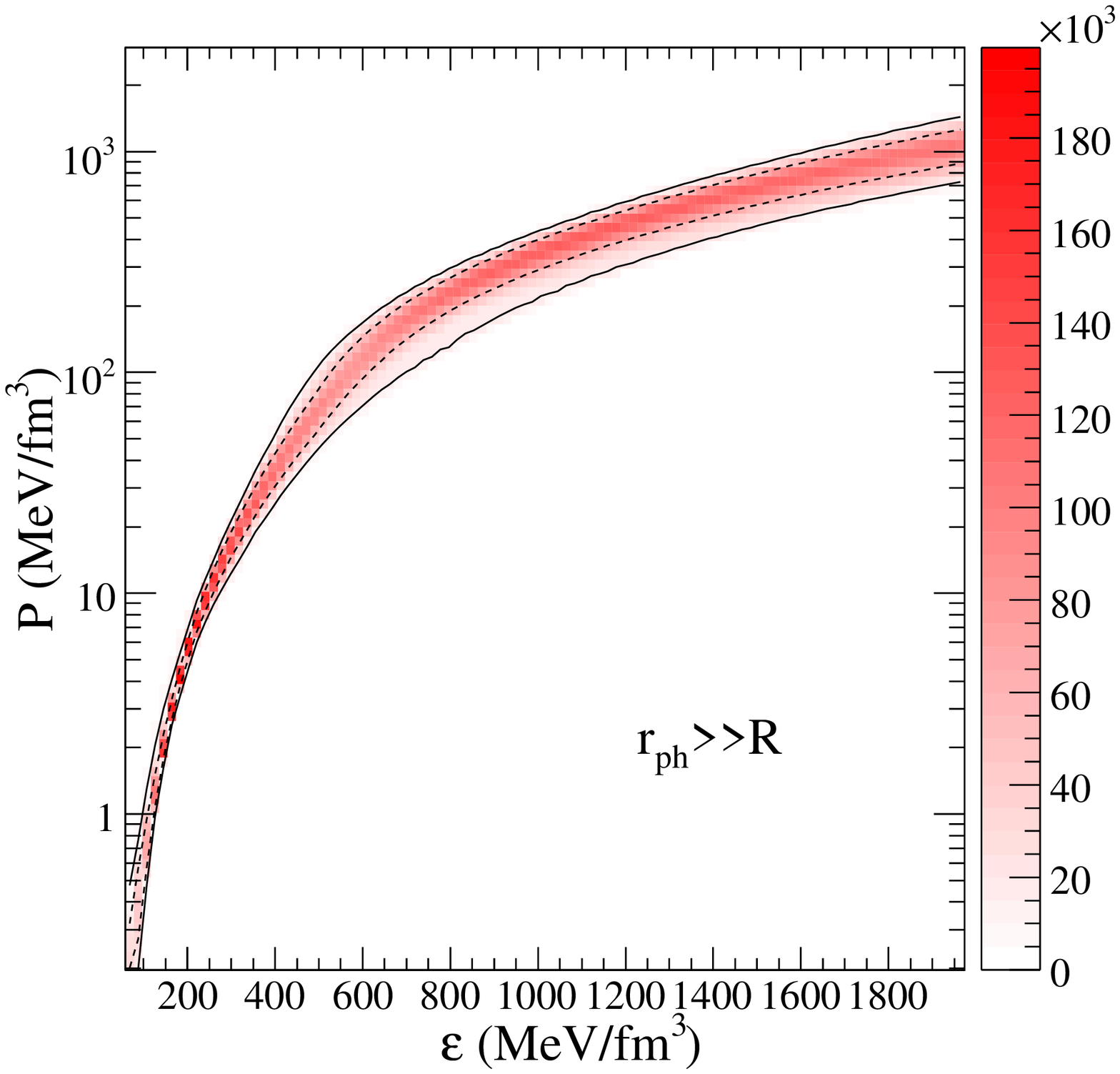}
\caption{Probability distributions from the Bayesian analysis of the
  the parametrized EOS. Upper panels show the ratio of the most
  probable EOS to the pressure of SLy4. The lower panels show the
  pressure of the most probable EOS. The left panels show results
  under the assumption $\rph = R$, and the right panes show results
  assuming $\rph \gg R$. In all panels, the \emph{solid}
  (\emph{dotted}) contour lines in each panel show the 2-$\sigma$
  (1-$\sigma$) contours implied by the data.}
\label{fig:2deos}
\end{figure}

Although the low-density EOS is not strongly affected by the
assumption that $\rph=R$, at higher densities this choice leads to
a softer EOS: the magnitude of the skewness $K^\prime$ and first
polytropic index $n_1$ are both larger for the case $\rph=R$
(Table~\ref{tab:posteos}). The differences between these predicted EOS
can be easily seen by referring to Fig.~\ref{fig:2deos}. Each panel
displays an ensemble of one-dimensional histograms over a fixed grid
in one of the axes (note that this is not quite the same as a
two-dimensional histogram). The bottom panels present the ensemble of
histograms of the pressure for each energy density. This was computed
in the following way: for each energy density, we determined the
histogram bins of pressure which enclose 68\% and 95\% of the total MC
weight. The locations of those regions for each 1-D histogram are
outlined by dotted and solid lines, respectively, and these form the
contour lines. These 1- and 2-$\sigma$ contour lines give constraints
on the pressure of the EOS as a function of the energy density as
implied by the 6 neutron star data sets and are presented in
Tables~\ref{tab:eoslimits} and \ref{tab:eoslimits2} along with the
most probable EOS for the cases $\rph\gg R$ and $\rph=R$,
respectively. The softer nature of the EOS in the $\rph=R$ case is
most apparent at the highest energy density, for which the pressure is
10\% less than in the $\rph\gg{R}$ case.

The upper panels of Fig.~\ref{fig:2deos} display the ensemble of
histograms of the ratio of the pressure of the EOS to the pressure of
SLy4 over a grid of energy density. The choice of the SLy4 EOS here is
essentially arbitrary, and just assists in plotting the results since
the pressure varies over a couple orders of a magnitude. The inferred
pressure ratio from several Skyrme models is also shown. These Skyrme
models were selected in \citet{Steiner09} because they are a
representative set of the recommended models from~\citet{Stone03} which
have symmetry energies which do not become negative at high density.
In both the $\rph=R$ and $\rph\gg R$ cases, there appears to be a
softening of the EOS at low densities which is incompatible with some
of the Skyrme models. Some models are apparently ruled out
independently of assumptions about the radius of the photosphere in
X-ray bursts: field-theoretical models like NL3~\citep{Lalazissis97},
which have stiff symmetry energies; FSUGold~\citep{ToddRutel05}, which
has a softer symmetry energy but becomes too soft at high density; and
APR~\citep{akmal98}, which becomes too stiff at high densities. The
nearly vertical nature of APR for energy densities just over $200\,
\mathrm{MeV\,fm^{-3}}$ is due to the phase transition in APR to matter
which includes a $\pi^0$ condensate. Even after applying a Gibbs phase
construction, the pressure increases only very slowly with density in
the small mixed phase region.

\begin{deluxetable}{cccccc}
\footnotesize
\tablecolumns{6} 
\tablewidth{0pc}
\tablecaption{Most probable values and 68 and 95\% confidence limits 
for the pressure as a function of energy density, for $\rph = R$.}
\tablehead{
\colhead{Energy density} & 
\colhead{2-$\sigma$ lower limit} & 
\colhead{1-$\sigma$ lower limit} &
\colhead{Most probable} &
\colhead{1-$\sigma$ upper limit} & 
\colhead{2-$\sigma$ upper limit} \\
\colhead{$\mathrm{MeV}/\mathrm{fm}^3$} & 
\multicolumn{5}{c}{$\mathrm{MeV/fm^{3}}$}
}
\startdata
150 & 1.97 & 2.02 & 2.16 & 3.77 & 4.11 \\
200 & 4.57 & 5.01 & 5.25 & 7.37 & 8.26 \\
250 & 8.03 & 9.24 & 11.99 & 12.63 & 13.89 \\
300 & 11.68 & 14.23 & 17.48 & 18.88 & 20.76 \\
350 & 16.23 & 20.00 & 22.47 & 25.95 & 28.72 \\
400 & 22.73 & 26.72 & 31.96 & 33.82 & 37.29 \\
450 & 30.51 & 34.73 & 40.75 & 43.69 & 48.43 \\
500 & 38.28 & 43.97 & 49.72 & 56.10 & 63.79 \\
550 & 47.44 & 53.93 & 61.06 & 71.02 & 84.56 \\
600 & 58.92 & 65.10 & 72.64 & 87.99 & 110.8 \\
650 & 70.68 & 77.15 & 82.31 & 107.2 & 139.4 \\
700 & 84.24 & 90.40 & 97.50 & 128.9 & 172.2 \\
750 & 99.90 & 105.2 & 117.6 & 153.4 & 207.4 \\
800 & 115.6 & 121.0 & 132.3 & 179.6 & 239.2 \\
850 & 132.5 & 136.5 & 155.8 & 206.1 & 270.3 \\
900 & 151.0 & 155.8 & 173.8 & 263.0 & 304.9 \\
950 & 169.7 & 177.5 & 191.6 & 302.2 & 338.5 \\
1000 & 190.8 & 197.9 & 207.7 & 337.3 & 371.1 \\
1050 & 213.2 & 220.1 & 236.5 & 364.2 & 406.2 \\
1100 & 234.8 & 243.0 & 253.5 & 389.8 & 439.3 \\
1140 & 257.2 & 268.1 & 277.5 & 423.2 & 476.7 \\
1200 & 280.2 & 294.5 & 308.3 & 459.7 & 516.3 \\
1250 & 300.5 & 321.1 & 343.4 & 462.8 & 551.2 \\
1300 & 323.4 & 346.2 & 379.2 & 504.2 & 585.8 \\
1350 & 352.0 & 369.9 & 411.0 & 528.2 & 622.8 \\
1400 & 381.2 & 398.4 & 436.4 & 561.5 & 666.9 \\
1450 & 402.4 & 432.6 & 453.1 & 612.5 & 712.3 \\
1500 & 427.1 & 460.4 & 502.1 & 643.9 & 749.2 \\
1550 & 459.3 & 484.6 & 518.8 & 676.0 & 799.6 \\
1600 & 480.0 & 520.7 & 550.8 & 718.2 & 848.6 \\
\enddata
\label{tab:eoslimits}
\end{deluxetable}

\begin{deluxetable}{cccccc}
\footnotesize
\tablecolumns{6} 
\tablewidth{0pc}
\tablecaption{Most probable values and 68 and 95\% confidence limits 
for the pressure as a function of energy density, for $\rph \gg R$.}
\tablehead{
\colhead{Energy density} & 
\colhead{2-$\sigma$ lower limit} & 
\colhead{1-$\sigma$ lower limit} &
\colhead{Most probable} &
\colhead{1-$\sigma$ upper limit} & 
\colhead{2-$\sigma$ upper limit} \\
\colhead{$\mathrm{MeV}/\mathrm{fm}^3$} & 
\multicolumn{5}{c}{$\mathrm{MeV/fm^{3}}$}
}
\startdata
150 & 1.84 & 2.00 & 2.15 & 2.52 & 3.25 \\
200 & 4.44 & 4.97 & 5.76 & 6.06 & 6.88 \\
250 & 8.10 & 9.14 & 10.67 & 11.50 & 12.71 \\
300 & 12.34 & 14.61 & 15.87 & 19.08 & 21.43 \\
350 & 18.19 & 21.52 & 24.43 & 29.16 & 34.17 \\
400 & 25.32 & 30.46 & 35.80 & 42.86 & 52.07 \\
450 & 34.56 & 41.54 & 49.11 & 60.99 & 77.97 \\
500 & 45.32 & 55.19 & 69.75 & 84.54 & 107.7 \\
550 & 57.47 & 73.14 & 89.83 & 114.1 & 138.1 \\
600 & 70.62 & 94.02 & 115.5 & 145.0 & 168.1 \\
650 & 85.07 & 116.8 & 149.7 & 175.8 & 199.8 \\
700 & 100.8 & 141.0 & 173.2 & 206.8 & 230.6 \\
750 & 115.7 & 165.8 & 203.3 & 237.8 & 263.8 \\
800 & 131.5 & 190.6 & 232.7 & 268.8 & 297.7 \\
850 & 154.0 & 214.9 & 254.1 & 299.9 & 330.9 \\
900 & 175.0 & 239.7 & 277.4 & 331.6 & 364.7 \\
950 & 196.7 & 264.4 & 310.0 & 364.2 & 399.7 \\
1000 & 217.0 & 289.9 & 336.2 & 397.8 & 437.6 \\
1050 & 236.5 & 316.5 & 371.9 & 432.7 & 474.1 \\
1100 & 260.3 & 343.5 & 414.4 & 468.9 & 509.8 \\
1140 & 282.1 & 369.6 & 458.2 & 504.7 & 547.6 \\
1200 & 307.3 & 395.4 & 457.7 & 541.2 & 592.7 \\
1250 & 333.2 & 424.7 & 494.3 & 581.2 & 639.6 \\
1300 & 356.8 & 454.5 & 550.4 & 622.4 & 680.9 \\
1350 & 381.4 & 481.1 & 603.9 & 660.6 & 726.2 \\
1400 & 407.7 & 512.5 & 608.5 & 704.7 & 777.2 \\
1450 & 434.2 & 542.8 & 672.0 & 749.1 & 824.9 \\
1500 & 460.5 & 572.1 & 661.5 & 791.8 & 877.5 \\
1550 & 487.5 & 606.4 & 732.5 & 840.2 & 932.8 \\
1600 & 514.9 & 635.2 & 734.9 & 884.5 & 979.8 \\
\enddata
\label{tab:eoslimits2}
\end{deluxetable}

\subsection{Mass and Radius Results from the Statistical Analysis}
\label{sec:bayes-results2}

\begin{figure}
\includegraphics[width=3.5in]{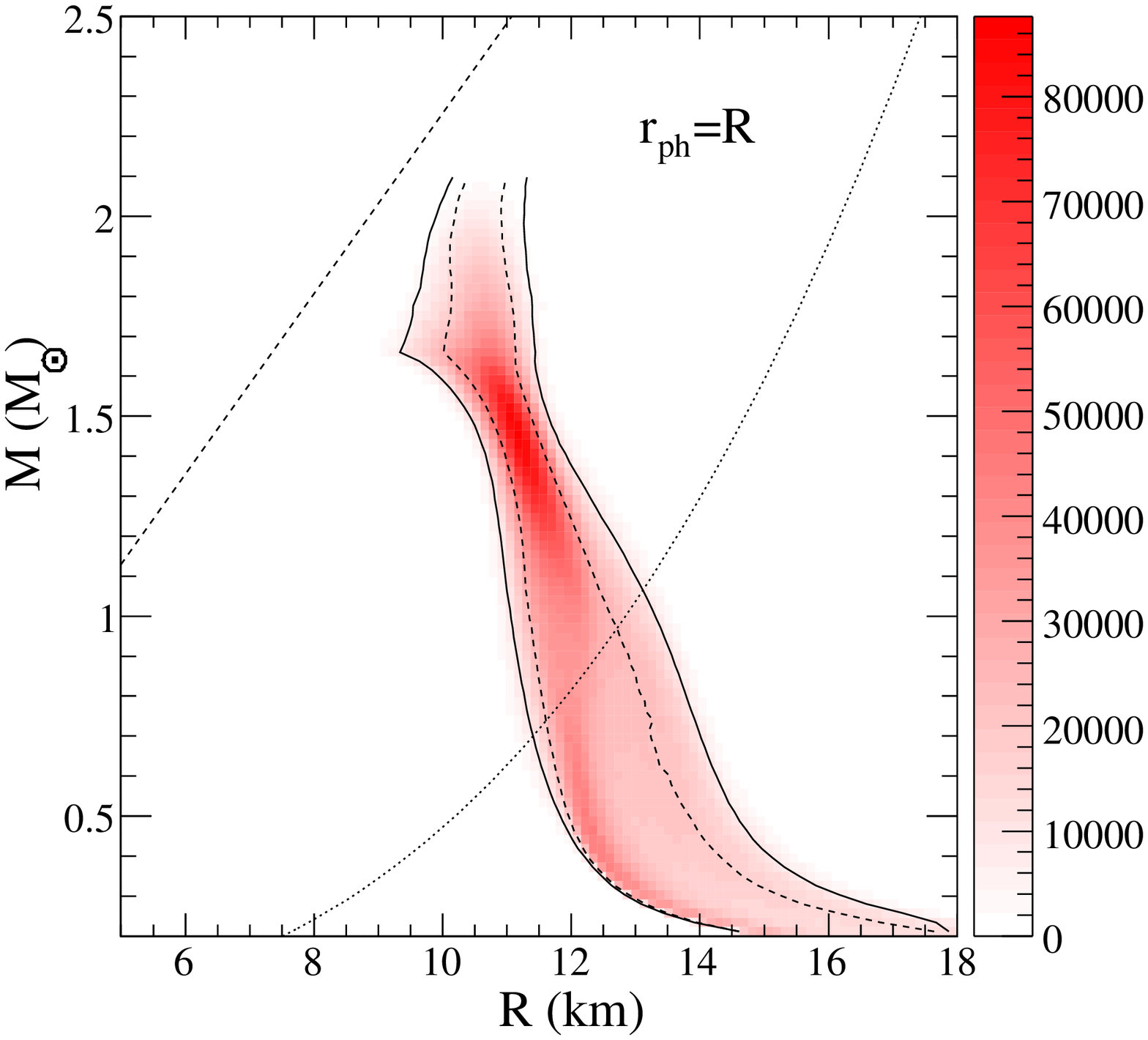}
\includegraphics[width=3.5in]{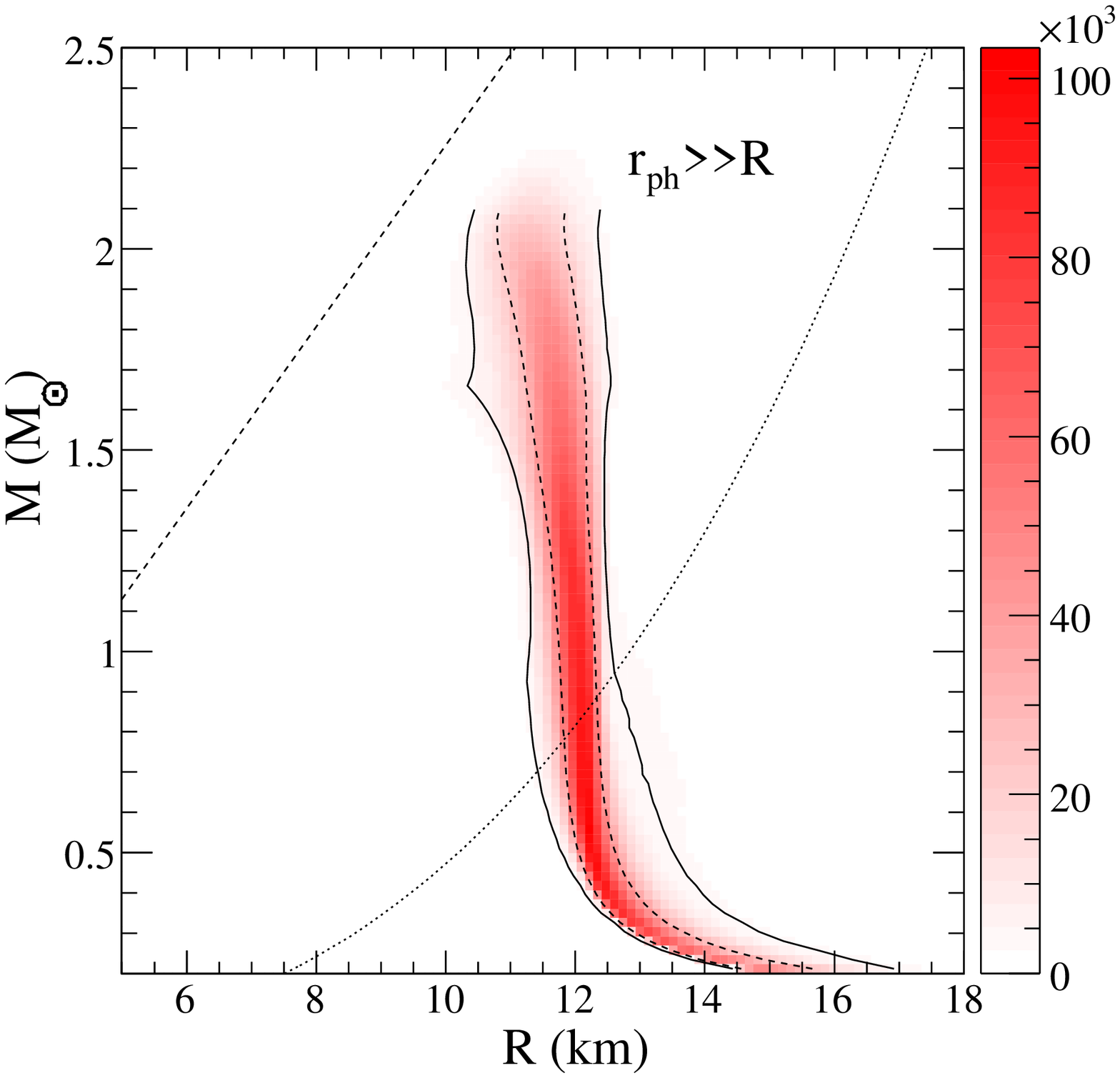}
\includegraphics[width=3.5in]{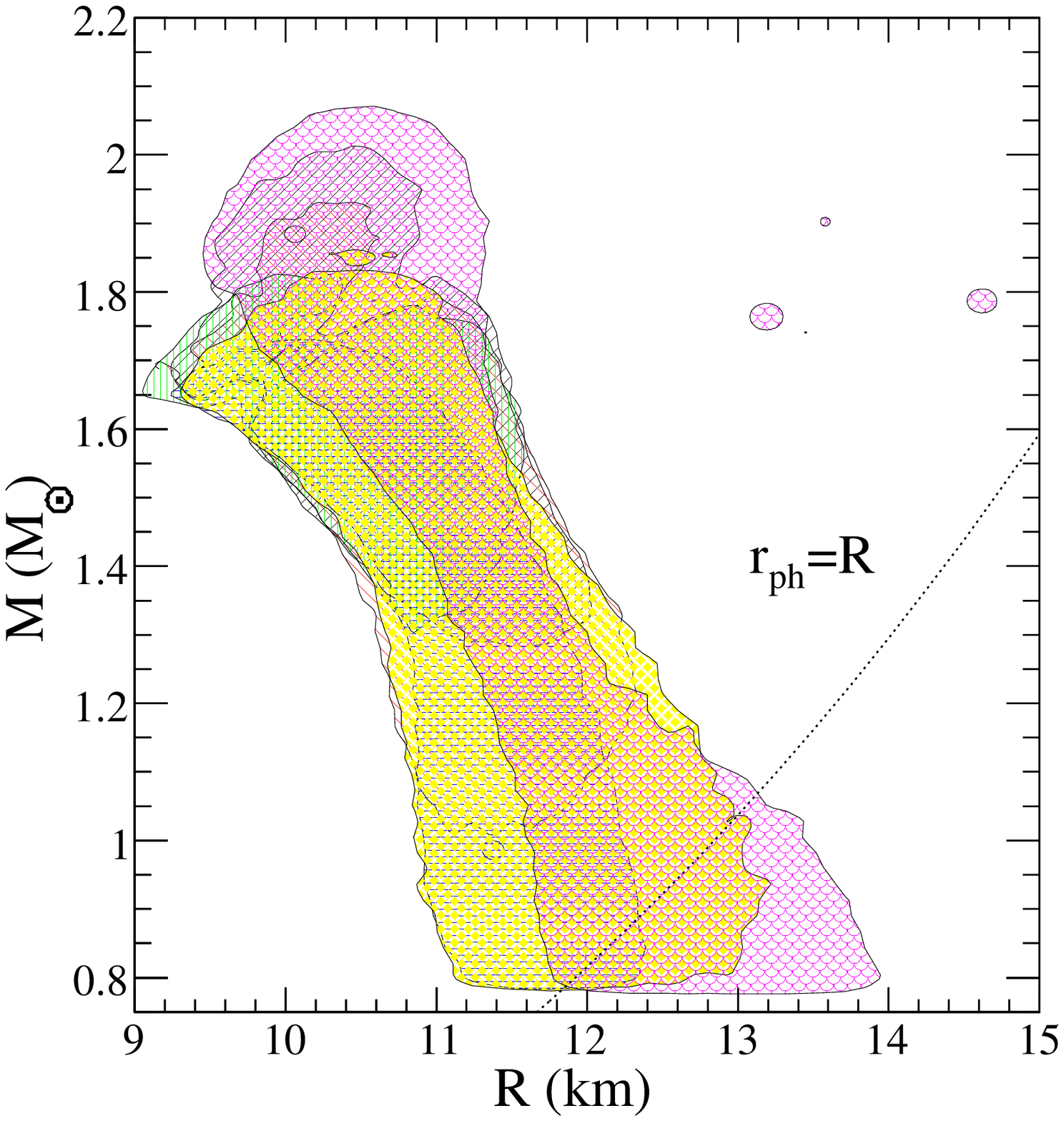}
\includegraphics[width=3.5in]{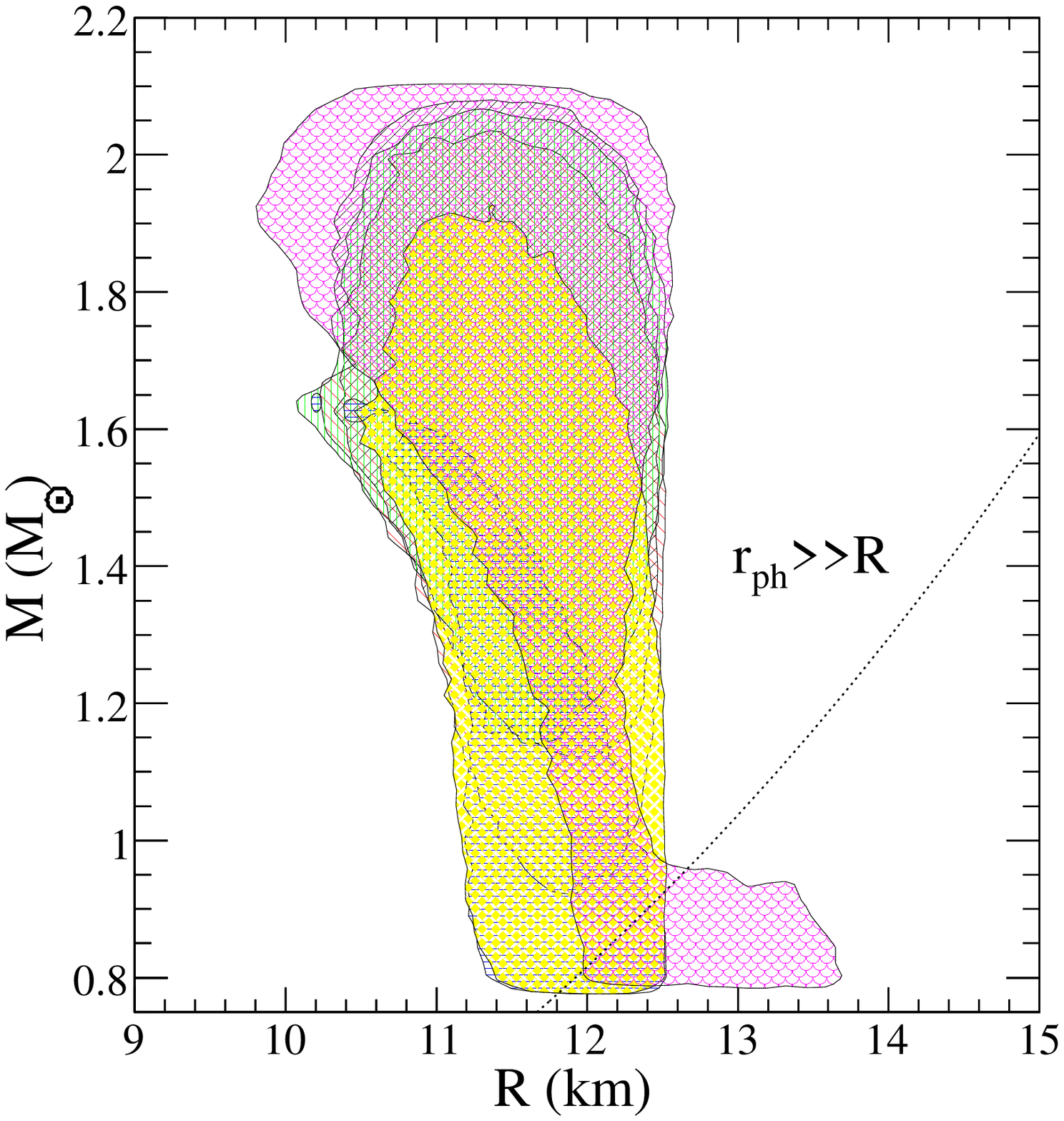}
\caption{ The upper panels give the probability distributions for the
  mass versus radius curves implied by the data, and the \emph{solid}
  (\emph{dotted}) contour lines show the 2-$\sigma$ (1-$\sigma$)
  contours implied by the data. The lower panes summarize the
  2-$\sigma$ probability distributions for the 6 objects considered in
  the analysis. The left panels show results under the assumption
  $\rph = R$, and the right panes show results assuming $\rph \gg R$.
  The dashed line in the upper left is the limit from causality. The
  dotted curve in the lower right of each panel represents the
  mass-shedding limit for neutron stars rotating at 716 Hz. }
\label{fig:2dmass}
\end{figure}

The upper panels in Figure~\ref{fig:2dmass} present our results for
the predicted mass-radius relation according to our two assumptions
regarding the photospheric radius for X-ray bursts. They give the
ensemble of histograms of the radius over a fixed grid in neutron star
mass. The width of the contours at masses below $1\,\Msun$ tends to
be large because the available neutron star mass and radius data
generally implies larger masses. In general, the
implied $M$--$R$ curve suggests relatively small radii, which is
consistent with our conclusions regarding the softness of the nuclear
symmetry energy in the vicinity of the nuclear saturation density.
Tables~\ref{tab:mrlimits} and \ref{tab:mrlimits2} summarize the 1- and
2-$\sigma$ contour lines from these panels, and give as well the most
probable $M$--$R$ curve. The assumption that $\rph = R$ implies
smaller radii for neutron star masses greater than about $1.3\,\Msun$
and perhaps very small radii for masses in excess of $1.5\,\Msun$.
This is suggestive of the onset of a phase transition above the
nuclear saturation density or perhaps simply the approach to the
neutron star maximum mass limit in this case. The mass and radius
contour lines, determined in the same way as for
Figure~\ref{fig:2deos}, are also given in the figure, and summarized
in Tables~\ref{tab:mrlimits} and \ref{tab:mrlimits2}. The radii in the
$\rph=R$ case average about 1 km less than in the $\rph\gg R$ case,
except around $1.4\,\Msun$, where the difference is about 0.4~km.

\begin{deluxetable}{cccccc}
\tablecolumns{6} 
\tablewidth{0pt}
\tablecaption{Most probable values and 68 and 95\% confidence limits 
for neutron star radii of fixed mass, for $\rph = R$.}
\tablehead{
\colhead{Mass } & \colhead{ 2-$\sigma$ lower limit} & 
\colhead{1-$\sigma$ lower limit} &
\colhead{Most probable} & 
\colhead{1-$\sigma$ upper limit} & 
\colhead{2-$\sigma$ upper limit} \\
\colhead{$\Msun$} & \multicolumn{5}{c}{km}}
\startdata
1.0 & 11.06 & 11.35 & 12.09 & 12.62 & 13.31 \\
1.1 & 10.97 & 11.28 & 11.87 & 12.35 & 13.00 \\
1.2 & 10.90 & 11.23 & 11.69 & 12.11 & 12.65 \\
1.3 & 10.81 & 11.13 & 11.42 & 11.85 & 12.28 \\
1.4 & 10.68 & 10.98 & 11.32 & 11.60 & 11.93 \\
1.5 & 10.42 & 10.75 & 11.04 & 11.37 & 11.65 \\
1.6 & 9.93 & 10.37 & 10.87 & 11.17 & 11.47 \\
1.7 & 9.44 & 10.05 & 10.65 & 11.12 & 11.42 \\
1.8 & 9.60 & 10.14 & 10.65 & 11.06 & 11.38 \\
\enddata
\label{tab:mrlimits}
\end{deluxetable}

\begin{deluxetable}{cccccc}
\tablecolumns{6} 
\tablewidth{0pt}
\tablecaption{Most probable values and 68 and 95\% confidence limits 
for neutron star radii of fixed mass, for $\rph \gg R$.}
\tablehead{
\colhead{Mass } & \colhead{ 2-$\sigma$ lower limit} & 
\colhead{1-$\sigma$ lower limit} &
\colhead{Most probable} & 
\colhead{1-$\sigma$ upper limit} & 
\colhead{2-$\sigma$ upper limit} \\
\colhead{$\Msun$} & \multicolumn{5}{c}{km}}
\startdata
1.0 & 11.29 & 11.75 & 12.09 & 12.30 & 12.57 \\
1.1 & 11.31 & 11.71 & 11.94 & 12.28 & 12.51 \\
1.2 & 11.30 & 11.65 & 11.96 & 12.24 & 12.47 \\
1.3 & 11.24 & 11.58 & 11.81 & 12.21 & 12.45 \\
1.4 & 11.13 & 11.49 & 11.83 & 12.18 & 12.45 \\
1.5 & 10.94 & 11.39 & 11.83 & 12.17 & 12.45 \\
1.6 & 10.63 & 11.30 & 11.70 & 12.17 & 12.49 \\
1.7 & 10.42 & 11.21 & 11.70 & 12.13 & 12.54 \\
1.8 & 10.43 & 11.10 & 11.57 & 12.05 & 12.47 \\
\enddata
\label{tab:mrlimits2}
\end{deluxetable}

The lower panels in Figure~\ref{fig:2dmass} summarize the output
probability distributions for $M$ and $R$ for the 6 neutron stars
which were used in this analysis. Note that the scales have been
modified, and these output probability distributions are much smaller
than the input distributions given in Figures~\ref{fig:pre},
\ref{fig:pre2}, and \ref{fig:other}. \emph{The combination of several
  neutron star mass and radius measurements with the assumption that
  all neutron stars must lie on the same mass--radius curve puts a
  significant constraint on the mass and radius of each object.} Note
that these output probability distributions closely match the implied
$M$ vs. $R$ curves presented in the upper panels of
Figure~\ref{fig:2dmass}. The tendency for smaller radii when
$M>1.3\textrm{--}1.4\,\Msun$ is apparent.

The $M$ and $R$ constraints for each object are given in
Table~\ref{tab:mrpostres}, with their corresponding 1-$\sigma$
uncertainties. The three PRE burst sources suggest masses near 1.5
$\Msun$ and the radii near 11 km in the case where $\rph=R$. We
have already observed that this agreement is due to the 
extreme restrictions placed on the acceptability of points during the
Monte Carlo sampling. In the case $\rph\gg R$, the PRE burst sources
are predicted to have a wider range of masses, from 1.3 to 1.6 solar
masses. The quiescent LMXB masses tend to be smaller, but are strongly
dependent on assumptions about the radius of the photosphere in the
PRE burst sources. Particularly uncertain is the mass for X7. The
observations imply a rather large value of $R_{\infty}$, which is
compatible with a small radius if the mass is large ($\rph\gg R$), or
a large radius if the mass is small ($\rph=R$). The stars in M13 and
$\omega$ Cen show the opposite trend. They have small values of
$R_\infty$ which is compatible with small radii in the $\rph\gg R$
case if the mass is relatively small. In general, the predicted radii
of all stars range from about 10 km to 12.5 km, with the exception of
X7 which may have a large radius. Note that a 13 km radius for an 0.8
$\Msun$ star is beyond the limit implied by rotation at 716 Hz and
thus if the neutron star in X7 is observed to spin rapidly enough, a
much larger mass and smaller radius would be implied instead.

\begin{deluxetable}{ccccc}
\tablecolumns{5} 
\tablewidth{0pt}
\tablecaption{Most probable values for masses and radii for 
neutron stars constrained to lie on one mass versus radius curve.}
\tablehead{
\colhead{Object} & \colhead{$M$ ($\Msun$)} & \colhead{$R$ (km)} 
& \colhead{$M$ ($\Msun$)} & \colhead{$R$ (km)}  \\
& \multicolumn{2}{c}{$\rph = R$} & 
\multicolumn{2}{c}{$\rph \gg R$} \\
}
\startdata
4U 1608--522 & 
$1.52^{+0.22}_{-0.18}$ &
$11.04^{+0.53}_{-1.50}$ &
\phn$1.64^{+0.34}_{-0.41}$ &
$11.82^{+0.42}_{-0.89}$ \\
EXO 1745--248 & 
$1.55^{+0.12}_{-0.36}$ &
$10.91^{+0.86}_{-0.65}$ &
\phn$1.34^{+0.450}_{-0.28}$ &
$11.82^{+0.47}_{-0.72}$ \\
4U 1820--30 & 
$1.57^{+0.13}_{-0.15}$ &
$10.91^{+0.39}_{-0.92}$ &
\phn$1.57^{+0.37}_{-0.31}$ &
$11.82^{+0.42}_{-0.82}$ \\
M13 & 
$1.48^{+0.21}_{-0.64}$ &
$11.04^{+1.00}_{-1.28}$ &
$0.901^{+0.28}_{-0.12}$ &
$12.21^{+0.18}_{-0.62}$ \\
$\omega$ Cen & 
$1.43^{+0.26}_{-0.61}$ &
$11.18^{+1.14}_{-1.27}$ &
$0.994^{+0.51}_{-0.21}$ &
$12.09^{+0.27}_{-0.66}$ \\
X7 & 
$0.832^{+1.19}_{-0.051}$ &
$13.25^{+1.37}_{-3.50}$ &
\phn$1.98^{+0.10}_{-0.36}$ &
$11.3^{+0.95}_{-1.03}$ \\
\enddata
\label{tab:mrpostres}
\end{deluxetable}

\begin{figure}
\begin{center}
\includegraphics[width=3.5in]{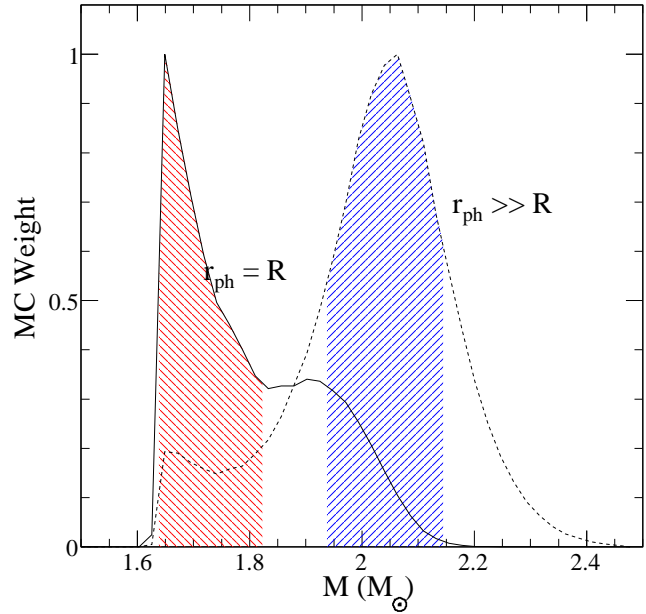}
\end{center}
\caption{The probability distributions for the maximum mass. Because
  of the observation of a neutron star of at least $1.66\,\Msun$,
  we reject all curves for which $M_{\mathrm{max}}<1.66\,\Msun$ and
  thus the probability is cutoff below this value. The shaded regions
  indicate the 1-$\sigma$ confidence regions.}
\label{fig:max}
\end{figure}

The largest difference between the predicted equations of state
between the $\rph=R$ and $\rph\gg R$ cases is the high-density
behavior. This leads to large differences in the predicted maximum
masses. The probability distributions for the maximum neutron star
mass are given in Figure~\ref{fig:max}, along with the associated
1-$\sigma$ confidence regions. The two probability distributions are
arbitrarily normalized so that their peak is unity. These results are
strongly dependent on assumptions of the photospheric radius at
touchdown. The two-peaked behavior in the case $\rph=R$ suggests a
possible phase transition could match the data and implies a maximum
mass very close to the observed limit of $1.66\,\Msun$. This result is
similar to that claimed by~\citet{Ozel10}, but there it is stated that
the results are incompatible with a nucleonic equation of state. Our
results do not support this extreme interpretation. Although the
neutron star radii implied by this analysis are small, they are not
small enough to require a phase transition; for neutron stars of mass
$1.4\,\Msun$, radii smaller than 10~km can be generated by purely
nucleonic equations of state~\citep{Steiner05}.

\section{Discussion}

In this paper, we have determined an empirical dense matter equation
of state from a heterogeneous dataset containing PRE bursts and quiescent
thermal emission from X-ray transients. Previous works
\citep{Ozel06,Ozel09,Guver10,Guver10b} have demonstrated the potential
utility of PRE bursts for determining neutron star masses and radii.
Their analysis assumes that the photosphere at touchdown has fully
retreated to the quiescent radius, and that the high-redshift solution
is favored. We argue that their model requires re-examination. First,
we find no grounds on which the high-redshift solution in the case
$\rph=R$ could be favored. Second, internal consistency (i.e., the
obligatory rejection of an overwhelming number of Monte Carlo
realizations) implies the assumption that $\rph=R$ is suspect and
should be generalized. We then explored an alternative, that the
radius of the photosphere is extended and does not recede until later
in the observed burst. A larger photosphere indeed provides internal
consistency without requiring strong cuts on the observed values for
flux, normalization, distance, and composition.

There are other sources of systematic errors in the PRE burst model.
\citet{Boutloukos10} found that \emph{RXTE}/PCA spectra of bursts from
4U~1820--30 were better fit by Planck or Bose-Einstein spectra rather
than Comptonized spectra with large color correction factors. If the
color correction factor is indeed of order unity, then the high color
temperatures would indicate a locally super-Eddington flux, even in
the tail of the burst \citep{Ebisuzaki1984Are-X-ray-burst}. In
addition, at near-Eddington fluxes there are other possible
complications, such as a photon bubble instability \citep{Hsu97} that
may impact the observed properties of a PRE X-ray burst.

A recent analysis of PRE bursts in 4U~1724--307
by~\citet{Suleimanov10} also found that the photosphere is somewhat
extended at touchdown, $\rph \approx 2 R$. Using a new model of the
atmosphere (the details of which are not yet published), they obtain
masses and radii that are both larger than we would have found. This
result is principally due to a larger value of the color correction
factor, $f_c$, predicted by their model atmosphere.  It is 10--15\%
larger than what one would obtain from the model of \citet{Madej04}
for the same values of $F/\Fedd$, surface gravity $g$, and
composition.  In this case, we find values of $\alpha\propto f_c^{-2}$
which are about 25\% larger and $R_\infty\propto f_c^2$ that are
about 25\% smaller than \citet{Suleimanov10} did.  Assuming
$\rph\gg R$, the predicted mass scales with $f_c^2$.  When we analyse
4U~1724--307 with the lower value of $f_c$, we get masses and radii
that are consistent with those from the other three PRE burst
sources. This highlights an important avenue for future work, namely
that a clearer understanding of the atmosphere and, in particular,
$f_c$ are essential.

Further progress in using PRE bursts to constrain neutron star masses
and radii clearly requires better models of the spectral evolution
during X-ray bursts. It is important to note that increases 
in the precision of mass and radius estimates are not
absolutely necessary, as we have shown that existing
errors, large as they are, do not inhibit placing interesting limits
to the equation of state when combined with other observations. It is
very important, however, to resolve the uncertainty regarding the
location of the photospheric radius at ``touchdown'' and to
characterize systematic uncertainties, including potential
correlations between $\Ftd$, $A$, $f_c$, and $X$ in the spectral
models. Also, should a larger range in $f_c$ be required, then the
uncertainties on masses and radii will increase accordingly.

Our results imply that the EOS in the vicinity of the nuclear
saturation density is relatively soft. As shown by \citet{Lattimer01},
this is primarily due to the weak dependence of the nuclear symmetry
energy with density. This conclusion is robust with respect to
variations in how PRE burst sources are modeled, and is strengthened
by the small values of $R_\infty$ deduced for the globular cluster
LMXB sources in M13 and $\omega$ Cen. This conclusion has immediate and
important ramifications for laboratory nuclear physics, in particular
for the scheduled parity-violating electron scattering measurement of
the neutron skin thickness of
$^{208}$Pb~\citep{Michaels00,Horowitz01b}. In the context of the
liquid droplet model, the ratio of the surface and volume symmetry
coefficients can be expressed as~\citep{Steiner05}
\begin{equation}
{S_s\over S_v}={9E_{s0}\over K}{I_2\over I_1},
\label{eq:sssv}
\end{equation}
where $E_{s0}\simeq19$ MeV is the surface energy coefficient for
symmetric nuclei.  The integrals $I_1$ and $I_2$ for the schematic
EOS described by equation~(\ref{eq:ldeos}) are given by
\begin{eqnarray}
I_1 &=& \int_0^1\sqrt{u}(1-u)\sqrt{1-a(1-u)}~du,\cr
I_2 &=& \int_0^1\sqrt{u}\left[{S_v\over S_ku^{2/3}+(S_v-S_k)u^\gamma}-1\right]
{1\over1-u}{1\over\sqrt{1-a(1-u)}}~du\,,
\label{eq:i1i2}
\end{eqnarray} 
where $a=K^\prime/(9K)$.
The predicted neutron excess in the center of a nucleus $(N,Z)$ is then
\begin{equation}
\delta={N-Z\over N+Z}\left(1+{S_s\over S_vA^{1/3}}\right)^{-1},
\label{eq:delta}
\end{equation}
and the neutron skin thickness is
\begin{equation}
\delta R=\sqrt{3\over5}{2r_0\over3}{S_s\over S_v}{\delta\over1-\delta^2}\,,
\label{eq:dr}
\end{equation}
where $r_0=(4\pi n_0/3)^{-1/3}$. With a value of $\gamma=0.26$ and
$a=-0.141$ and errors as established in Table~4, we
deduce $S_s/S_v\simeq1.5\pm0.3$ and $\delta
R(^{208}\mathrm{Pb})\simeq0.15\pm0.02$ fm, a value at the lower end of
expectations.

We also conclude that in our preferred model, in which the
photospheric radius is not restricted to be the stellar radius, the
EOS at high densities is stiff enough to support a maximum mass of
order $2\, \Msun$ or greater. This result is supported by the
simultaneous observations of LMXB sources with both small and large
values of $R_\infty$, if one rejects the possibility of the existence
of neutron stars with masses less than about $1\,\Msun$. The latter condition
seems to be borne out by models of massive star supernovae, which
strongly suggest that proto-neutron stars are born with high trapped
lepton fractions and moderate entropies. Hydrostatic stability
requires, in this case, that protoneutron stars are bound only if
their mass exceeds about $1\,\Msun$~\citep{Strobel99}, ruling out the
existence of cold, catalyzed neutron stars of smaller masses. Had this
assumption been used as a prior condition in our analysis, the LMXB
sources with small values of $R_\infty$ would further support
solutions with EOS parameters favoring large maximum masses.

Obviously, more neutron star observations with mass and radius
constraints would enable one to improve our constraints. It is a
particular advantage of our methodology that new observations can be
integrated into the formalism easily, if estimates of a
two-dimensional probability distribution of mass and radius can be
made. As pointed out by~\citet{Heinke06}, it is particularly important
that atmosphere models treat the surface gravity self-consistently. In
particular, spectral features would be very constraining, and although
the most recent result from~\citet{Cottam02} has not been verified in
longer observations \citep{Cottam2008The-Burst-Spect}, future
determinations of the surface gravity may provide strong constraints.
\citet{Ho09} also demonstrated that constraints on the mass and radius of 
the Cas A supernova remnant may be obtained. These constraints
are roughly consistent with our $\rph \gg R$ results described above.

Although we did not include the estimate of the mass and radius
($M=1.7\pm0.3\, \Msun$ and $R=11.5\pm1.2\,\mathrm{km}$) of
RX~J1856--3754 \citep{Pons02,Walter10} in our baseline fit, we found
them to be remarkably consistent with those of the other 6 neutron
stars. All of the results we obtained for the parameters of the EOS,
the pressure versus energy density curve, the mass versus radius
curve, our estimate of the maximum mass, and the predicted individual
masses and radii for all 6 other sources were unchanged to within
1-$\sigma$. For example, the most probable radius of a $1.4\,M_\odot$
star would change by less than $0.01\,\mathrm{km}$ and the most probable pressure
at an energy density of $1000\,\mathrm{MeV\,fm^{-3}}$ would increase by 0.03\%.

In addition to X-ray bursts, quiescent low-mass X-ray binaries, and
thermally-emitting isolated neutron stars, other potential methods for
determining neutron star masses and radii exist. These include neutron
star seismology~\citep{Samuelsson07,Steiner09}, pulse profiles in
X-ray
pulsars~\citep{Leahy2009Constraints-on-,Morsink2009Multi-epoch-Ana},
and moment of inertia measurements in relativistic
binaries~\citep{Lattimer05}. Gravitational wave signals of neutron
star mergers may also provide significant
constraints~\citep{Ferrari10}. Measurements of the thickness of the
neutron star crust, which controls the crust cooling of transiently
accreting
LMXBs~\citep{Shternin2007Neutron-star-co,Brown2009Mapping-Crustal} as
well as the distribution of observed cooling curves compared to a
minimal cooling model \citep[e.g.,][]{Page09}, could
also constrain neutron star masses and radii.

The Bayesian analysis in Section~\ref{sect:bayes} is a novel procedure
for combining mass and radius constraints from disparate objects
to form new constraints on the mass versus radius curve and
the EOS of dense matter. The construction of the EOS from
astrophysical observations has also been recently addressed
in~\citet{Read09},~\citet{Ozel09b}, and \citet{Ozel10}. \citet{Read09}
showed that piecewise polytropes with relatively few parameters could
effectively describe more sophisticated models for a high-density EOS,
but confined attention to constraints stemming from observations
limiting the neutron star maximum mass and maximum spin rate. They did
not attempt reconstruction of the EOS from simultaneous mass and
radius measurements. We therefore obtain stronger constraints on the
EOS, but have utilized observational data which contains significant
systematic uncertainties. \citet{Ozel09b} examined the constraints on
the EOS obtained from a synthetic data set obtained with simultaneous
mass and radius measurements of either 5\% or 10\% accuracy for three
separate objects; they found that although individual EOS parameters
were difficult to estimate precisely, significant correlations between
them could be obtained. The Jacobian technique employed in
\citet{Ozel09b} is a simple approximation of our full Markov Chain
Monte Carlo method. That Jacobian technique requires that the number
of EOS parameters and the number of neutron stars are equal, and
thus provides only an incomplete marginalization over the EOS
parameters.

One could imagine various alternatives to the statistical procedure in
Section~\ref{sect:bayes}. One possibility is the use of prior
distributions to represent either astrophysical or nuclear input.
Neutron star masses in some double-pulsar systems are well-measured, and
could provide a prior mass distribution. It is not clear, however,
that either low-mass X-ray binaries or isolated neutron stars follow
this same initial mass function. As mentioned above, there are strong
theoretical reasons to suspect that neutron stars cannot be formed
with less than about $1.0\,\Msun$. In the analysis above, we chose
$0.8\,\Msun$ to ensure that the boundary does not interfere with
results near the proto-neutron star minimum mass. There are several
nuclear physics observables which can be used to constrain the EOS
through measurements of properties like the compressibility and the
symmetry energy. Several authors have also computed the EOS of neutron
matter directly from nucleon-nucleon
interactions~\citep{Tolos07,Hebeler09}, and we have found that our
predicted EOS is consistent with these studies. In the context of this
work, information from calculations of the neutron matter EOS, neutron
skin thicknesses, the surface symmetry-volume symmetry energy
correlation observed from mass formula fits~\citep{Steiner05},
information from giant dipole resonances, and so forth, could also
provide constraints for the schematic EOS parameters described above.
We have chosen not to use this information in this work, in part to
ensure that our results are free from extra model dependencies.
Future work on implementing more
nuclear physics input into the analysis of the observational data is
certainly warranted.

One might also question, for the high-density EOS, whether or not our
prior distributions for the polytropic indicies ought to be treated as
uniform, as a uniform distribution in the polytropic index is
different than, for example, a uniform distribution in the pressure at
any fixed energy density, or in the polytropic exponent. Finally, one
could consider reformulating the model directly in terms of the
observables like flux and distance instead of the ``two-step''
procedure we have used above which uses a MC simulation to generate
masses and radii from the observables for each neutron star, and then
afterwards performs a Bayesian analysis from the output of these
initial MC results. This alternative would strongly disfavor $\rph=R$,
since so many MC realizations must be rejected in this picture.

It is important to note that the constraints we have obtained are a
guide, but will likely require revision in the future. Because we only
used 6 mass and radius measurements a single additional measurement
could have a significant impact on our results. Alternatively, if one
of the 6 input probability distributions used above changes
significantly, because, for example, of a new understanding of
systematic uncertainties, our final constraints on the EOS would
change accordingly. We have already noted that there is significant
tension on our results from assumptions about the photospheric radius
of X-ray bursts, and also from the fact that the neutron star in M13
has a small value of $R_{\infty}$ while X7 in Terzan 5 has a much
larger value of $R_{\infty}$. 

It is possible that there exist extreme models
which live in very small regions of parameter space and are not fully
sampled in this work. One example of such an EOS would be those which
exhibit the ``twinning'' phenomenon, i.e. the presence of a turning
point in the mass vs. radius curve which admits a new family of
compact neutron stars~\citep{Glendenning00,Schaffner-Bielich02}
Such solutions should be addressed in future work.

\citet{Ozel10} claim that constraints from the PRE burst sources imply
that equations of state which contain only nucleonic degrees of
freedom are inconsistent with data. We disagree with their conclusion.
Although we find the EOS must be quite soft at moderate densities, we
do find many nucleonic models which are consistent with the data. More
importantly, however, we find that the conclusion of extreme softening
evaporates if we use a slightly different model for the PRE burst
sources that accounts for systematic uncertainties. The inclusion of
mass and radius data from other neutron star sources supports our
interpretation of the PRE burst sources. While our results do not rule
out a phase transition at supernuclear densities, extreme softening of
the EOS is not compatible with observations. Rather, the implication
is that the maximum mass is likely large, greater than $1.8\,\Msun$.

\acknowledgments

We thank D. Galloway, A. Kundu, J. Linnemann, M. Prakash, S. Reddy,
and R. Rutledge for useful discussions and critical comments on this
manuscript. AWS and EFB are supported by the Joint Institute
for Nuclear Astrophysics at MSU under NSF PHY grant 08-22648 and by
NASA ATFP grant NNX08AG76G. JML acknowledges research support
from the U.S. DOE grant DE-AC02-87ER40317. The TOV solver, benchmark
EOSs, fitting and data analysis routines were obtained from the
O$_2$scl library found at \url{http://o2scl.sourceforge.net}. AWS,
EFB, and JML acknowledge the kind hospitality of the ECT*, where this
paper was started. EFB and JML also thank the Yukawa Institute for
Theoretical Physics for a providing a stimulating environment where
this work received extensive discussion among the participants.

\bibliographystyle{apj}
\bibliography{paper,master}

\appendix
\section{Observations of Type I X-Ray Bursts}\label{s.observations}

In this section, we describe in detail our probability distributions
for the angular emitting area, or normalization, the distance, the
touchdown flux, and the photospheric hydrogen mass fraction for the
three sources with PRE bursts used in this analysis. We also note,
where appropriate, how the distances compare to the value obtained by
assuming the maximum flux is less than the Eddington value,
\begin{equation}\label{e.standard-candle-distance}
 D < \left(\frac{GMc}{\kappa F_{\max,\infty}}\right) = 
17.1\,\mathrm{kpc}\left[\left(\frac{M}{1.4\,\Msun}
\right) 
\left(\frac{10^{-8}\,\mathrm{ergs\,cm^{-2}\,s^{-1}}}
{F_{\max,\infty}}\right)\right]^{1/2}.
\end{equation}

\subsection{EXO 1745--248}\label{s.exo1745}

{\it Normalization:} The normalization factors obtained from the two
PRE bursts in~\citet{Ozel09} are $A/(1\,\Aunit) = 1.04 \pm 0.01$ and
$1.30 \pm 0.01$. These two measurements differ significantly, and this
means that either the color correction factors for the two bursts
differ by roughly 6\% or the geometry of the two bursts is different.
It is not clear how to resolve this conflict. \citet{Ozel09} use a
boxcar distribution with $A=1.16\,\Aunit$ and $\Delta A=0.13\,\Aunit$,
which represents the choice with the minimum possible uncertainty. We
instead choose a Gaussian centered at $A=1.17\,\Aunit$ with
$\sigma_A=0.13\,\Aunit$, which ensures that the two
  observations are included to within 1 $\sigma$. This choice is
consistent with the objects below, which also have normalization
factors which are simulated with Gaussian distributions, but with
notably smaller uncertainties.

{\it Distance: } EXO~1745--248 is located in the globular cluster
Terzan 5. \citet{Ortolani07} analyzed the distance to Terzan 5 using
both the NICMOS instrumental magnitudes and the calibrations from
\citet{Stephens00} and~\citet{Cohn02}. The combined distance
estimation is $D=5.9 \pm 0.9$~kpc, where the uncertainty has been
obtained from the standard deviation of the three different distance
measurements, suggesting a Gaussian with the same standard deviation.
On the basis that the distance measurement from the NICMOS
instrumental magnitudes was independent of photometric calibrations
and thus more accurate,~\citet{Ozel09} used the NICMOS distance of
$D=6.3$~kpc with $\Delta D=0.32$~kpc. This choice assigns, however, a
zero probability to a distance of 5.9~kpc, the central value suggested
in~\citet{Ortolani07}. Until this distance measurement is more clearly
determined, we choose a Gaussian distribution with $D=6.3$~kpc and
$\sigma_D=0.6$~kpc.

{\it Touchdown flux: } We employ the result from~\citet{Ozel09}, which
is a Gaussian distribution with $\Ftdi=6.25\,\fluxunit$ and
$\sigma_{F}=0.2\,\fluxunit$.

{\it Hydrogen mass fraction: }~\citet{Galloway08} noted that
EXO~1745--248 has exhibited long Type I bursts with estimated values
of $\int\! dt\; F_{\mathrm {persistent}}/\int\! dt\;
F_{\mathrm{burst}} \approx GM/R/E_{\mathrm{nuc}} \approx
20\textrm{--}46$. These long bursts do not always show a strong
thermal evolution (Galloway, private communication); but if they are
indeed thermonuclear in origin, then the low ratio of persistent to
burst fluence indicates H-rich fuel and larger values of $X$. In
contrast to the determination of $X$ from these long bursts, this
object has also been identified as an ultracompact binary
in~\citet{Heinke03}, through a phenomenological assessment of its
spectral properties, suggesting that X is small, $0<X<0.1$. The radius
expansion bursts have short durations, which would be consistent for
ignition of a pure He layer (i.e., the hydrogen is consumed by the
stable hot CNO cycle). We retain the full range of hydrogen
composition, $0<X<0.7$ .

\subsection{4U 1608--522}\label{s.4u1608}

{\it Normalization: } The normalization factors for the 4 PRE bursts
are $A/(1\,\Aunit) = 3.267\pm 0.047$, $3.302\pm0.049$,
$3.258\pm0.054$, and $3.170\pm0.047$~\citep{Guver10}. We use the 
 average from~\citet{Guver10}, $3.246\pm0.024$.

{\it Distance: }~\citet{Guver10} give a Gaussian distribution with
$D=5.8\,\mathrm{kpc}$ and $\sigma_D=2.0\,\mathrm{kpc}$ with a cutoff
below $3.9\,\mathrm{kpc}$. We also employ this result. We note that
if the flux is indeed less than the Eddington value, then $D < 4.36\,
\mathrm{kpc}(M/1.4\,\Msun)^{1/2}$ for the central value of the
touchdown flux.

{\it Touchdown flux: } Two of the four PRE bursts gave a value for the
touchdown flux, $\Ftdi=(15.58 \pm 0.82)\,\fluxunit$ and $(15.14 \pm
1.05)\,\fluxunit$~\citep{Guver10}. The value
$(15.41\pm0.65)\,\fluxunit$ was obtained from the fit in
\citet{Guver10}.

{\it Hydrogen mass fraction: } The bursts in 4U 1608--522 suggest an
accretion rate in of 3\%-5\% $\dot{M}_{\mathrm{Edd}}$, which suggests
H ignition~\citep{Galloway08}; the brighter bursts from this system
are of short duration, however, so it is likely that much of the
hydrogen is consumed via the HCNO cycle prior to He ignition. As with
EXO~1745--248, we use the full range $0<X<0.7$.

\subsection{4U 1820--30}
\label{s.4u1820}

{\it Normalization: } The three PRE bursts for which a normalization
was obtained give $A/(1\,\Aunit)=0.8886\pm 0.0373$, $0.9668\pm0.0339$
and $0.9040\pm 0.0200$~\citep{Guver10b}. We use the value quoted in
\citet{Guver10b}, $0.9198 \pm 0.0186$.

{\it Distance: } 4U~1820--30 is in the globular cluster NGC~6624.
\citet{Guver10b} uses a boxcar distribution from $6.8\,\mathrm{kpc}$
to $9.6\,\mathrm{kpc}$, to reflect two distance measurements of
$(7.6\pm 0.4)\,\mathrm {kpc} $~\citep{Kuulkers03} and $(8.4\pm
0.6)\,\mathrm{kpc}$~\citep{Valenti2007Near-Infrared-P}. We employ a
Gaussian distribution centered at $8.2\,\mathrm{kpc}$ with $\sigma_D =
0.7\,\mathrm{kpc}$ since the 95\% confidence regions for this Gaussian
is the same as the range suggested by the boxcar in~\citet{Guver10b}.
A more recent determination~\citep{Dotter2010The-ACS-Survey-} gives an
apparent distance of $10.2\,\mathrm{kpc}$, which when corrected for
extinction gives $8.1$ kpc, consistent with our value described above.

{\it Touchdown flux: } Five of the bursts have a measured touchdown
flux: $\Ftdi/(10^{-8}\,\fluxunit) = 5.33\pm0.27$, $5.65\pm0.20$,
$5.12\pm0.15$, $5.24\pm0.19$, and $5.42\pm0.16$~\citep{Guver10b}. The
fit in~\citet{Guver10b} gives $(5.31 \pm 0.10)\times
10^{-8}\,\fluxunit$, which we use here.

{\it Hydrogen mass fraction: } This object is likely an ultra-compact
binary with a H-poor donor~\citep{King86,Stella87}. Although
evolutionary models do not exclude the possibility that the envelope
could contain some H ($X\lesssim 0.1$;
\citealt{podsiadlowski.ea:evolutionary}), a comparison of burst
properties with theoretical ignition models suggests H-poor
fuel~\citep{Cumming03}. We fix $X=0$ for this source.

\end{document}